\documentclass[journal]{IEEEtran}
\pdfoutput=1
\usepackage{graphicx}
\usepackage{subfigure}
\usepackage{float}
\usepackage{multirow}
\usepackage{epstopdf}
\usepackage{amsmath}
\usepackage{cite}
\usepackage{caption}
\usepackage{xcolor}
\usepackage{url}
\usepackage{hyperref}
\usepackage{array}
\usepackage{tabularx}

\begin{document}

\title{Viewport Adaptation-Based Immersive Video Streaming: Perceptual Modeling and Applications}

\author{Shaowei Xie, Qiu Shen, Yiling Xu, Qiaojian Qian, Shaowei Wang, Zhan Ma, and Wenjun Zhang}



\maketitle

\begin{abstract}
Immersive video offers the freedom to navigate inside virtualized environment. Instead of streaming the bulky immersive videos entirely, a viewport (also referred to as field of view, FoV) adaptive streaming is preferred. We often stream the high-quality content within current viewport, while reducing the quality of representation elsewhere to save the network bandwidth consumption. Consider that we could refine the quality when
focusing on a new FoV, in this paper, we model the perceptual impact of the quality variations (through adapting the quantization stepsize and spatial resolution) with respect to the refinement duration, and yield a product of two closed-form
exponential functions that well explain the joint quantization and resolution induced quality impact. Analytical model is cross-validated
using another set of data, where both Pearson and Spearman's rank correlation coefficients are close to 0.98. Our work is devised to optimize the adaptive FoV streaming of the immersive video under limited network resource. Numerical results show that our proposed model significantly improves the quality of experience  of users, with about 9.36\% BD-Rate (Bjontegaard Delta Rate) improvement on average as compared to other representative methods, particularly under the limited bandwidth.
\end{abstract}


\IEEEpeerreviewmaketitle

\section{Introduction} \label{sec:introduction}
The vivid world projected on our human visual retina can be represented using the immersive video which offers the ultra high spatial resolution (i.e., gigapixel~\cite{david_giga}), panoramic viewing range and flexible focus depth. For a typical 30 frames per second (FPS) High-Definition (HD) video at $1920\times1080$ (1080p) spatial resolution, Netflix suggests a 5 Mbps ($\approx$150x compression ratio using the well-known H.264/AVC~\cite{H264}) connection speed for the broadcasting quality. Let us assume an immersive video at 32K$\times$16K spatial resolution (to cover the panoramic sphere scene), 120 FPS, and 25 different focus depths, it requires more than 10 Gbps stably from the underlying network to sustain the high quality delivery for a single connection. This is indeed unbearable, even for the emerging 5G communication networks. Besides, immersive video also demands an ultra-low latency for interaction, which again imposes quite stringent requirements for the underlying network due to a great amount of data exchanged when delivering entire immersive videos at high quality.

\begin{figure*}[!t]
 \centering
 \includegraphics[width=0.95\linewidth]{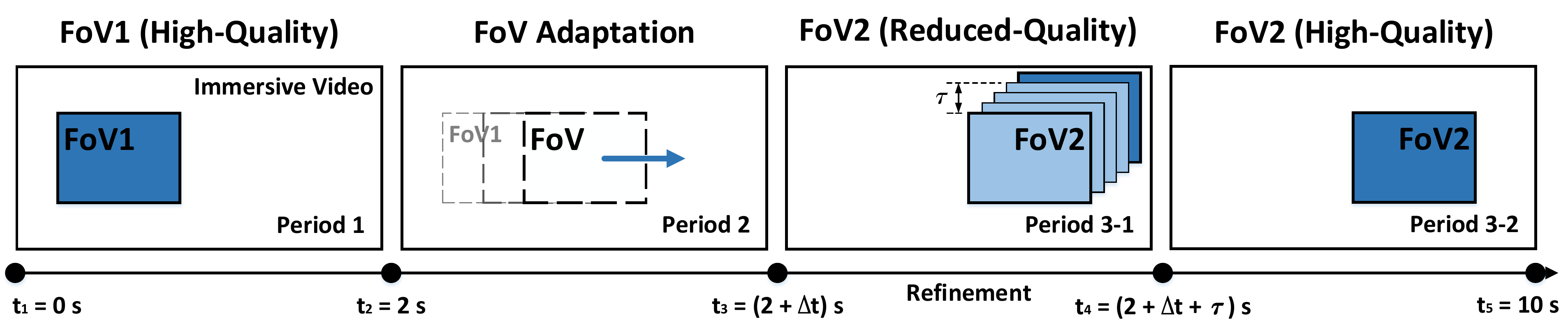}
 \caption{Illustration of the viewport adaptation for immersive video streaming. Different color shades represent various quality levels of current FoV: the darker blue means the original high-quality version, while the lighter one for reduced-quality copy. $\Delta t$ is the time consumed for navigation from FoV1 to FoV2, and $\tau$ is the refinement duration of the content from its reduced-quality copy to the original high-quality one. The total length of each PVS is 10 seconds.}
 \label{ViewportAdaptation}
\end{figure*}

Leveraging the characteristics of the human visual system (HVS) where our human being only can perceive the event
in the front with about binocular 220$^\circ$ viewing range horizontally in reality~\cite{jov_peri_vision_review}, we then could apply the adaptive viewport or field of view (FoV)\footnote{We use viewport or FoV interchangeably throughout this work.}
streaming instead of delivering the bulky immersive video entirely~\cite{DynamicVR,OptimizingCell,Two-tier360,HEVCsr}. Furthermore, user often navigate inside the virtualized environment, resulting in the FoV movement from time to time. To avoid the blackout caused by switching the FoV suddenly, a popular strategy is setting the content within current FoV at the highest quality (and the largest bit rate) but reduced quality (less bit rate) elsewhere \cite{Two-tier360,Viewportadaptive,AdaptiveStreaming,UltraWideView}, as the {\it period \#1} shown in Fig.~\ref{ViewportAdaptation}. This allows the user to immediately perceive the scene content when adapting his/her FoV. Often, we refer the content within current FoV to as ``effective data" and those outside of current FoV to as ``redundant data".

Since we apply unequal quality levels for different content blocks (e.g., inside and outside of current FoV through different compression or quality scales), it typically involves the refinement from a reduced quality version to a corresponding high-quality one when users navigate their focus to a new FoV, as shown in {\it period \#2 and \#3} of Fig.~\ref{ViewportAdaptation}. Ideally, it demands the seamless adaptation without perceiving any noticeable quality degradation and impairing the user experience, which intuitively depends on the gap between quality scales (i.e., how much is the difference between the reduced quality version and corresponding the highest quality for the same content) and the refinement duration $\tau$ (i.e., how fast does the refinement last). This urgently calls for a quality model to quantify the perceptual impact of the quality variations between consecutive FoVs, with respect to the $\tau$. Therefore, in this paper, we attempt to model this perceptual quality using the mean opinion score - MOS, and to the best of our knowledge, this shall be the first work discussing the subjective quality impacts for adaptive FoV streaming of immersive video.

The quality of compressed image/video content block is usually determined by the associated quantization stepsize $q$ (or equivalent quantization parameter QP, $q=2^{\frac{{\rm QP}-4}{6}}$\cite{QuantizationScheme}) and spatial resolution $s$, independently or jointly~\cite{pv_mobileQSTAR,Ma_RQModel,VQMTQ}. For simplicity, $q$ (or $s$) induced quality variation is referred to as the $q$-impact (or $s$-impact). Strictly speaking, the joint impacts of quantization and spatial resolution on the perceptual quality is generally not separable~\cite{pv_mobileQSTAR,VQMTQ}. However, we still enforce the separable effects to simplify the model derivation. This is also very helpful for future application driven implementation.
With extensive subjective assessments, the overall perceptual model can be represented using a product of two exponential functions that well explain the $q$ or $s$-induced quality impact with respect to the $\tau$, respectively. Model parameters are $q$ and $s$ dependent and can be accurately represented by a set of fixed closed-form functions.

To validate the accuracy of our proposed model, another set of data is chosen randomly. Note that in the phase of the cross-validation, we compare the measured and predicted MOS, where the measurements are performed through the independent subjective quality assessments, and
the predictions are calculated through our proposed model. The results have shown that both the Pearson correlation coefficient (PCC) and Spearman's rank correlation coefficient (SRCC) are close to 0.98.

We then devise this model to guide the bandwidth constrained immersive video streaming. It can be formulated as an
optimization problem to maximize the subjective quality under the rate constraint. Three scenarios are discussed assuming
both continuous $s$ and $q$,  few typical $s$ discretely but still continuous $q$ and finally both discrete $s$ and $q$ for a practical
implementation. With our developed model, the optimization problem is analytically
tractable, yielding the significant improvement of the subjective quality when compared with the heuristic method, particularly for the low bitrate
scenario.

The rest of this paper proceeds as follows: In section~\ref{sec:related work}, the state-of-the-art researches in the literature are discussed, including
the FoV dependent streaming as well as the FoV prediction. In section~\ref{sec:perceptual_modeling}, we explain how to measure the perceptual impact of immersive videos when adapting user's FoV, and propose the analytical models to quantify the $q$ and $s$ induced quality impacts with respect to $\tau$, as well as the independent model cross-validation. In Section~\ref{sec:application}, the quality-bandwidth optimized streaming application is introduced, and our model is applied to derive the optimal solution analytically. Finally, we draw a conclusion in Section~\ref{sec:conclusion}.

\begin{figure*}[t]
\centering
 \subfigure[KiteFlite*\dag]{ \includegraphics[width=1.3in, height=0.65in]{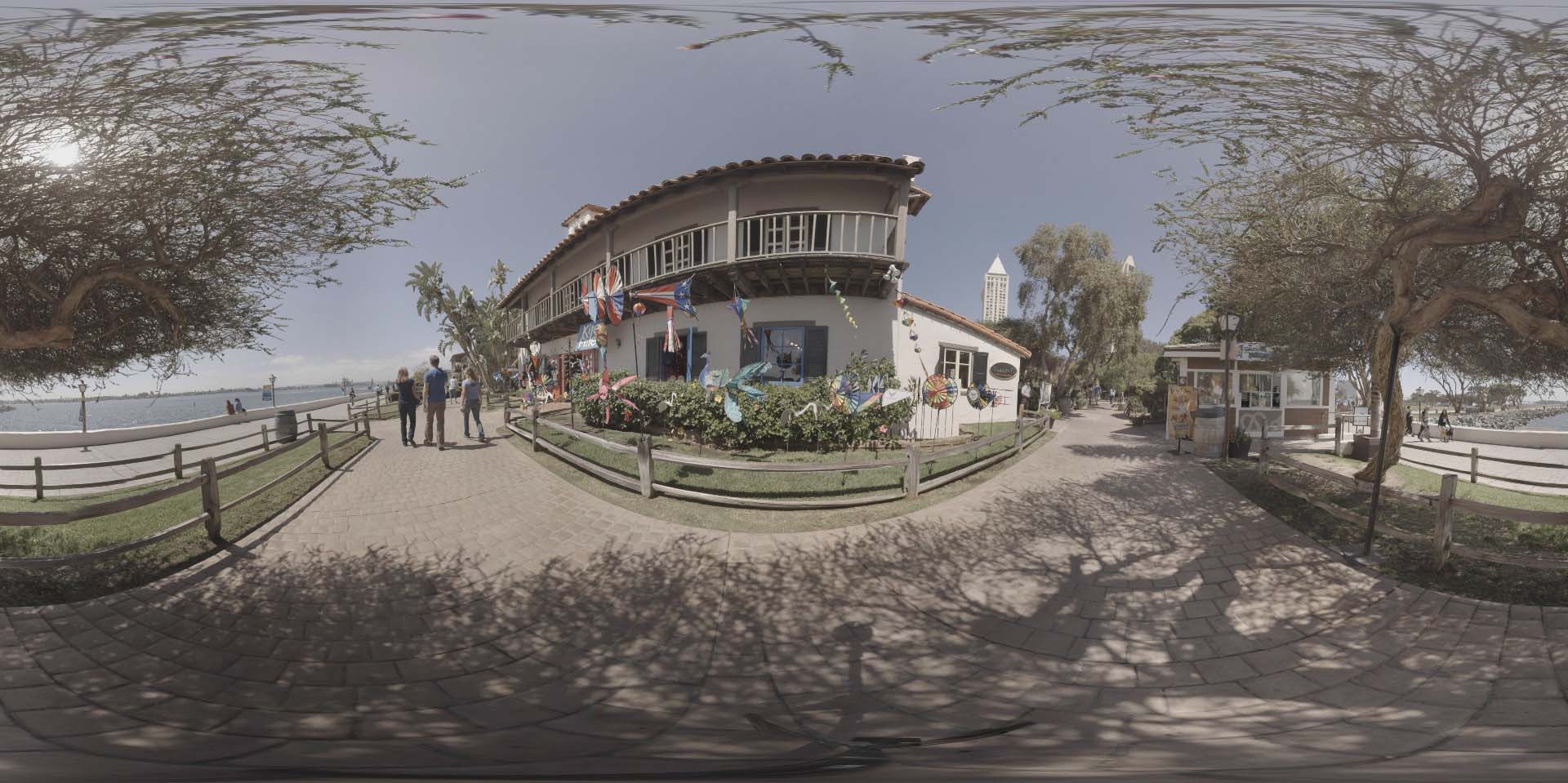}}
 \subfigure[AerialCity*]{ \includegraphics[width=1.3in, height=0.65in]{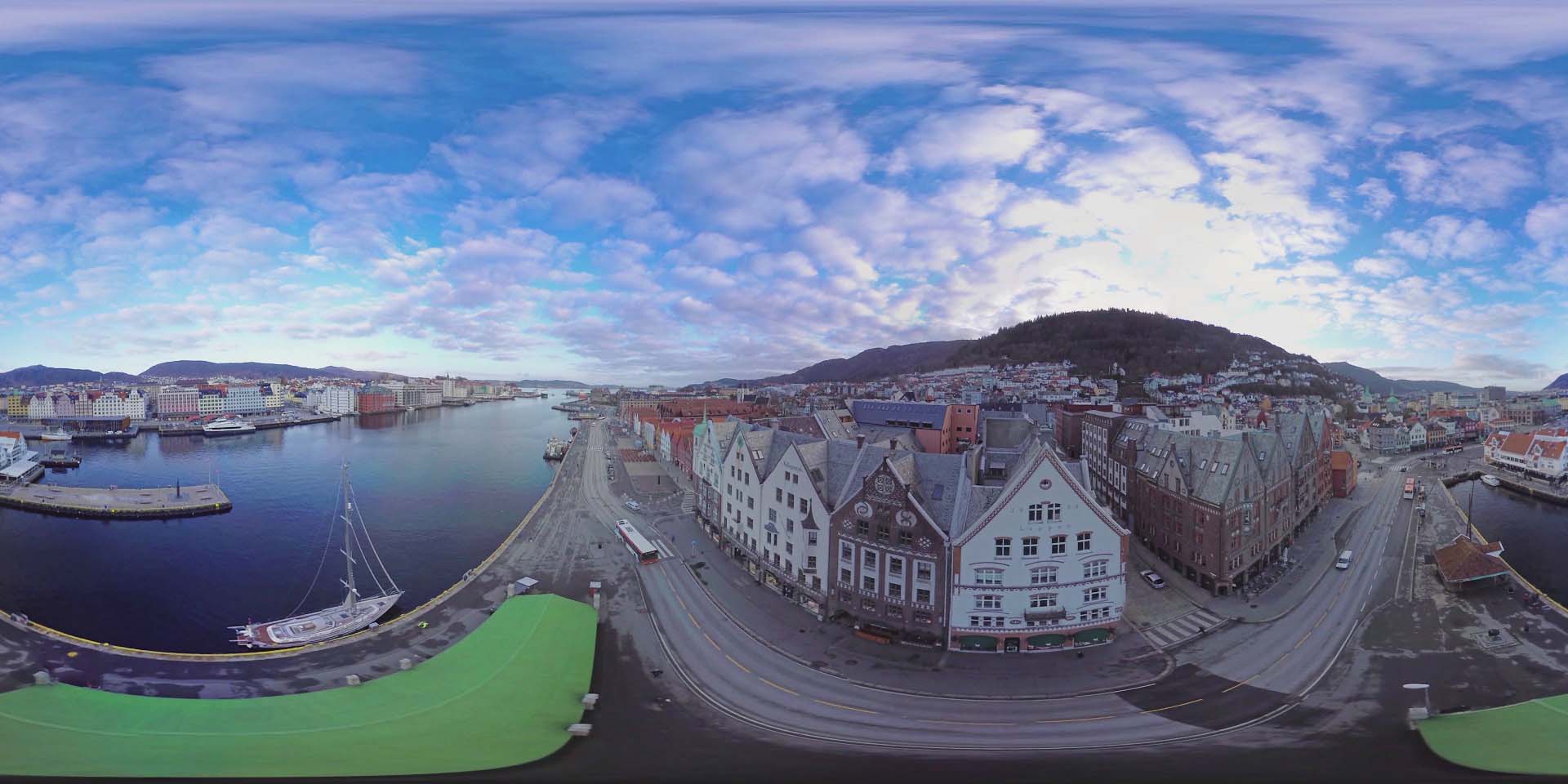}}
 \subfigure[Gaslamp*]{ \includegraphics[width=1.3in, height=0.65in]{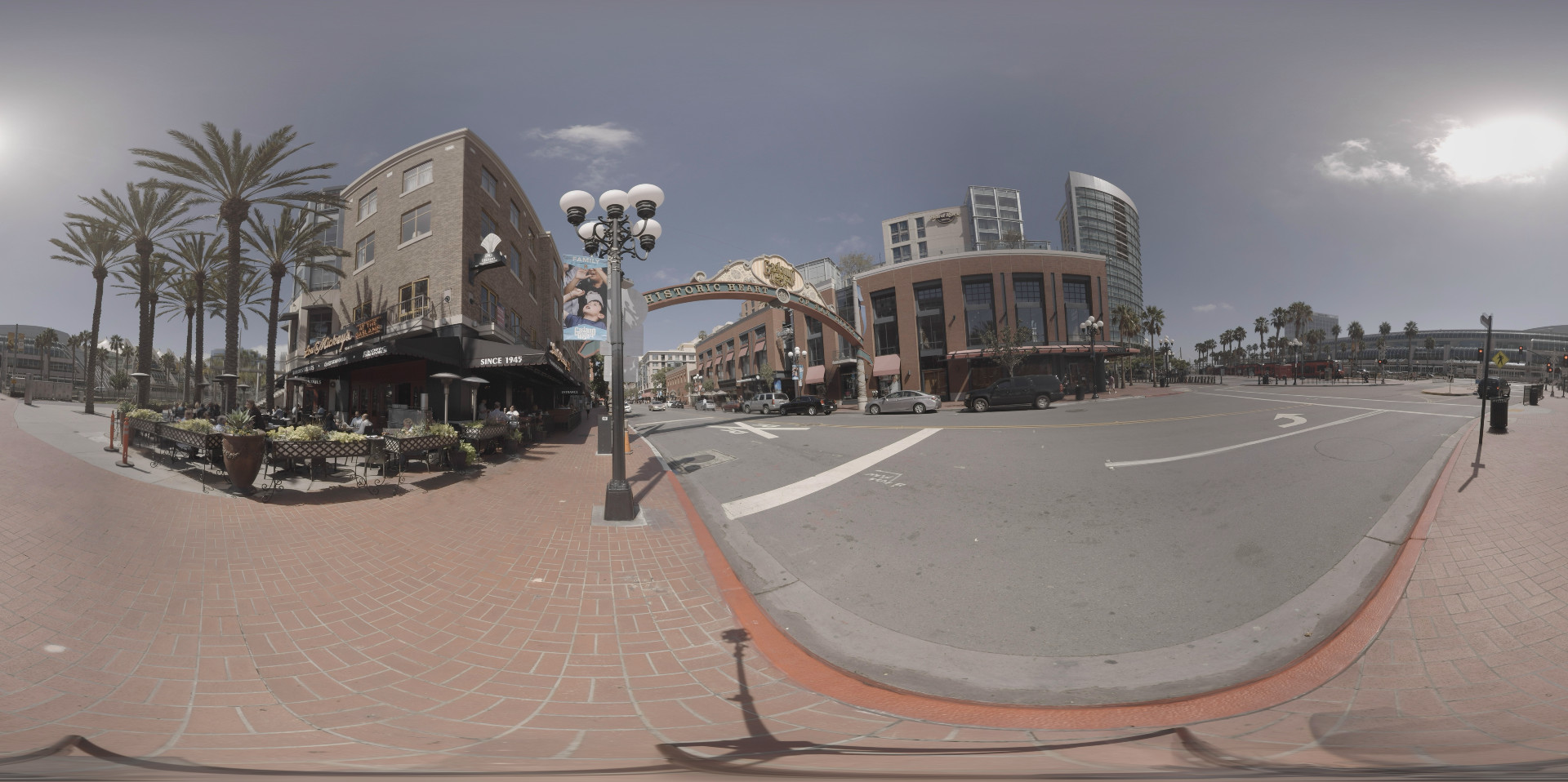}}
 \subfigure[Harbor*]{ \includegraphics[width=1.3in, height=0.65in]{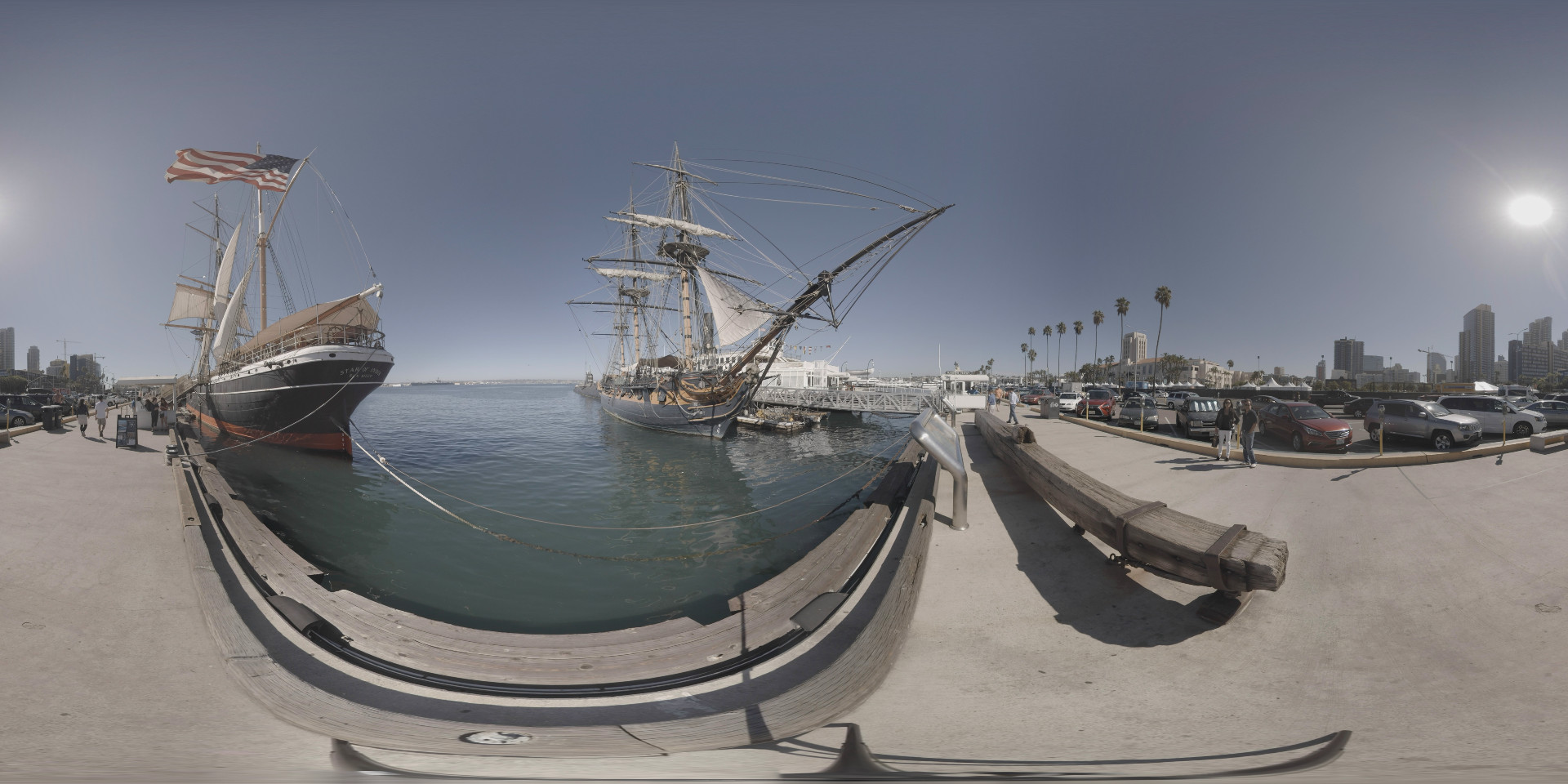}}
 \subfigure[Trolley*]{ \includegraphics[width=1.3in, height=0.65in]{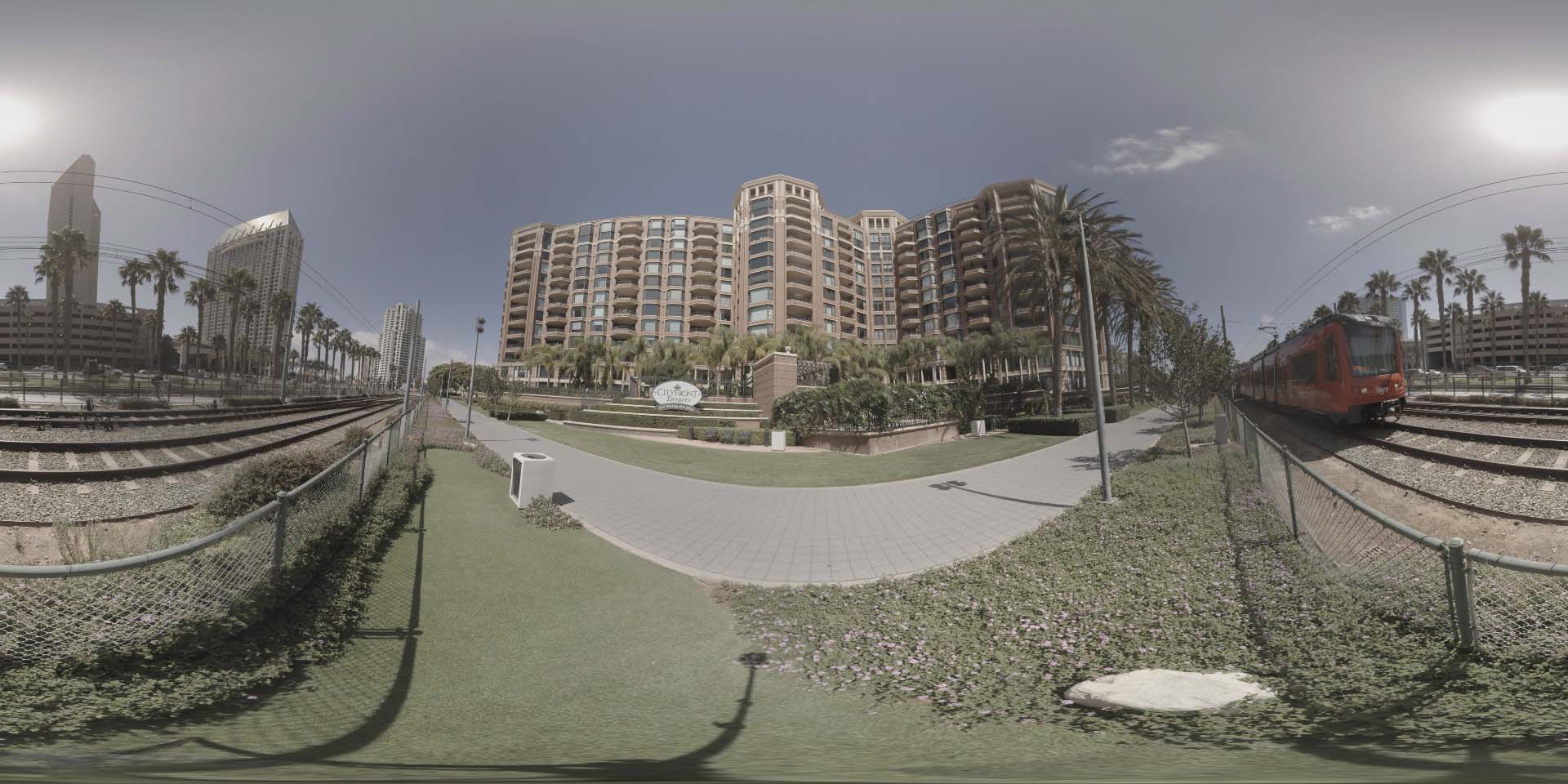}}
 \subfigure[Elephants]{ \includegraphics[width=1.3in, height=0.65in]{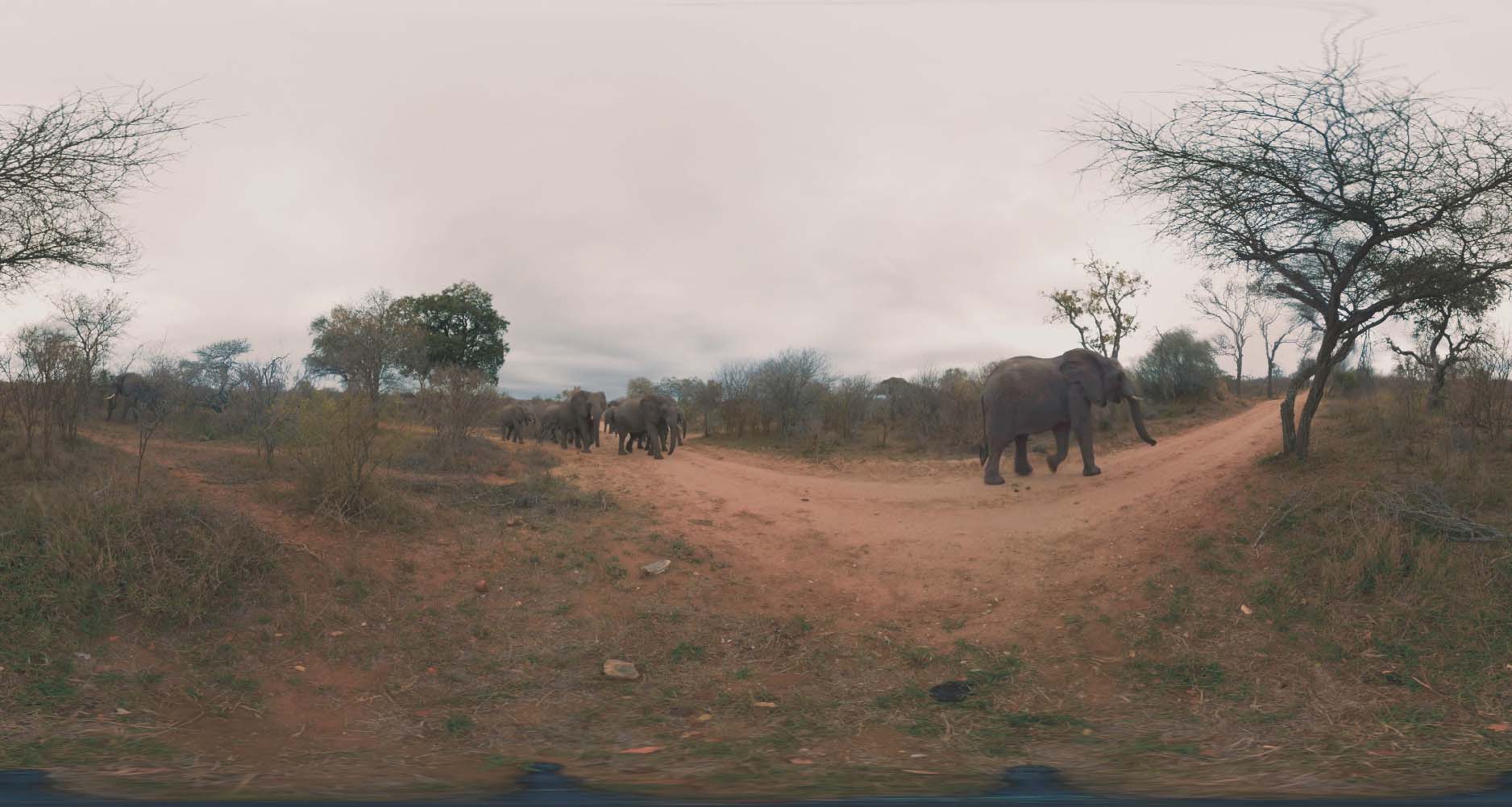}}
 \subfigure[Rhinos]{ \includegraphics[width=1.3in, height=0.65in]{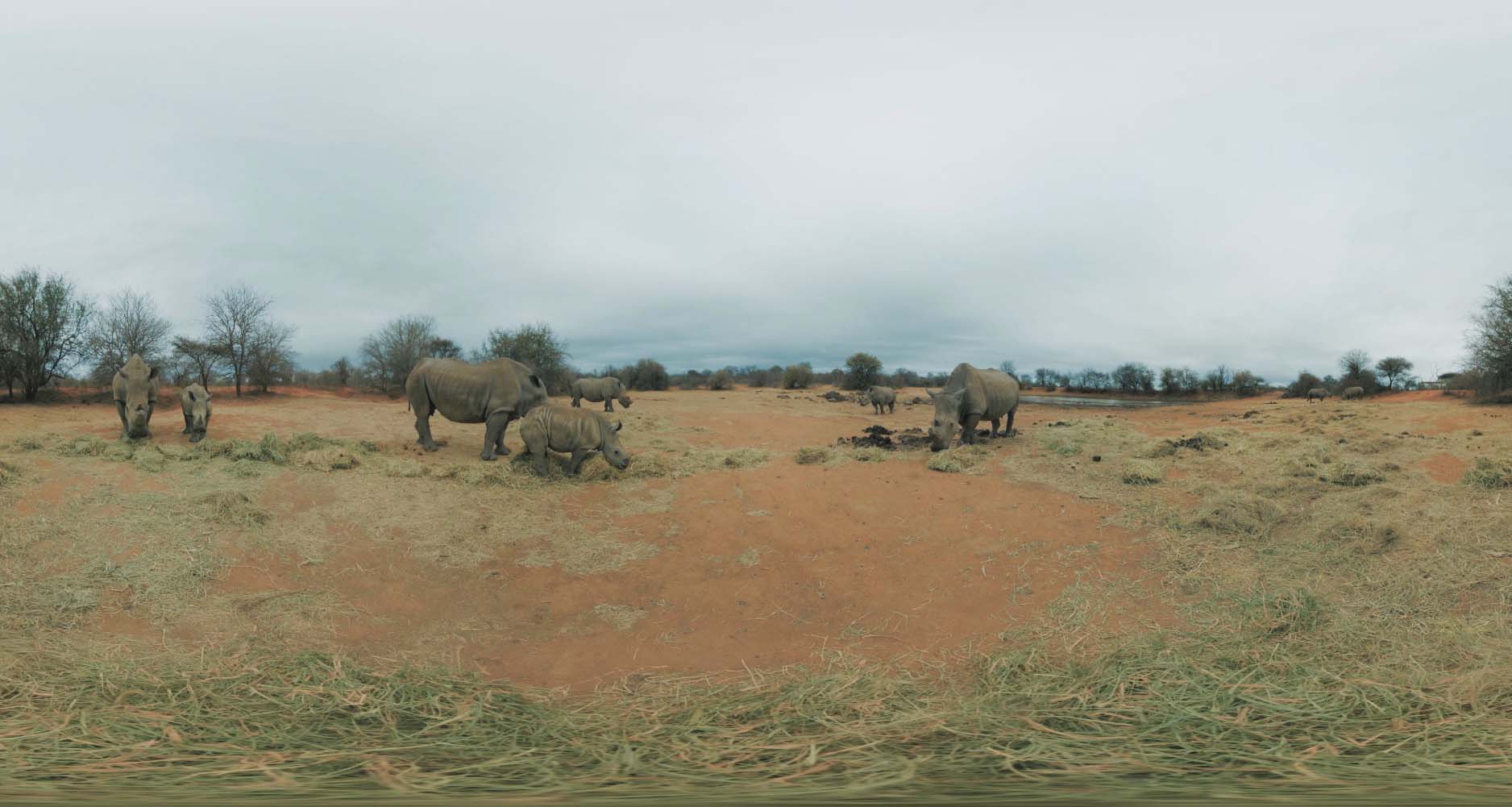}}
 \subfigure[Diving]{ \includegraphics[width=1.3in, height=0.65in]{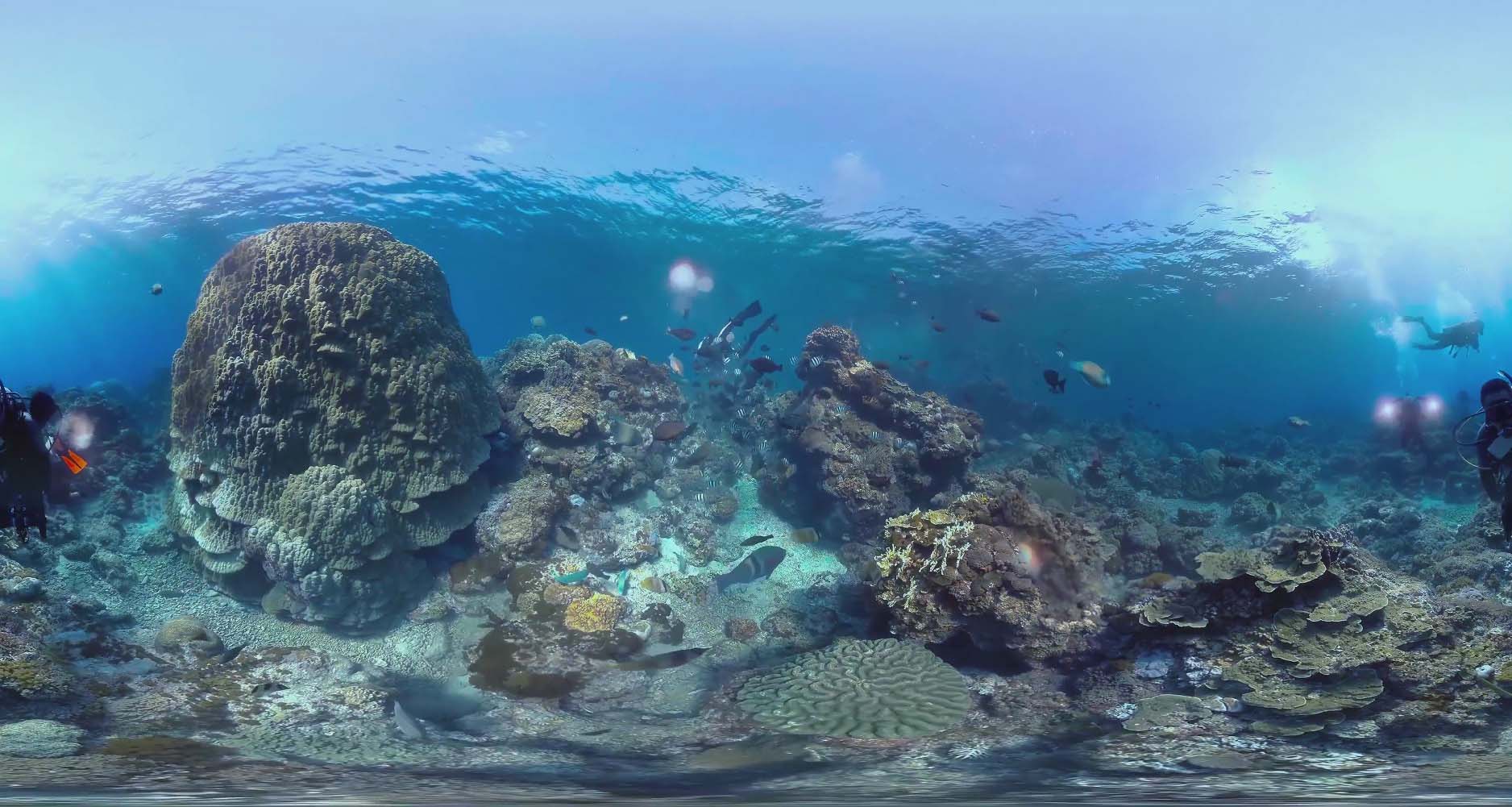}}
 \subfigure[Venice]{ \includegraphics[width=1.3in, height=0.65in]{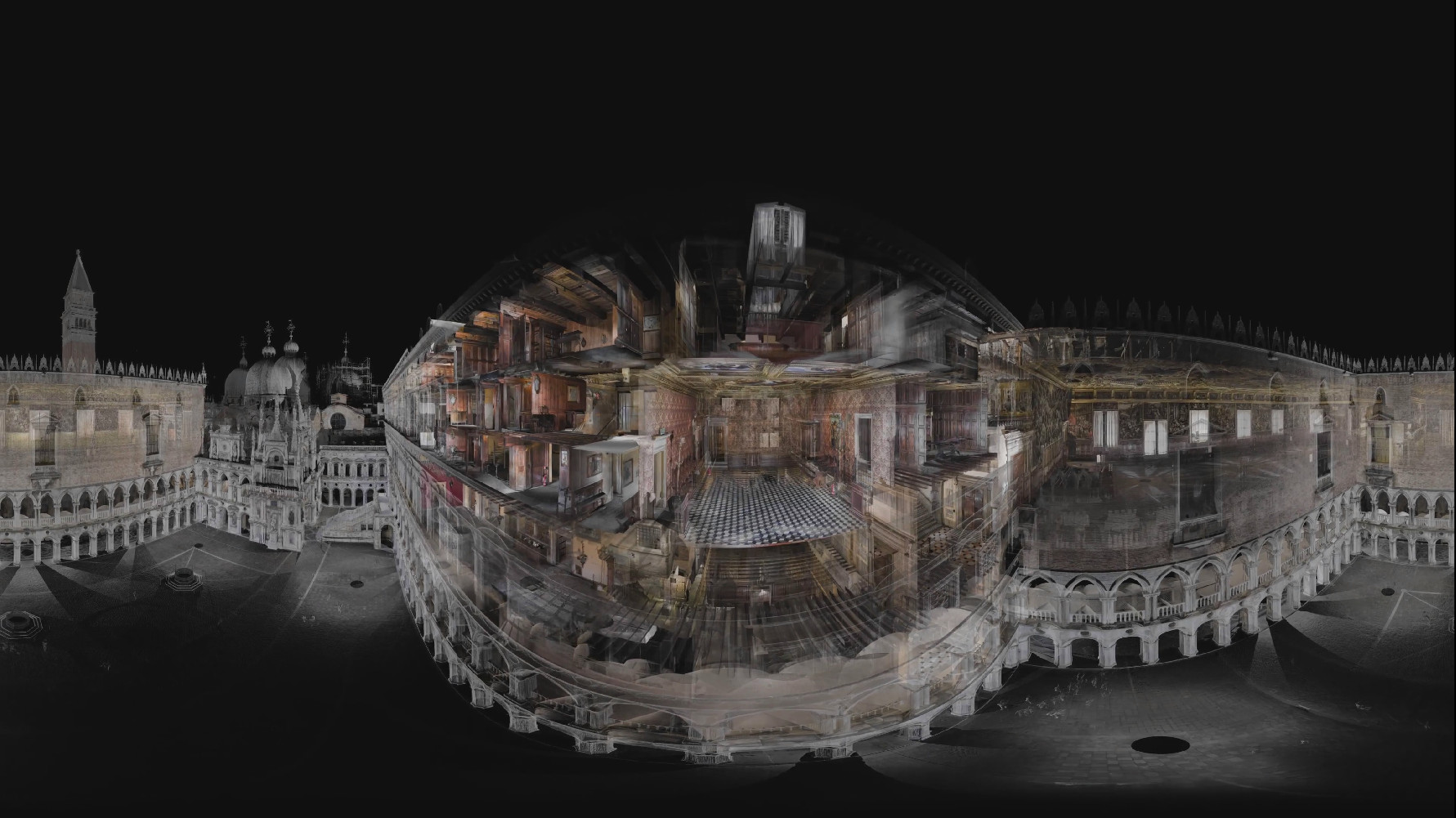}}
\caption{Illustration of sample frames of selected immersive videos. ``KiteFlite" (marked with \dag) is used to train the participants to familiarize themselves in this subjective assessment, and the rest eight videos are used for perceptual tests.}
\label{ImmersiveImages}
\end{figure*}

\section{Related Work} \label{sec:related work}

The viewport or FoV adaptive streaming of the immersive video has been extensively studied recently.
The basic idea is the way of transmitting only the visible content covering user's FoV in the front by partitioning a panoramic
video scene into multiple independent and non-overlapped spatial tiles~\cite{RoIcrop,SpatialSeg}. Nevertheless, in order to
avoid the blackout when performing the FoV navigation, it is indispensable to deliver redundant reduced-quality tiles outside of
current high-quality FoV~\cite{InteractiveOmni}. Alternatively, to simplify the system implementation, Ochi, Duanmu, and {\it et al.}
have proposed to deliver a high quality FoV tile as well as a complete panoramic video with reduced quality~\cite{LiveOmni,TwoTierSystem}.

We could further only stream the data of next predicted FoV, instead of delivering all other reduced-quality content alongside current FoV.
Intuitively, the more accurate prediction, the less data exchange.  Typically, user's movement (i.e., head, eye, and body) would impose
a strong temporal correlation in a small time scale (i.e., hundreds of milliseconds to a few seconds). Thus, it is natural to predict the next FoV
via the current and historical movement statistics sensed through the meters equipped in head mounted display (HMD) or similar devices~\cite{OptimizingCell,ShootingMove}. This part could be referred to as the user behavior.

Besides, user is often attracted by the salient area in a video/image. This is also
verified in a comprehensive report by investigating the visual stimuli, head orientation and gaze direction when exploring the virtualized
environment~\cite{SaliencyinVR}. Similarly, this could be referred to as the content characteristics.
Furthermore, Fan and {\it et al.} have proposed to fuse the statistics from both the user behavior and content characteristics, and
apply the neural networks to predict the future viewing fixation \cite{FixationPrediction}.

Either way aforementioned, i.e., delivering the data belong to the next predicted FoV and
the entire tiles outside of the current FoV, it is ``redundant" and is often set at a reduced-quality scale to save the network bandwidth.
User would first perceive the reduced-quality content block when focusing and stabilizing their attention to a new FoV. This incurs the quality
refinement to restore the content from its reduced-quality version to the original high-quality one.
Thus, it is highly desired to have an analytical model to
quantify the perceptual impacts. However, to the best of our knowledge, we have not seen a systematic efforts yet on this matter.

The most relevant works on the perceptual quality modeling of the immersive image/video are our previous attempts to study the joint impacts of the quantization and spatial resolution on an entire immersive image~\cite{imageModel_rongbin,Xiaokai_TIP}, and the scenario considering
the peripheral vision impact in practice~\cite{peiyao_VCIP, peiyao_TIP}, as well as the preliminary results on the perceptual quality assessments
when performing the FoV adaption~\cite{shaowei_ISCAS}.


\section{Perceptual Modeling of the Viewport Adaptation for Immersive Video} \label{sec:perceptual_modeling}

In this section, we investigate the perceptual impacts of the viewport adaptation for immersive video application. More specifically, we evaluate the subjective opinions with various quality gaps (i.e., through different $q$ and $s$) and refinement durations ($\tau$s), and develop a closed-form analytical model to well address the joint impacts of the quantization and spatial resolution on the perceptual quality. Towards this goal, we first perform the subjective quality assessments to collect the MOSs.

\subsection{Subjective Assessment and Data Processing} \label{sec:subjective_assessment}
We have chosen nine immersive videos, as shown in Fig.~\ref{ImmersiveImages}, where five of them are 360$^\circ$ test sequences from the common test dataset
selected by the international standard organization - Joint Video Exploration Team (marked with *)~\cite{JVET}, and the rest four are from the popular YouTube 360$^\circ$ videos~\cite{YouTube}.

These experimental videos are chosen to cover different use cases and a wide range of spatio-temporal activities. In the meantime, we also ensure that the videos contain sufficient saliency regions~\cite{HMVR}, each of which could possibly belong to a distinct FoV. Usually, user's viewport adapts among these salient FoVs~\cite{SaliencyinVR}.

\begin{figure}[b]
 \centering
 \includegraphics[width=0.95\linewidth]{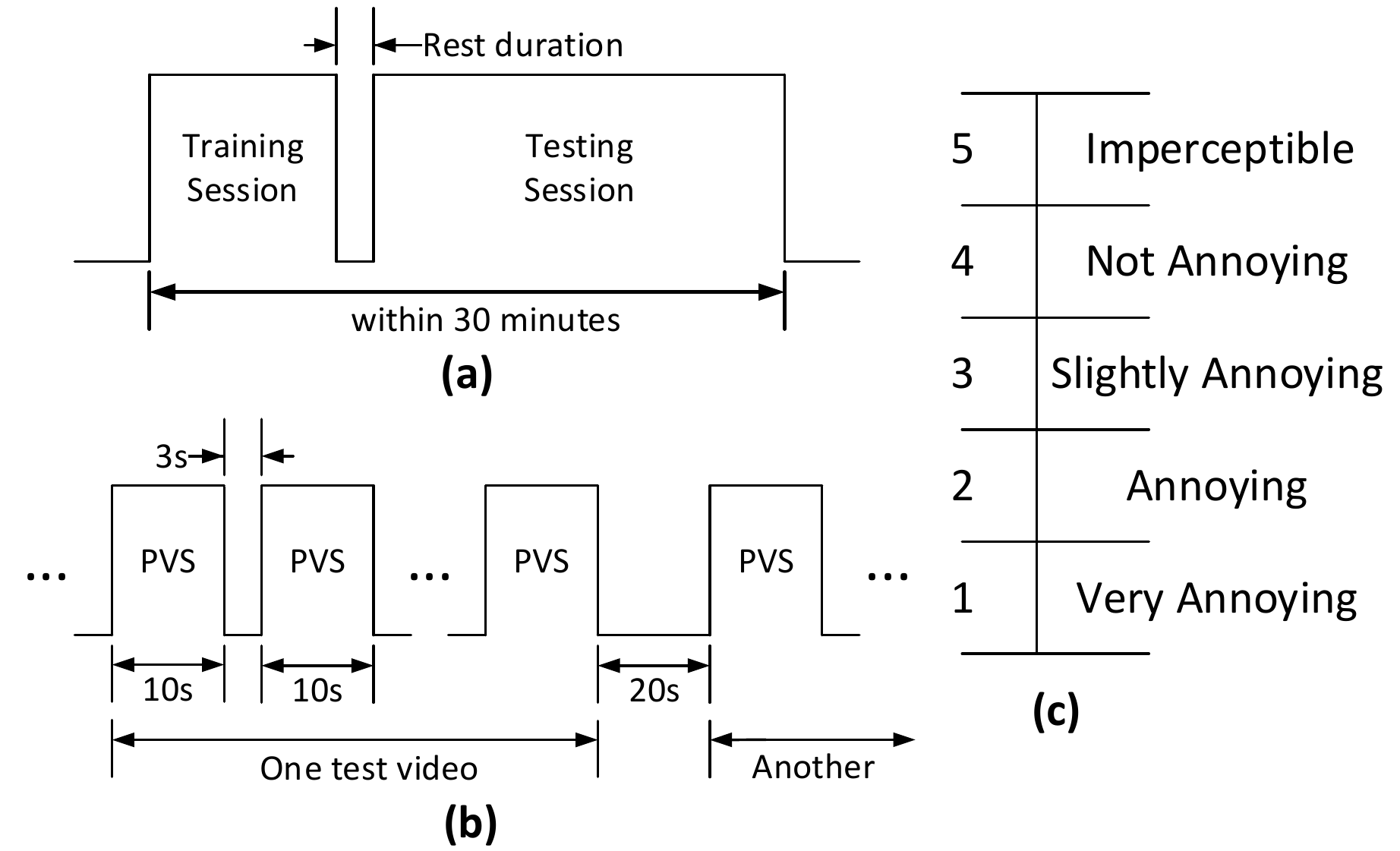}
 \caption{Illustration for subjective assessment process: (a) training and testing session, (b) consecutive rating procedure setup for each test video, (c) MOS rating scale.}
 \label{AssessmentProcedure}
\end{figure}

\begin{figure}[t]
\centering
 \subfigure[AerialCity*]{ \includegraphics[width=1.5in]{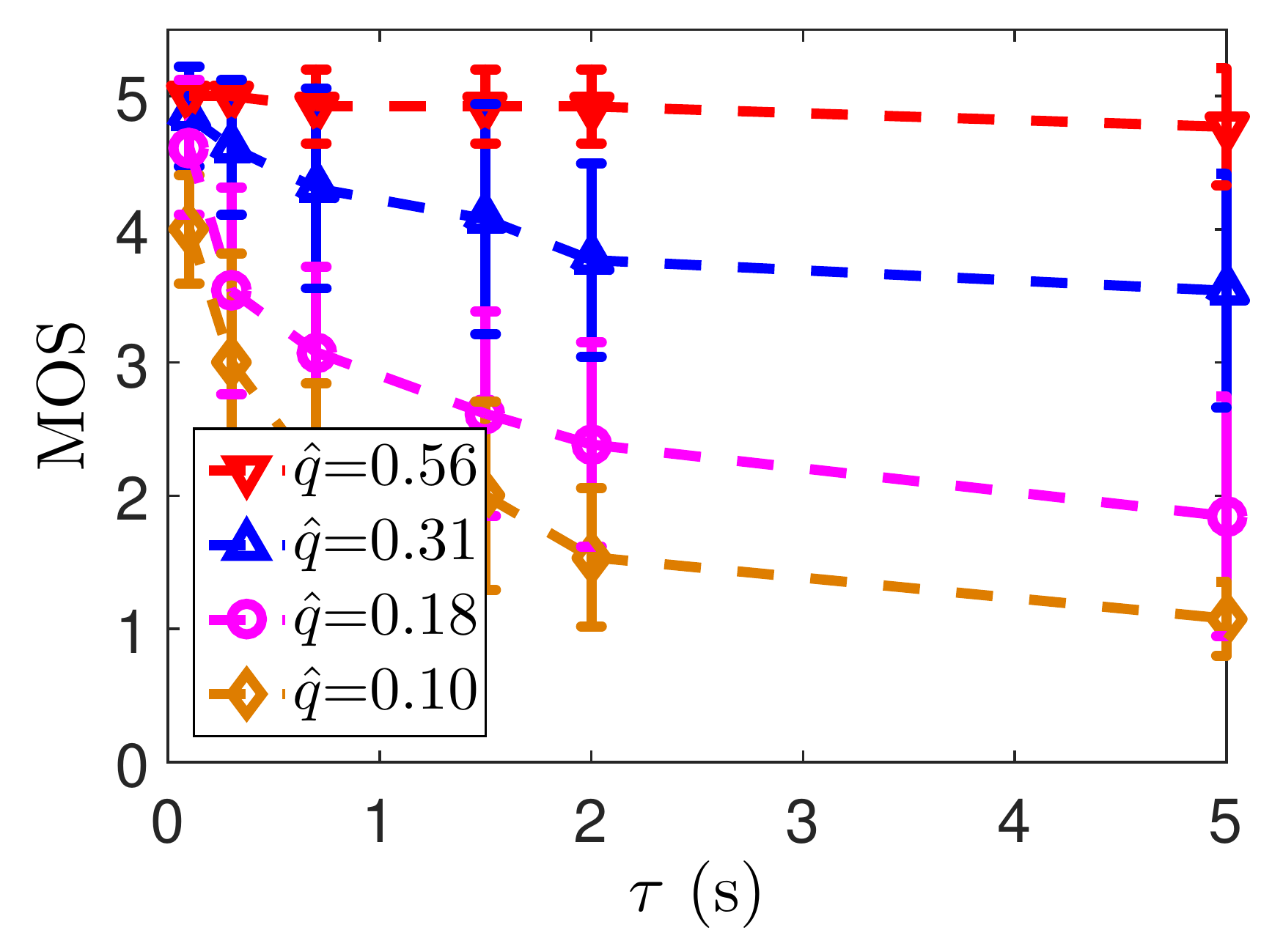}}
 \subfigure[Gaslamp*]{ \includegraphics[width=1.5in]{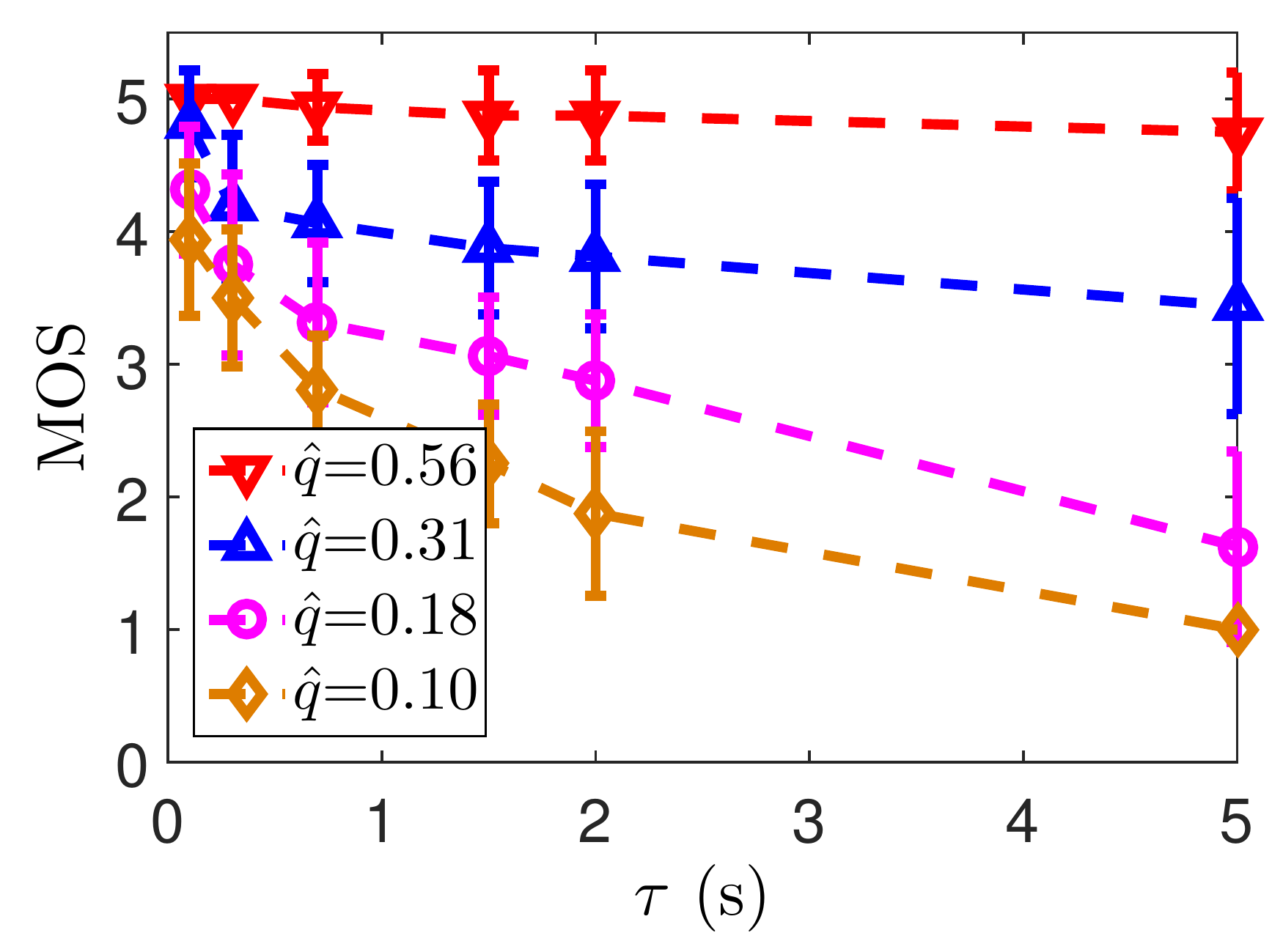}}
 \subfigure[Harbor*]{ \includegraphics[width=1.5in]{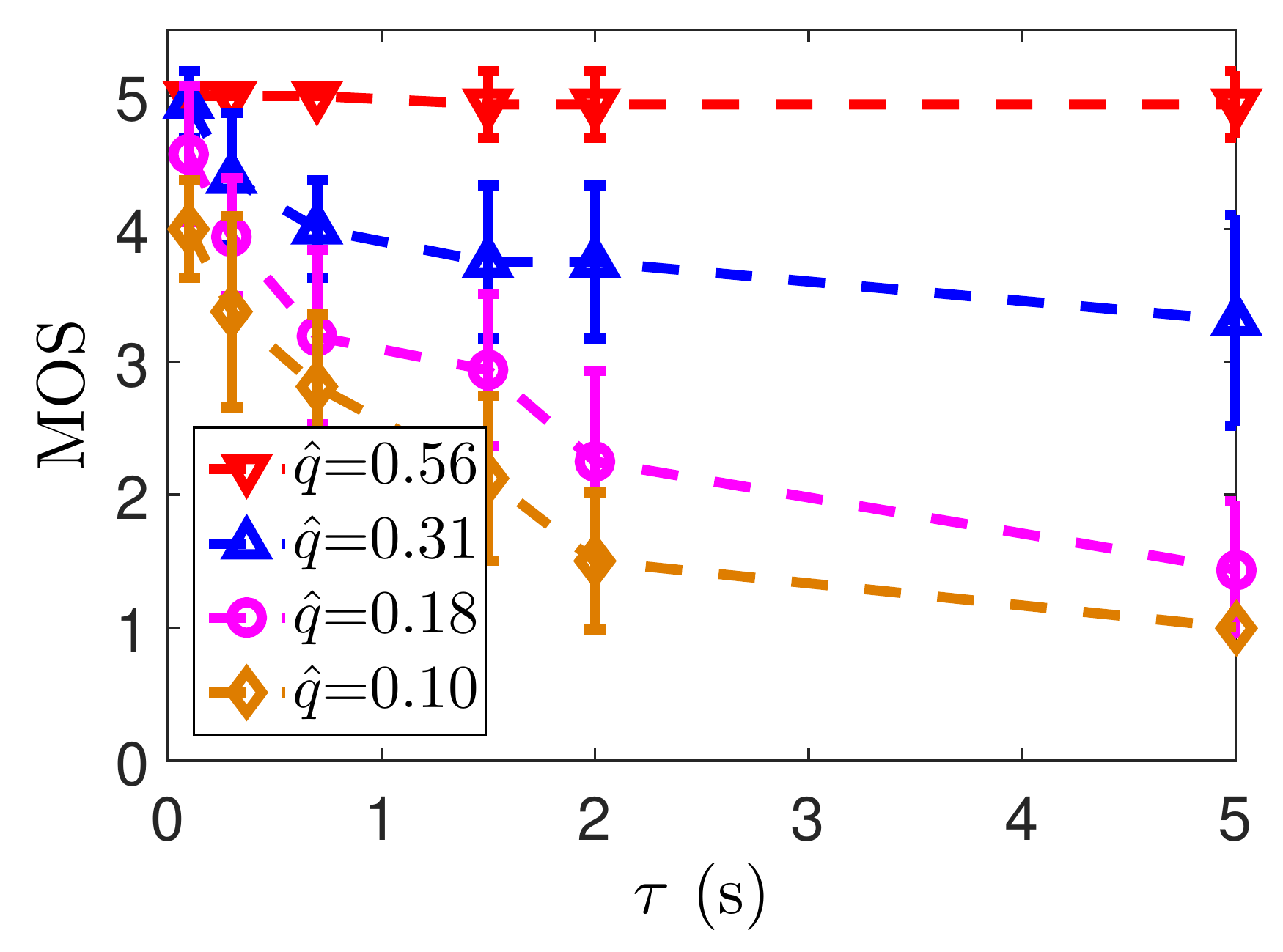}}
 \subfigure[Trolley*]{ \includegraphics[width=1.5in]{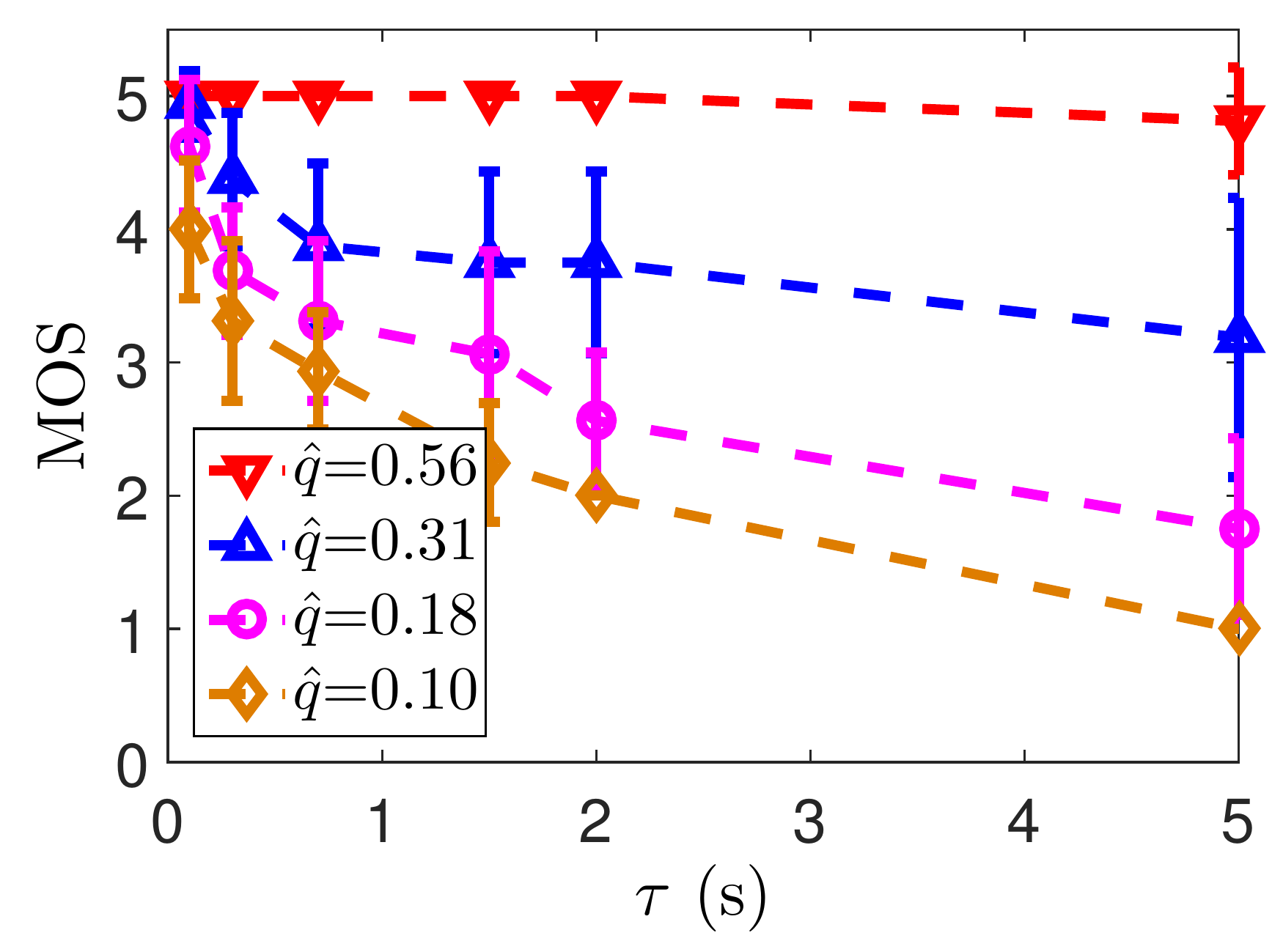}}
 \subfigure[Elephants]{ \includegraphics[width=1.5in]{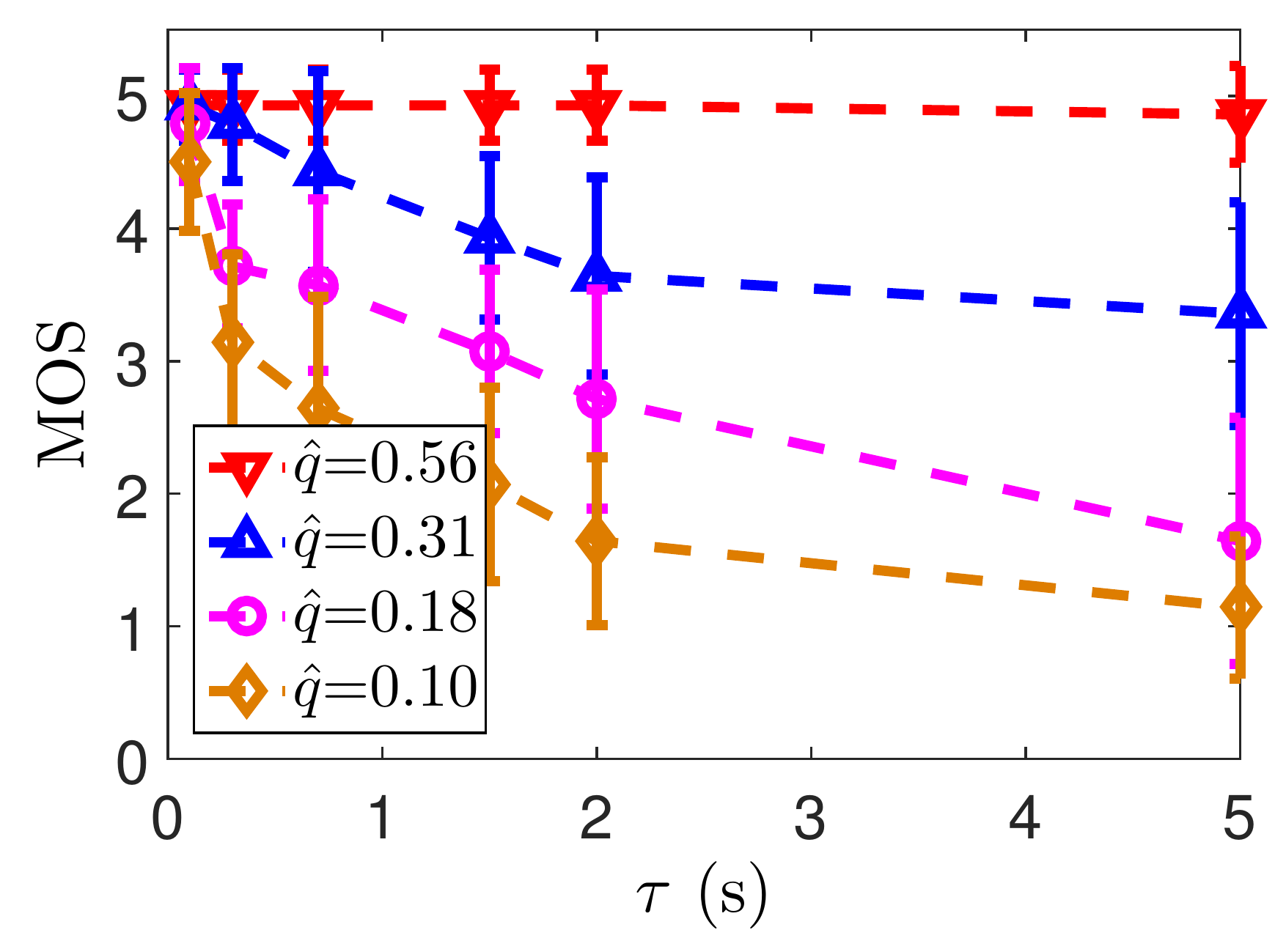}}
 \subfigure[Rhinos]{ \includegraphics[width=1.5in]{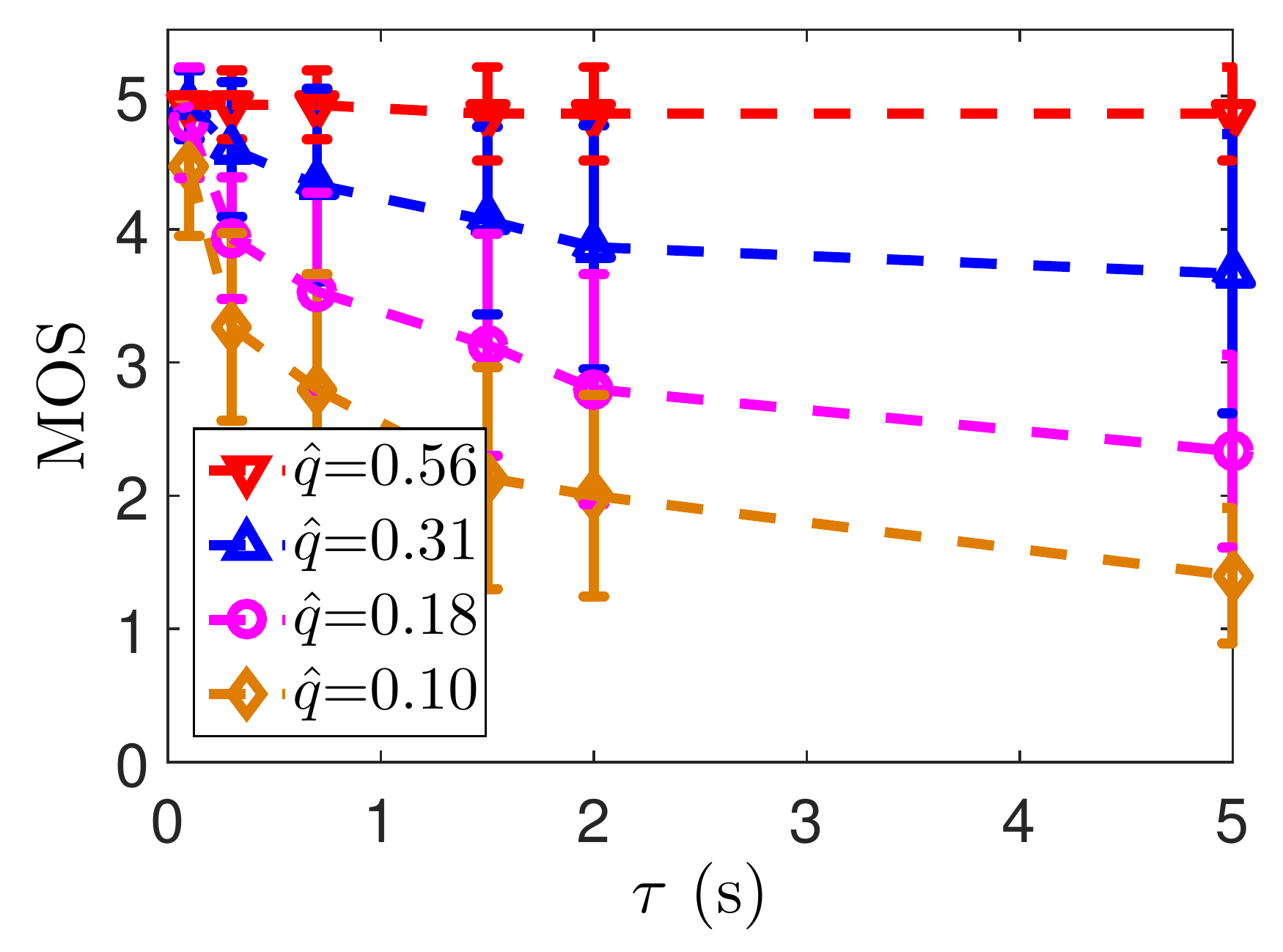}}
 \subfigure[Diving]{ \includegraphics[width=1.5in]{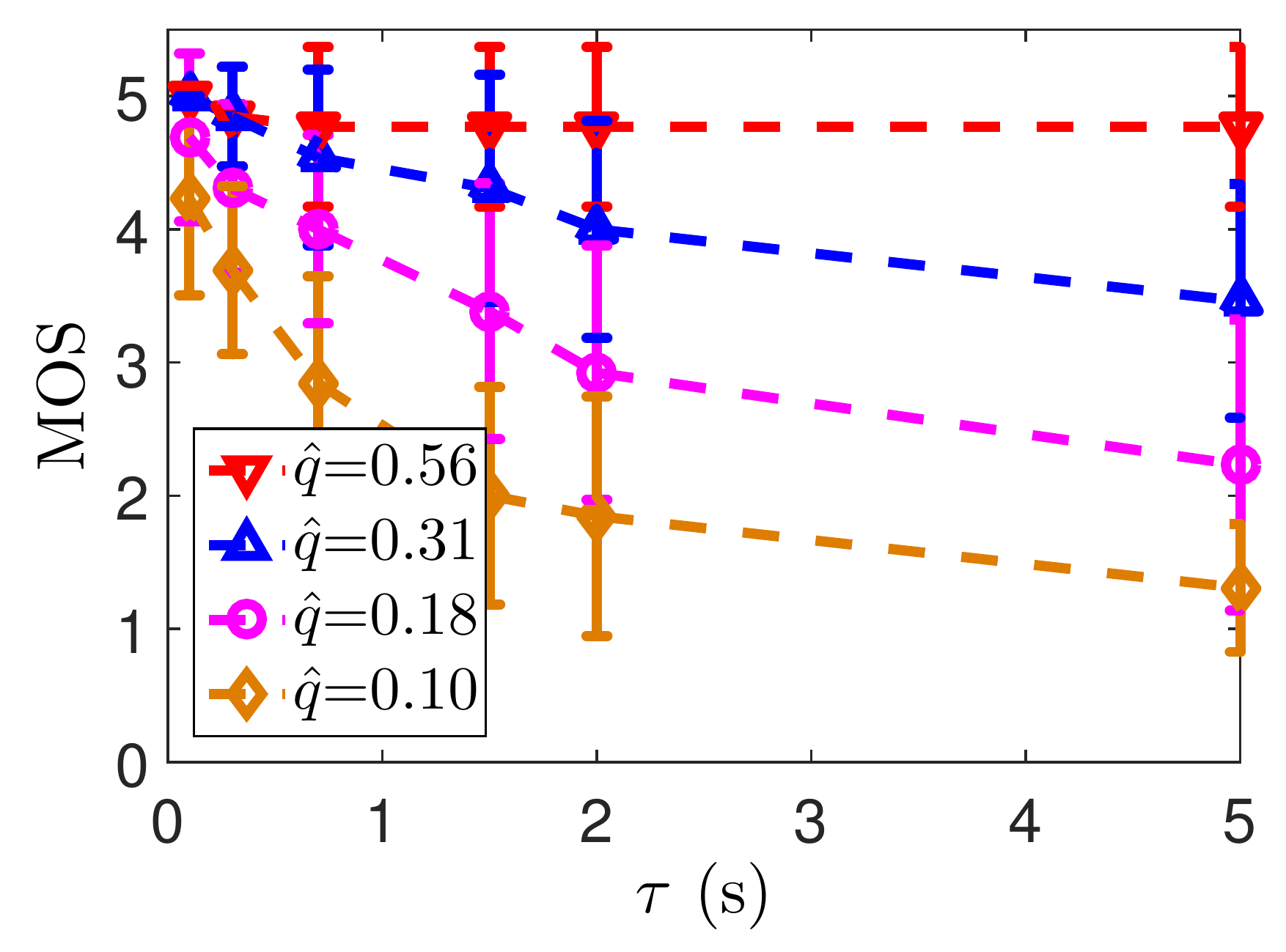}}
 \subfigure[Venice]{ \includegraphics[width=1.5in]{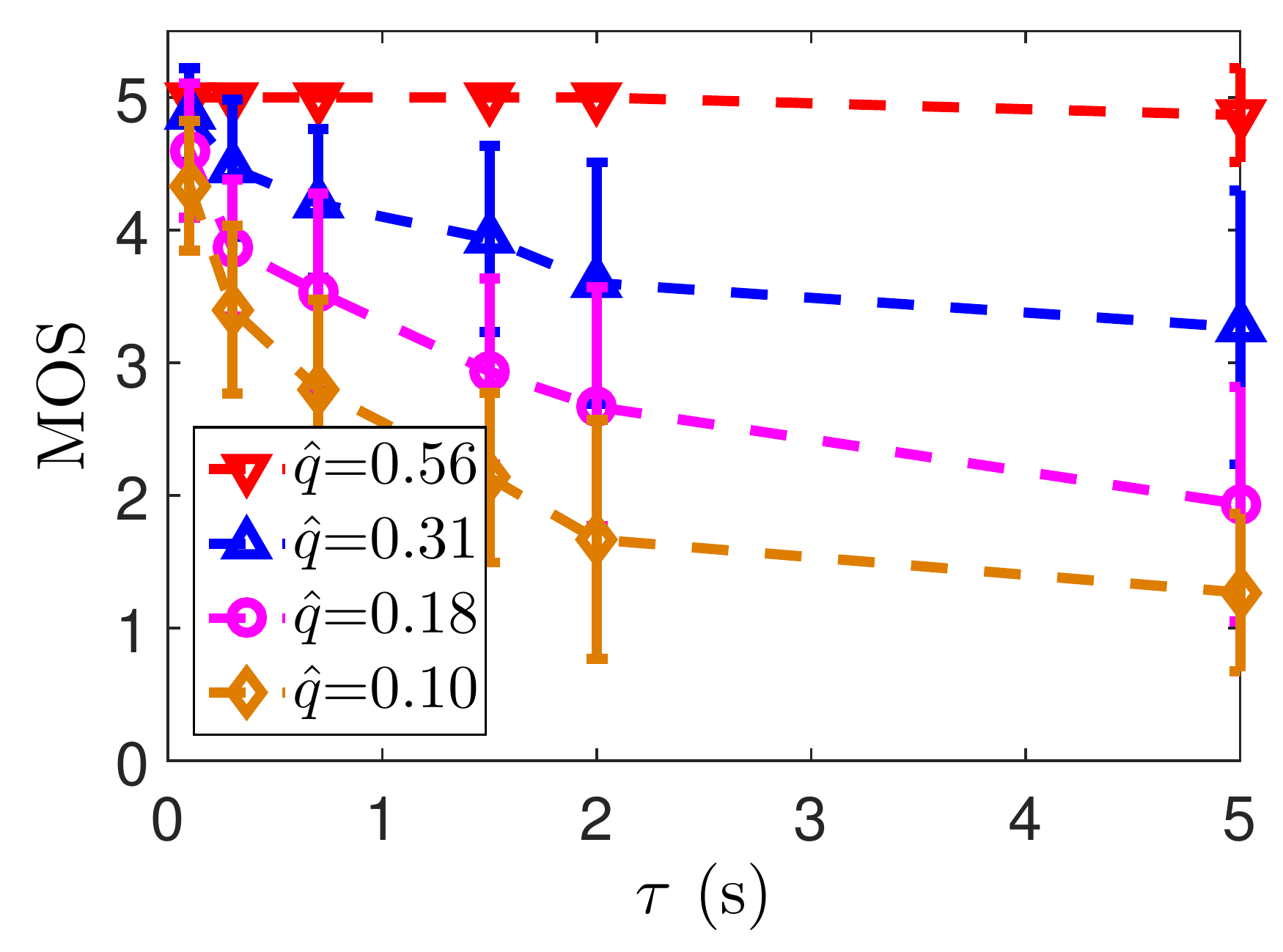}}
\caption{$q$ induced MOS variation versus refinement duration $\tau$}
\label{MOS_QP}
\end{figure}

\begin{figure}[t]
\centering
 \subfigure[AerialCity*]{ \includegraphics[width=1.5in]{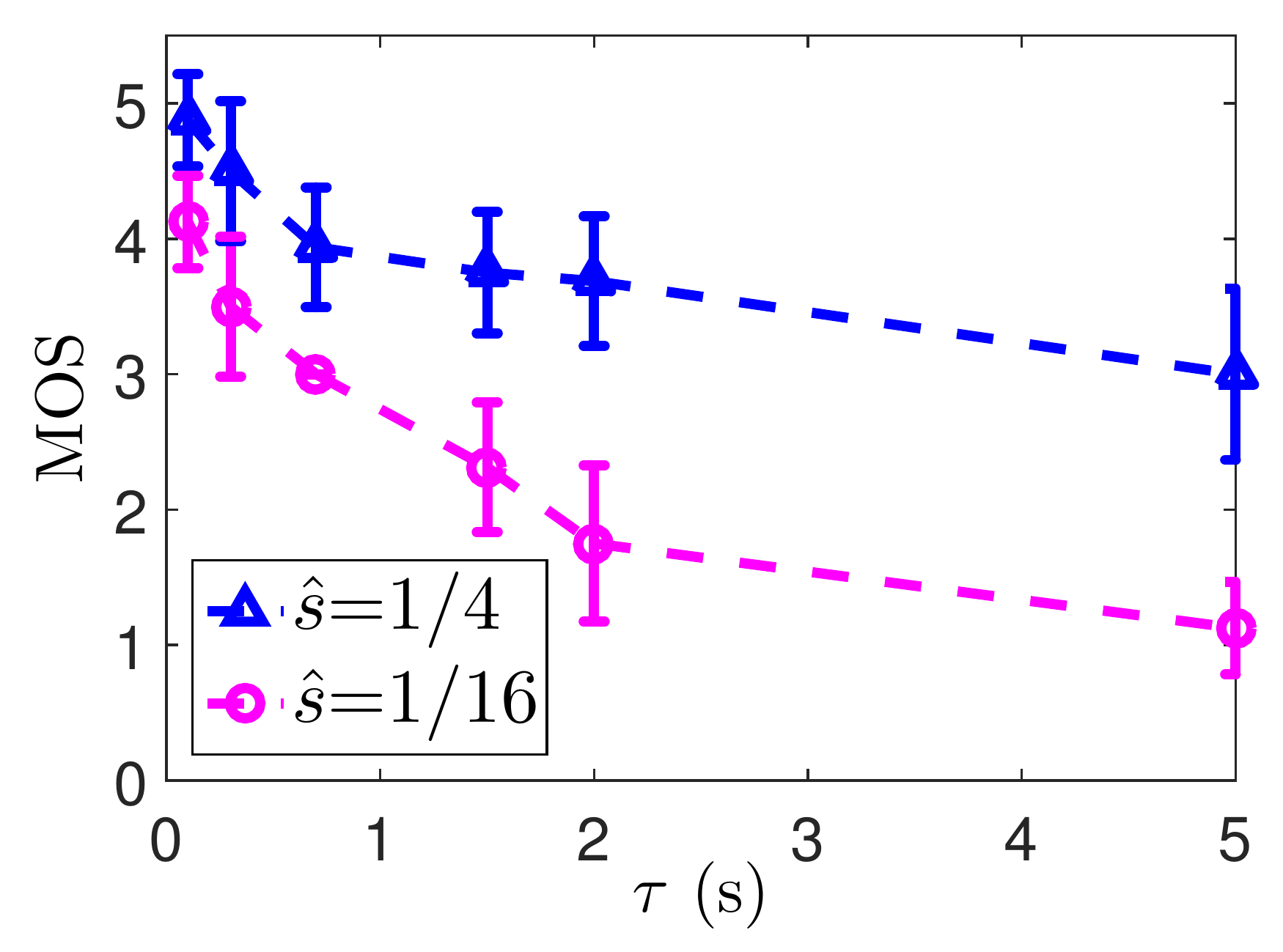}}
 \subfigure[Gaslamp*]{ \includegraphics[width=1.5in]{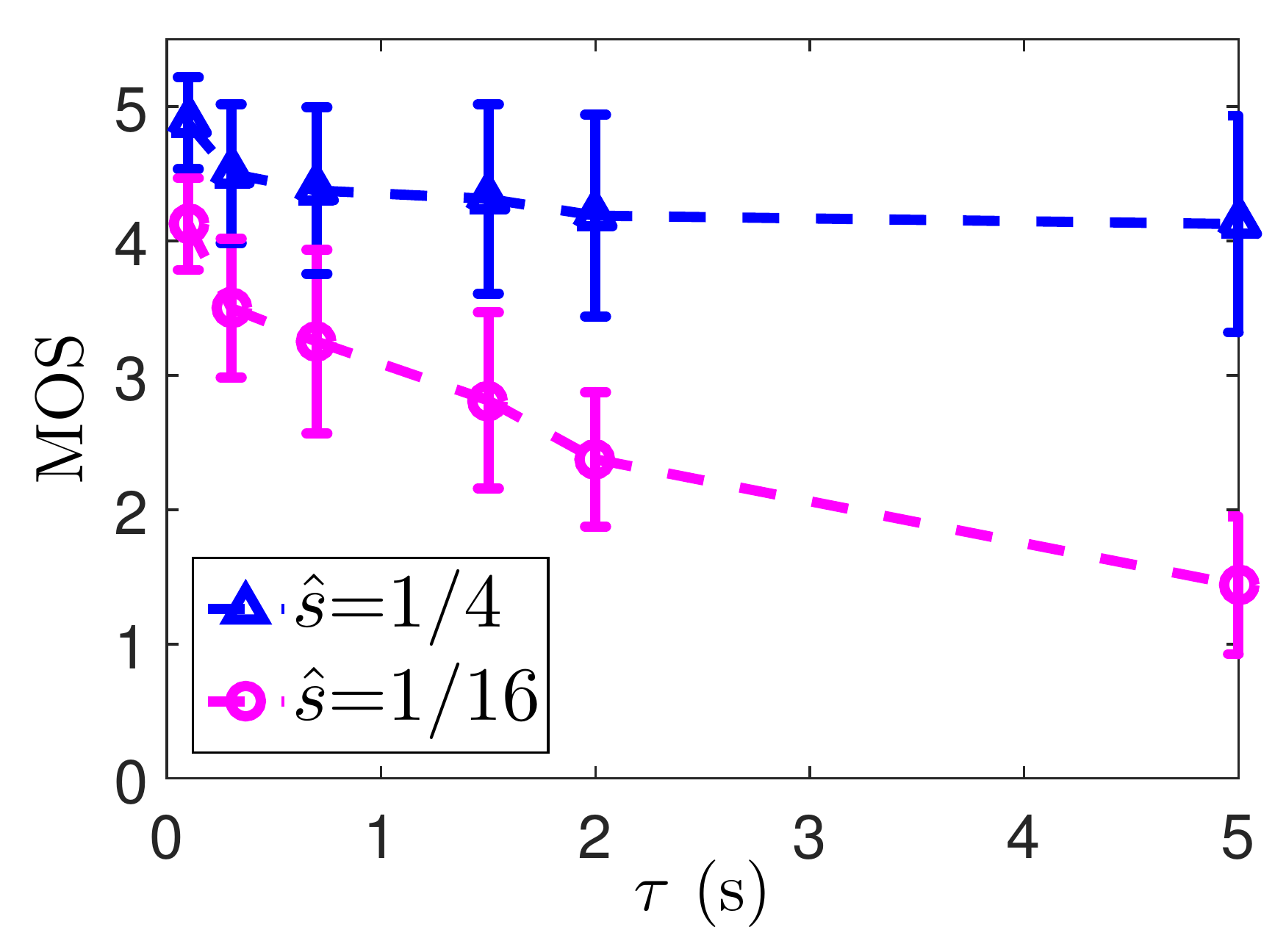}}
 \subfigure[Harbor*]{ \includegraphics[width=1.5in]{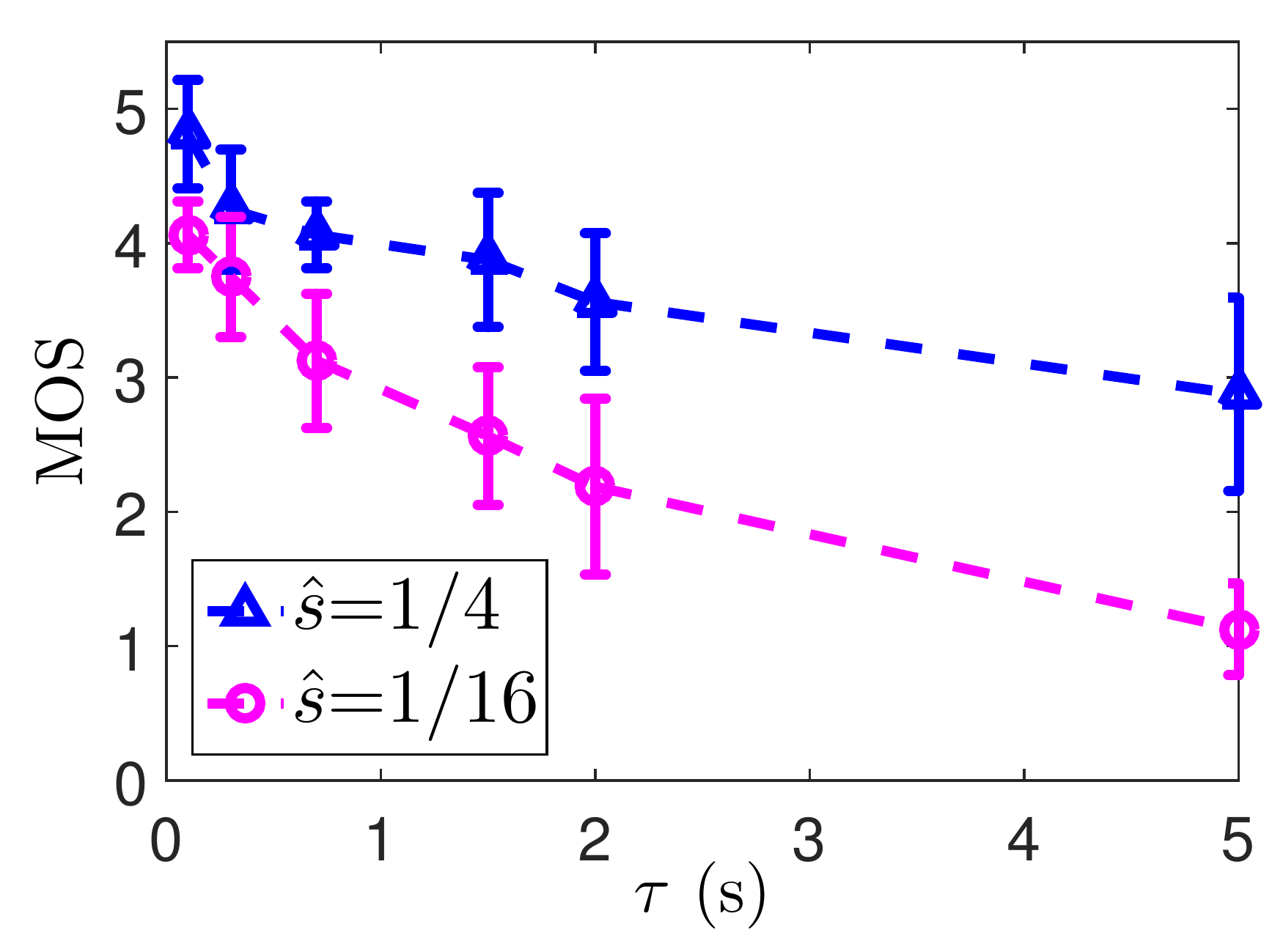}}
 \subfigure[Trolley*]{ \includegraphics[width=1.5in]{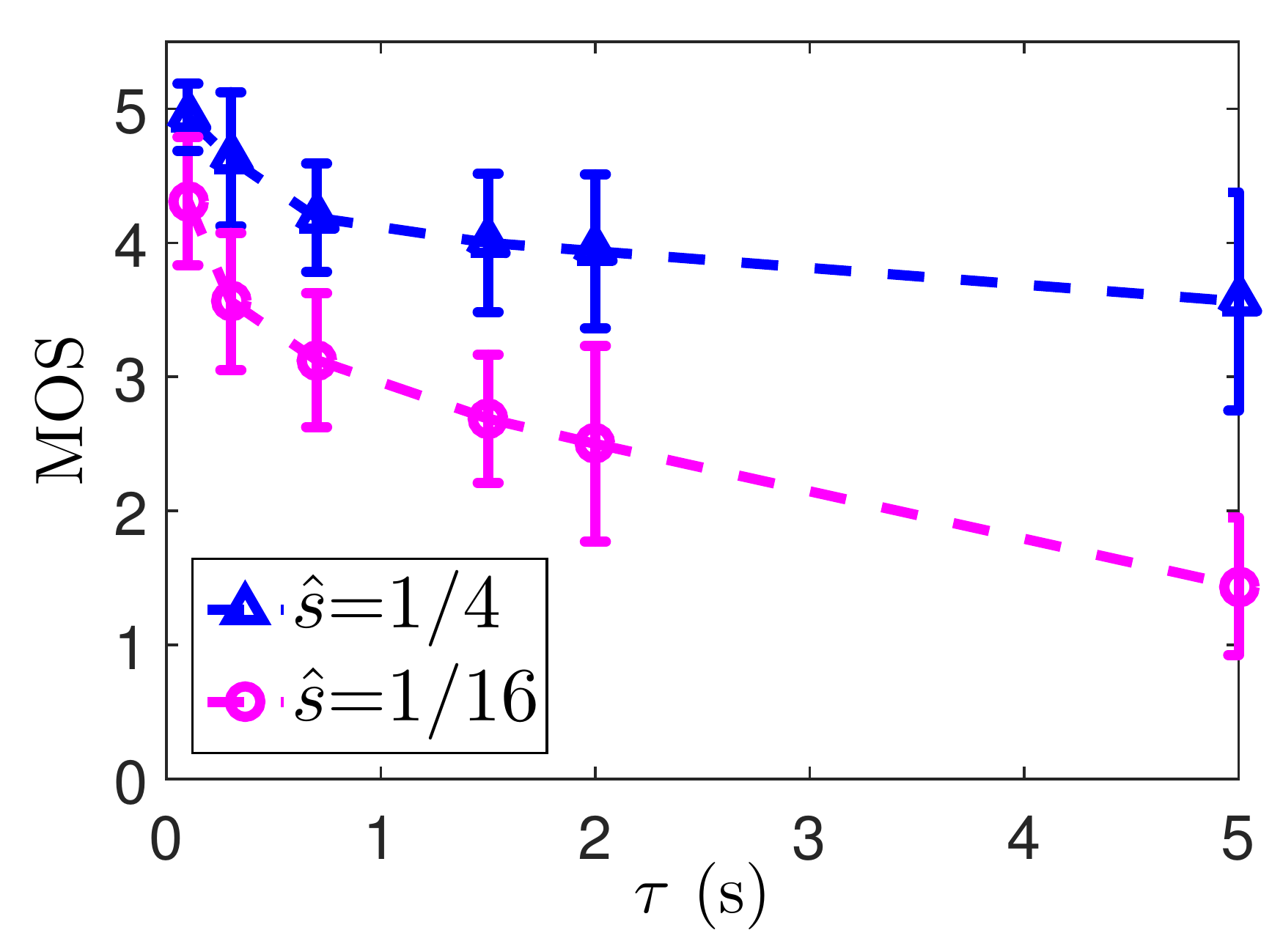}}
 \subfigure[Elephants]{ \includegraphics[width=1.5in]{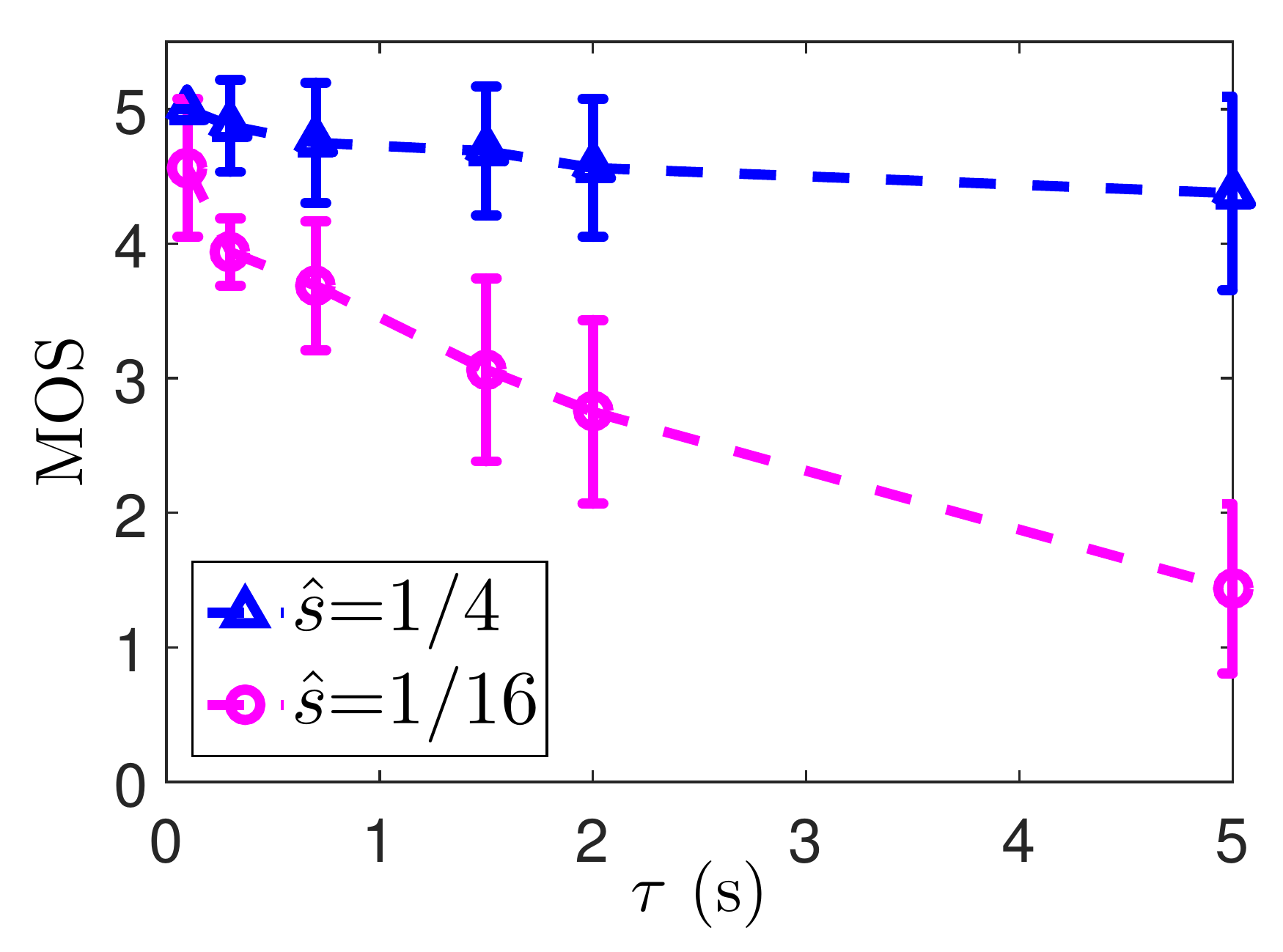}}
 \subfigure[Rhinos]{ \includegraphics[width=1.5in]{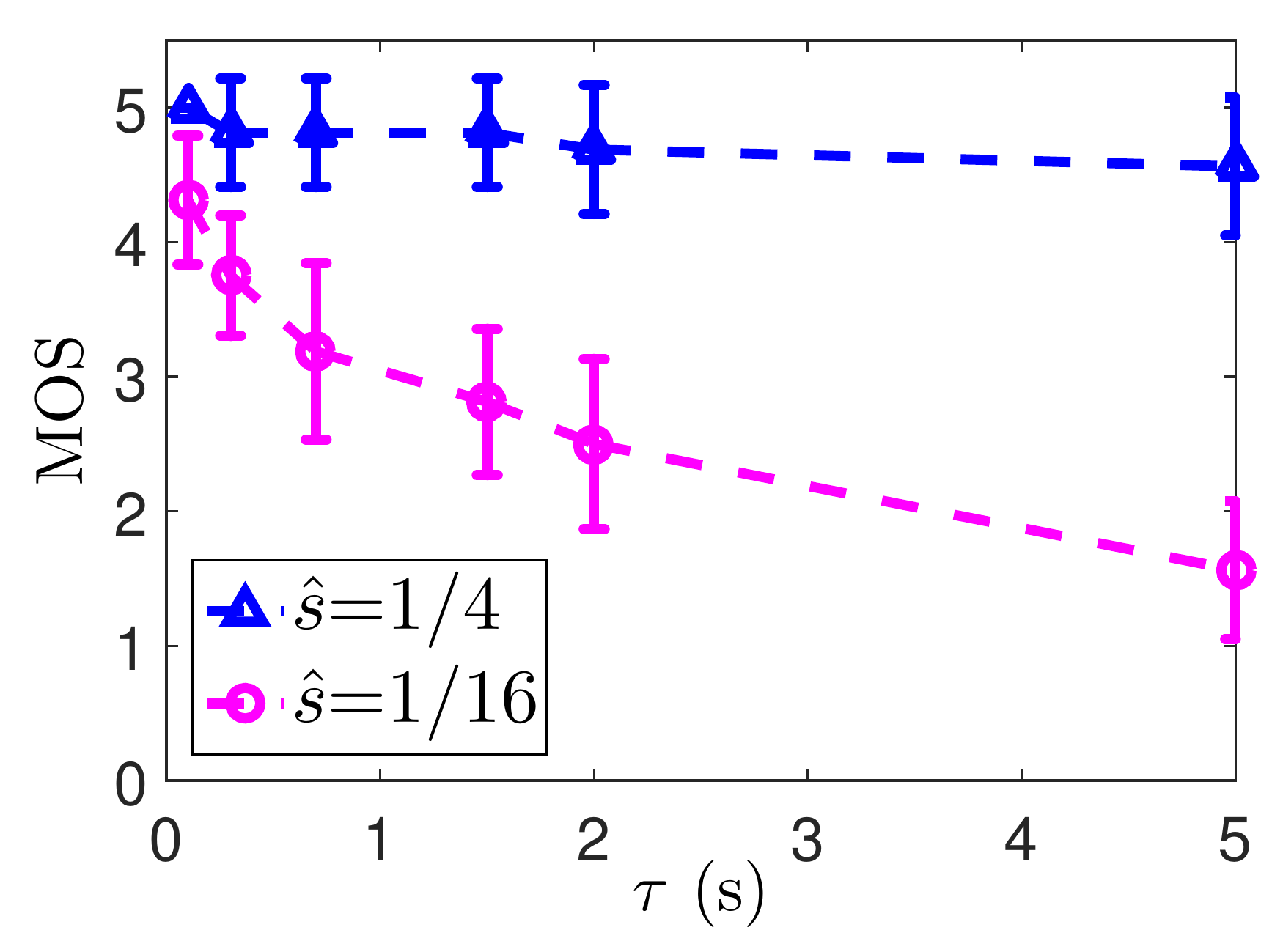}}
 \subfigure[Diving]{ \includegraphics[width=1.5in]{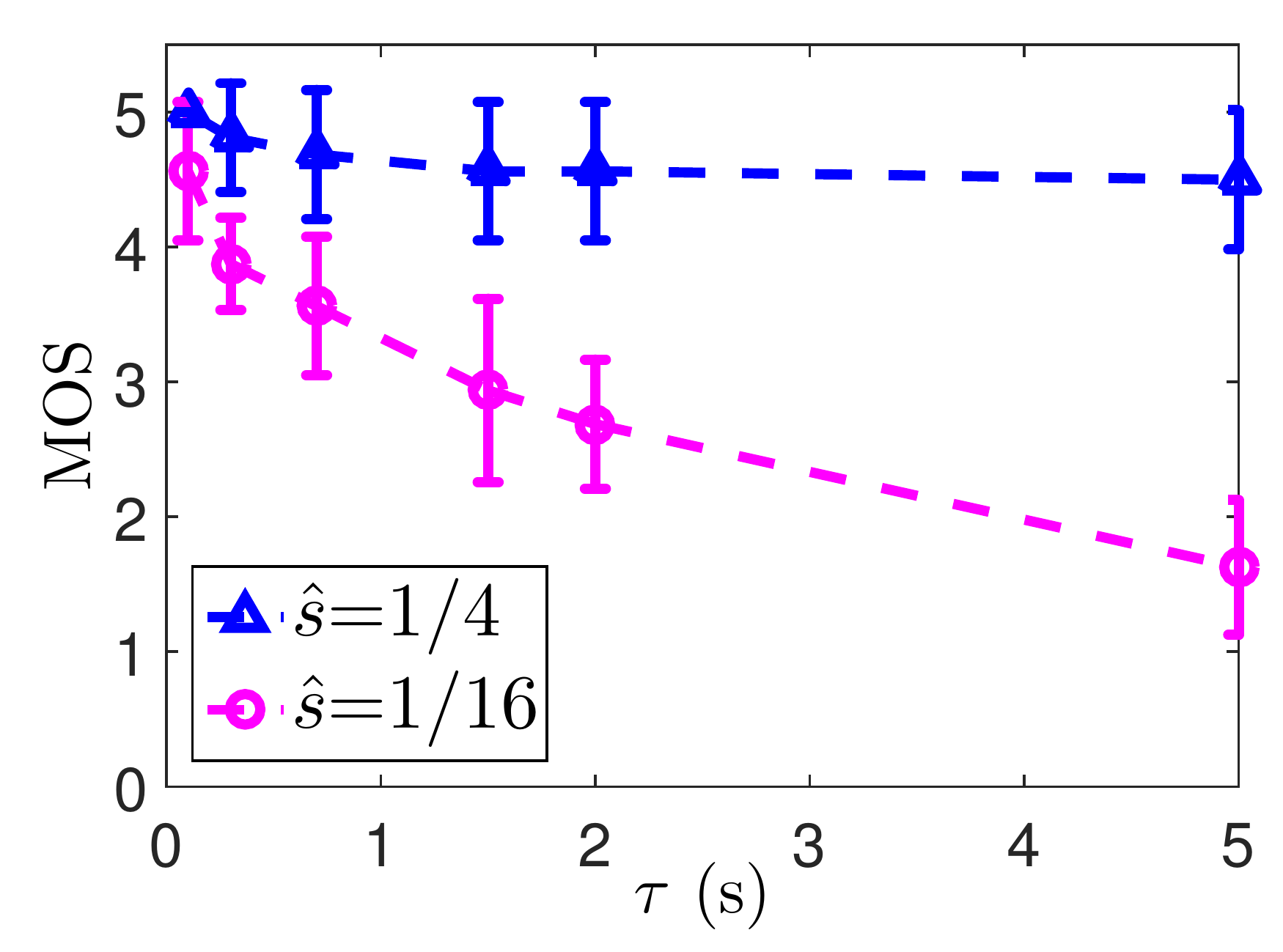}}
 \subfigure[Venice]{ \includegraphics[width=1.5in]{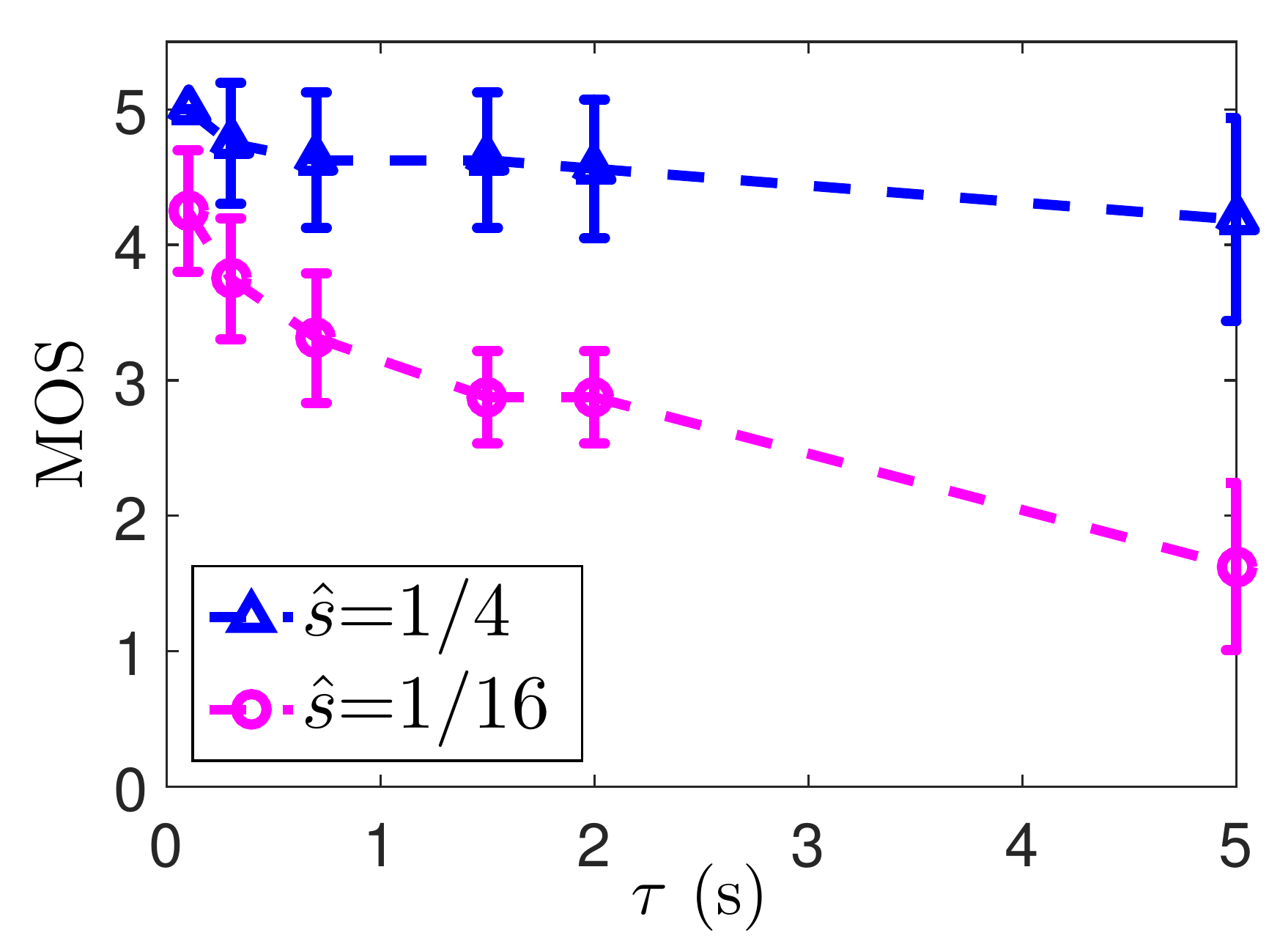}}
\caption{$s$ induced MOS variation versus refinement duration $\tau$}
\label{MOS_SR}
\end{figure}

\begin{figure}[t]
\centering
 \subfigure[AerialCity*]{ \includegraphics[width=1.5in]{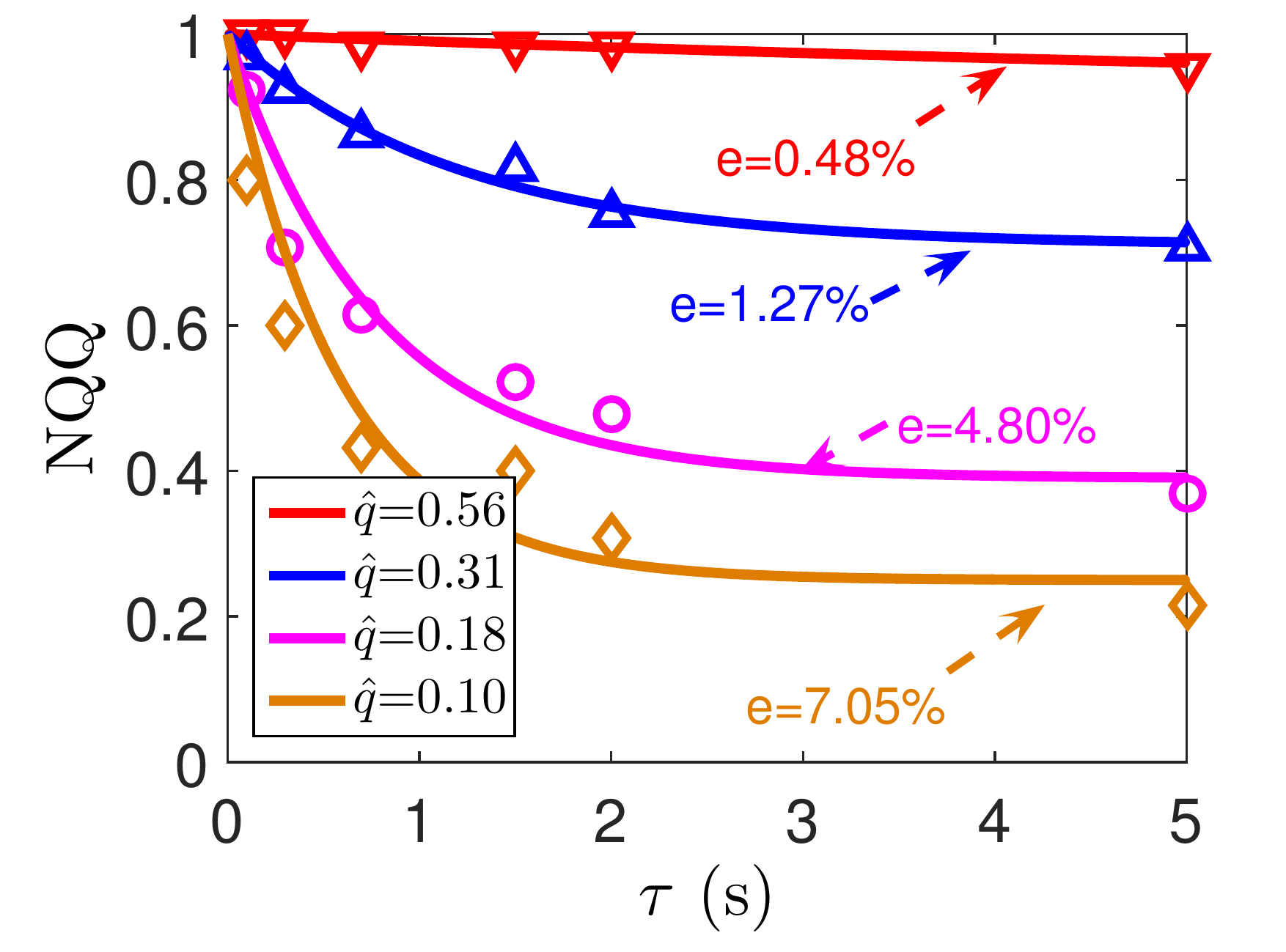}}
 \subfigure[Gaslamp*]{ \includegraphics[width=1.5in]{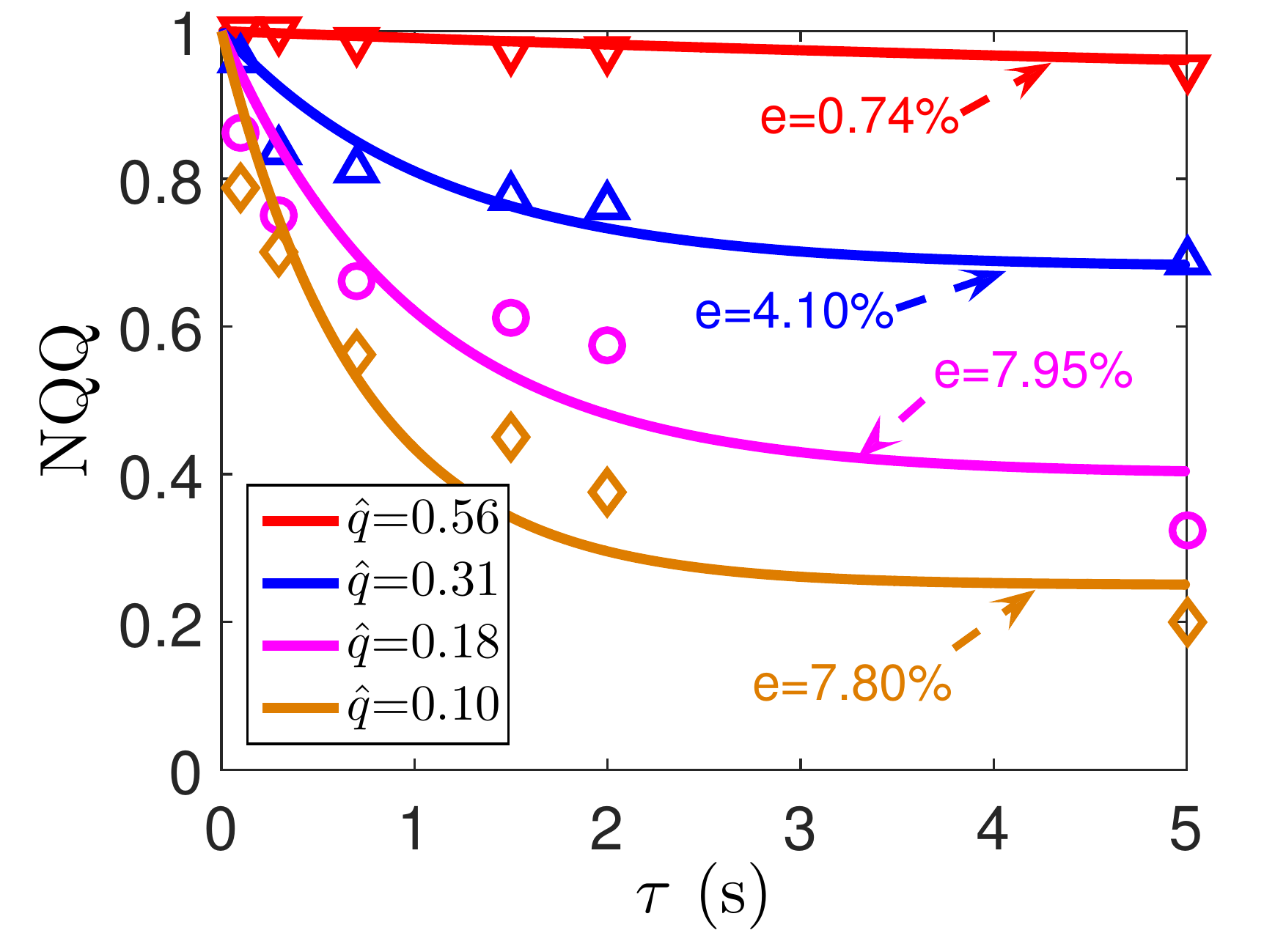}}
 \subfigure[Harbor*]{  \includegraphics[width=1.5in]{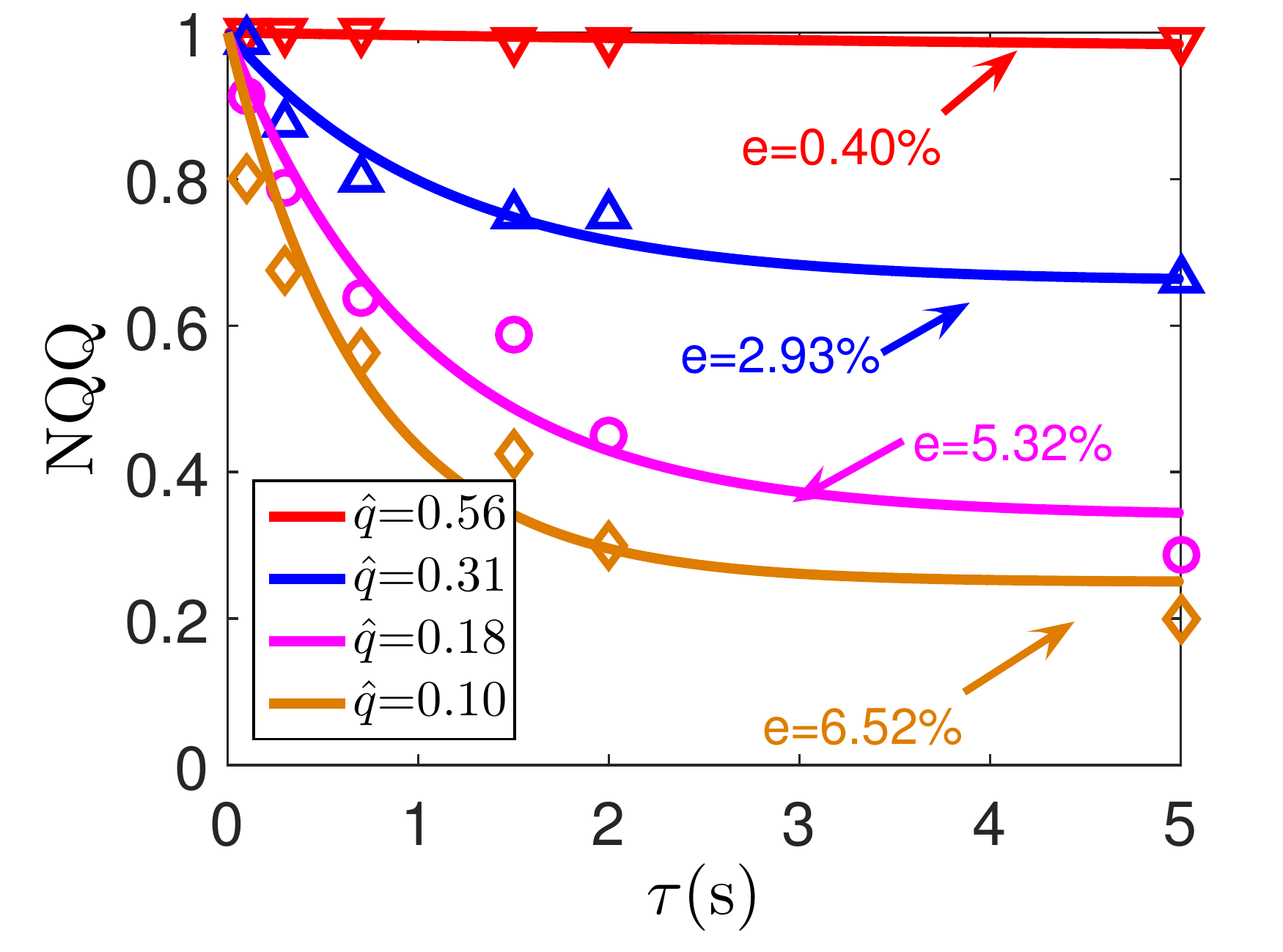}}
 \subfigure[Trolley*]{ \includegraphics[width=1.5in]{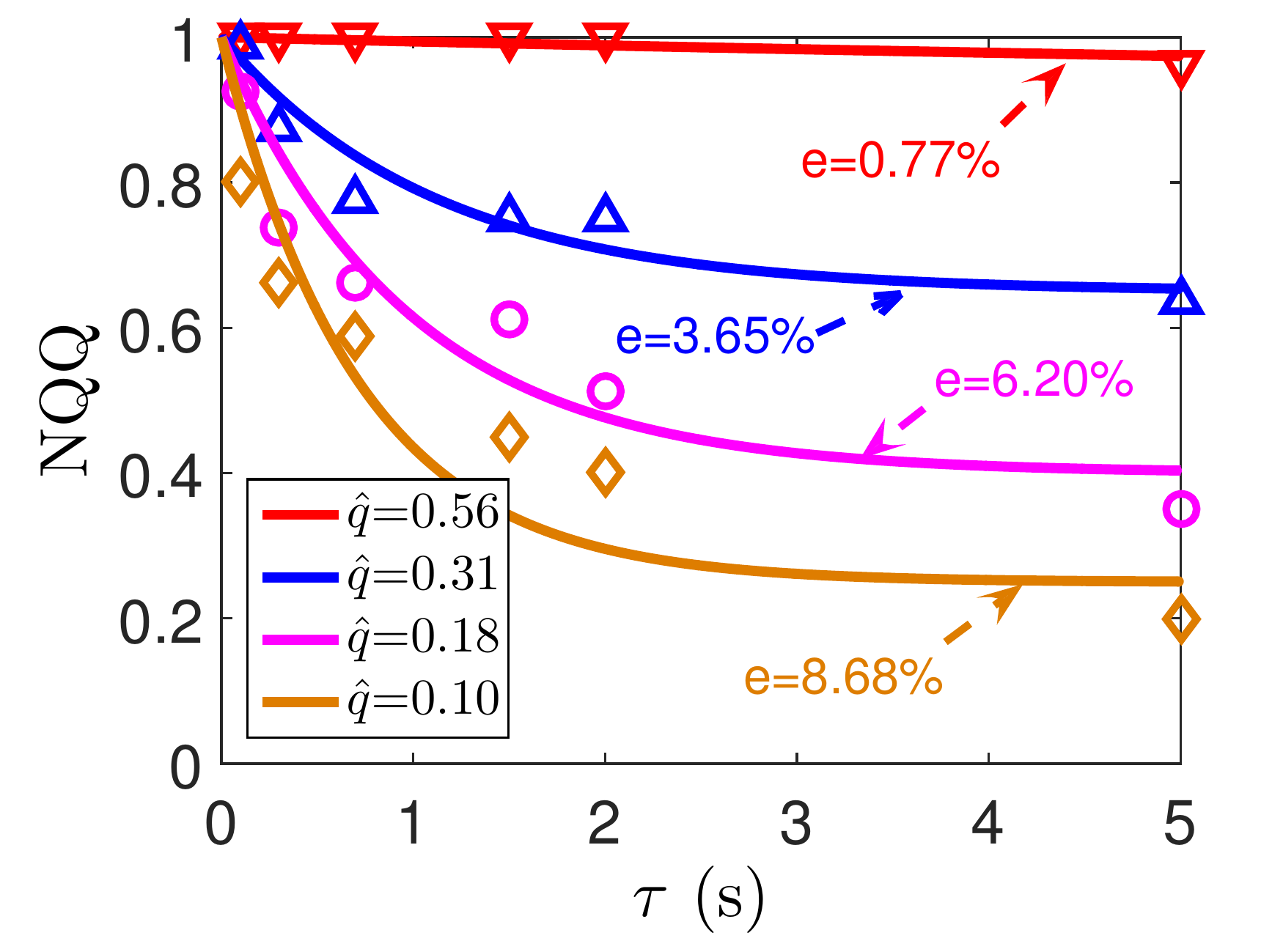}}
 \subfigure[Elephants]{ \includegraphics[width=1.5in]{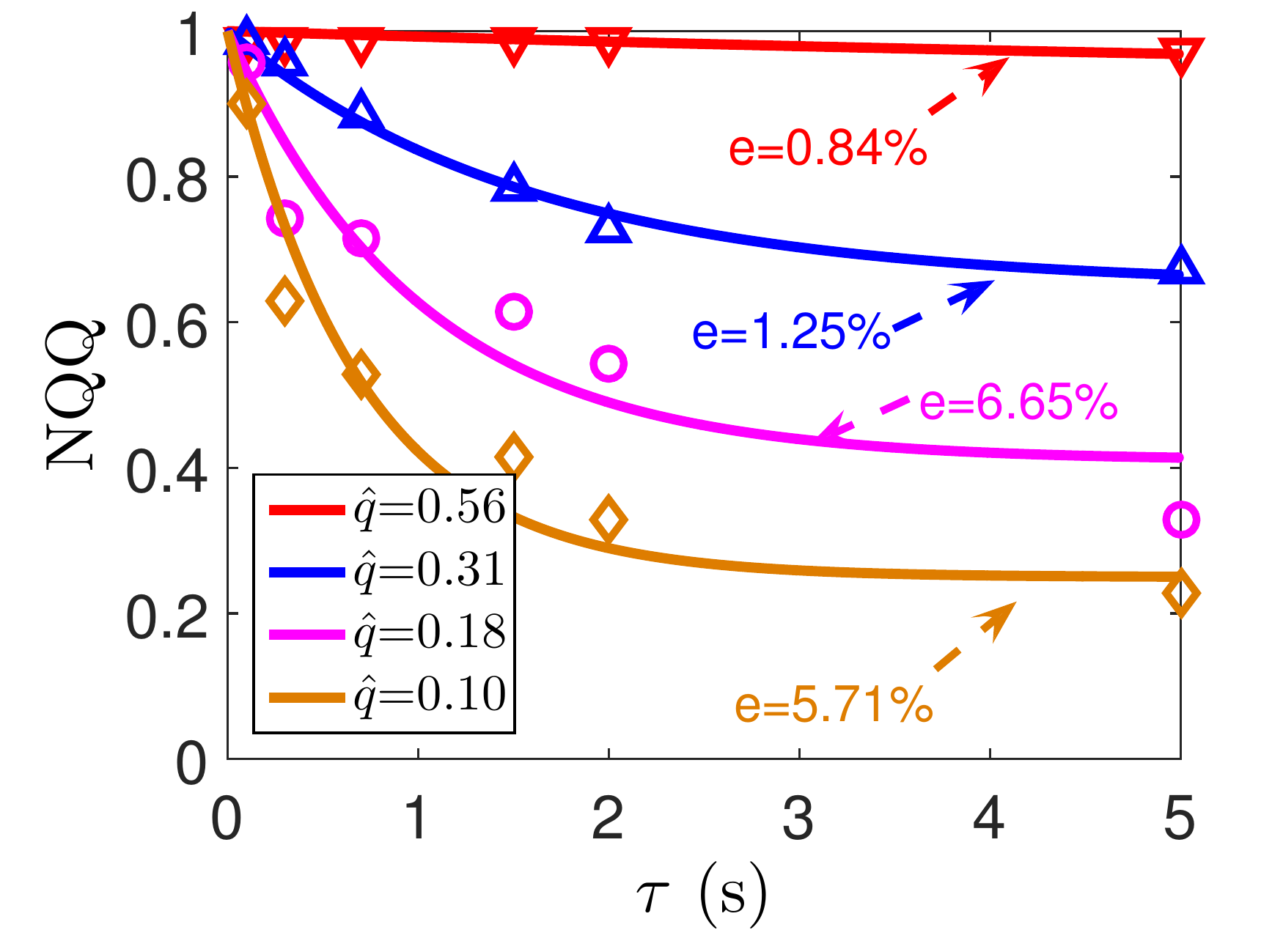}}
 \subfigure[Rhinos]{ \includegraphics[width=1.5in]{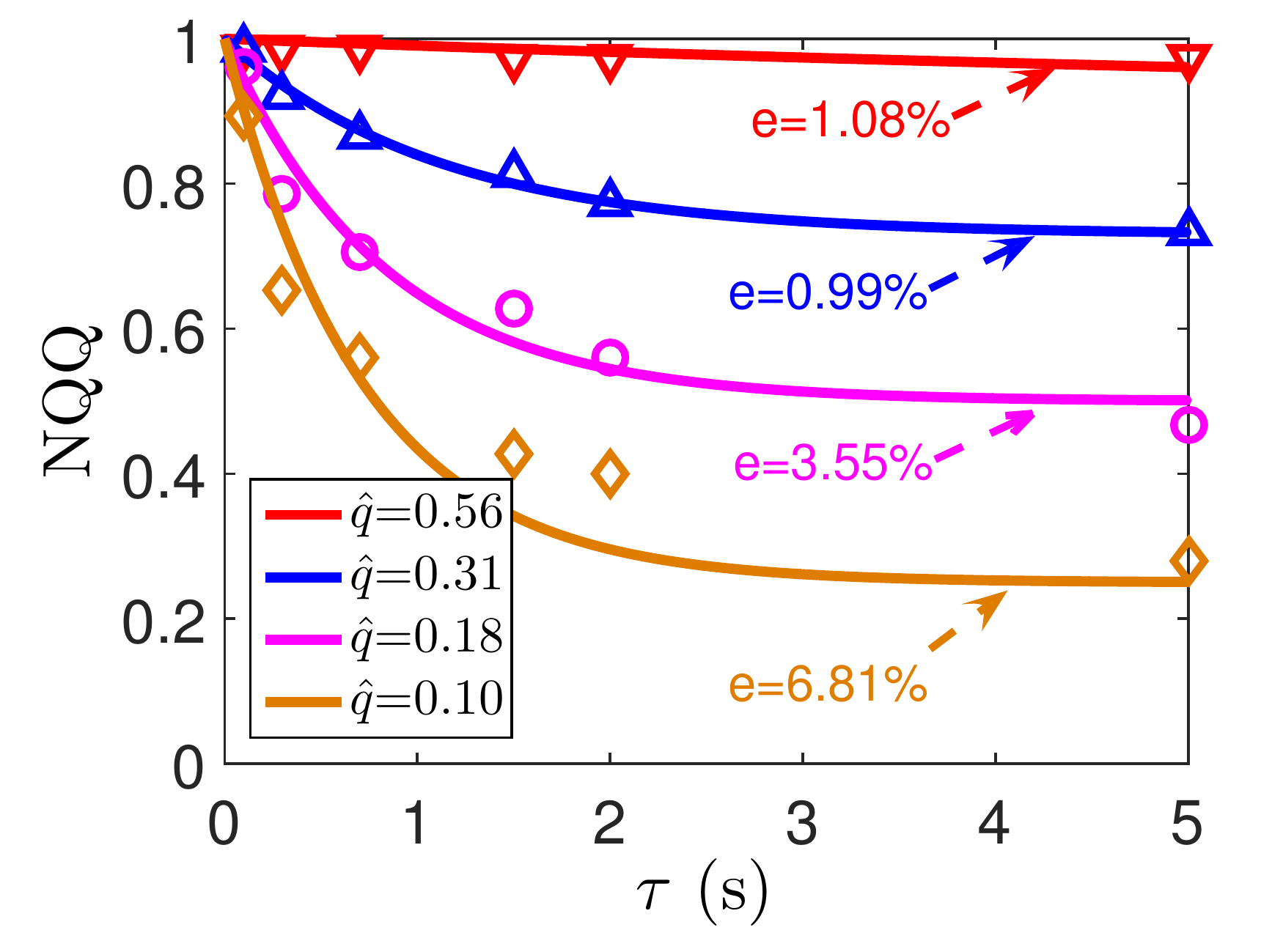}}
 \subfigure[Diving]{ \includegraphics[width=1.5in]{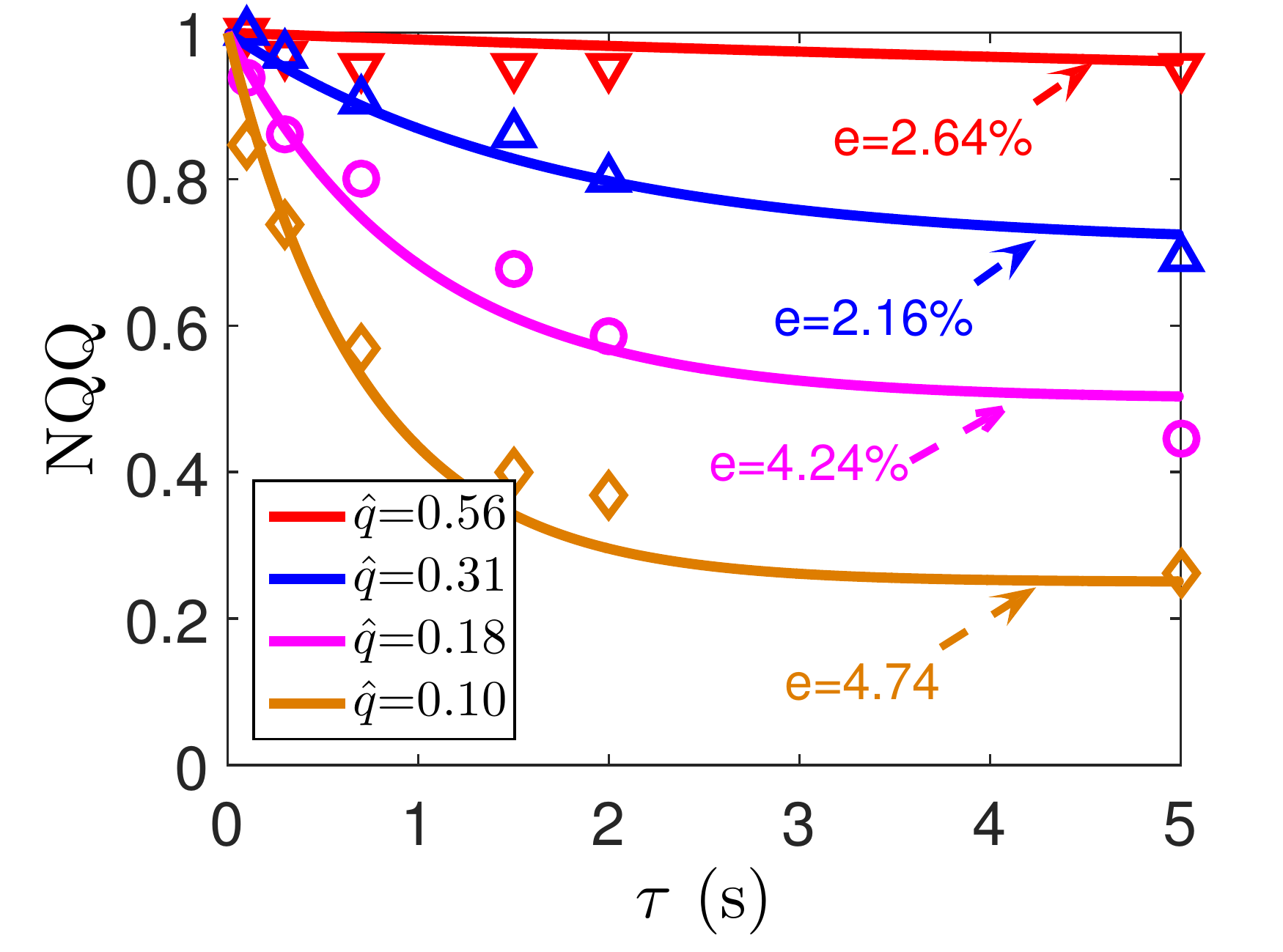}}
 \subfigure[Venice]{ \includegraphics[width=1.5in]{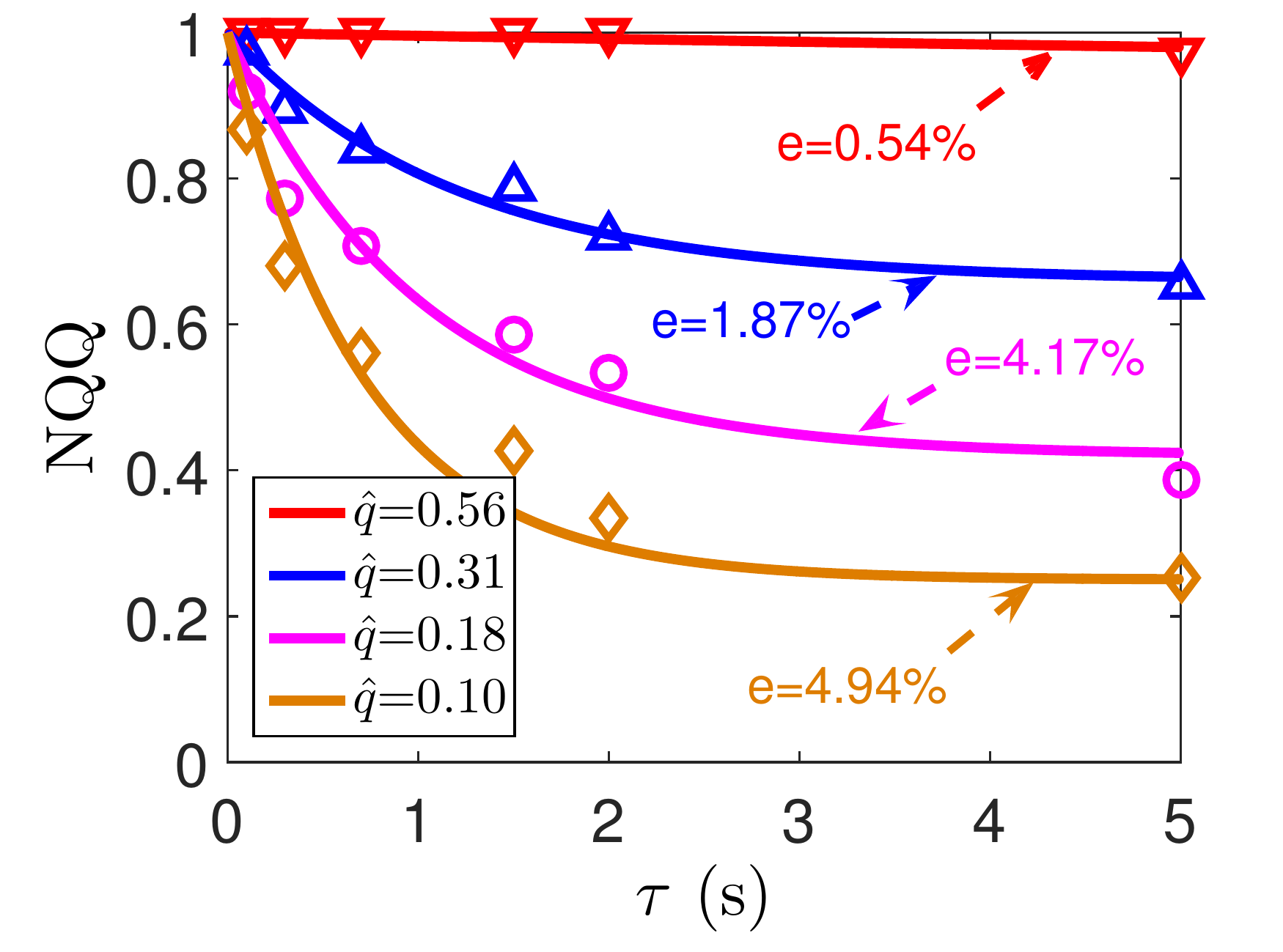}}
\caption{Normalized quality of $q$-impact (NQQ) with respect to the refinement duration $\tau$: points are collected MOS and curves are fitted using the analytical model~\eqref{Qrate}}
\label{parametersFitting_QP}
\end{figure}

\begin{figure}[t]
\centering
 \subfigure[AerialCity*]{ \includegraphics[width=1.5in]{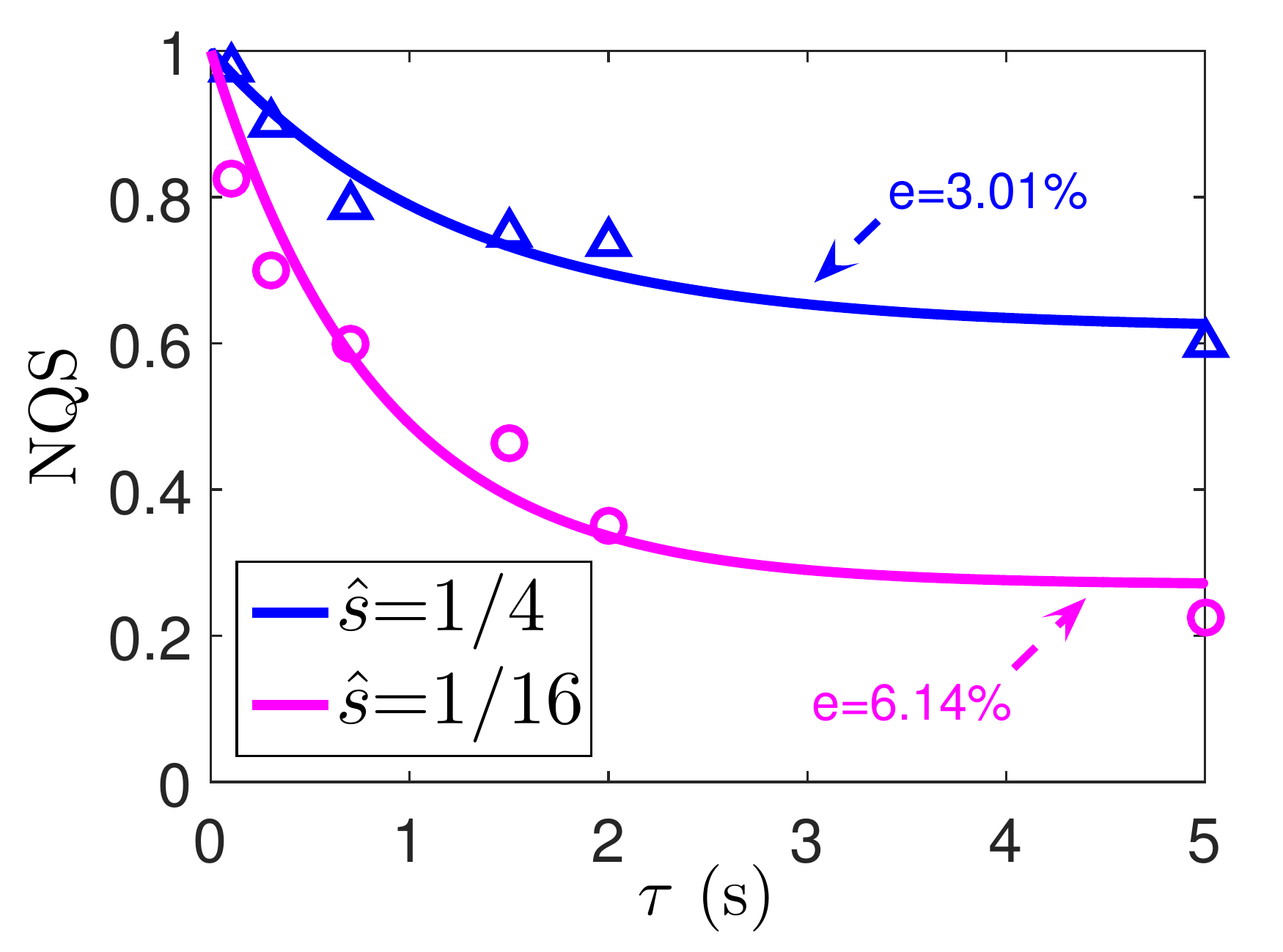}}
 \subfigure[Gaslamp*]{ \includegraphics[width=1.5in]{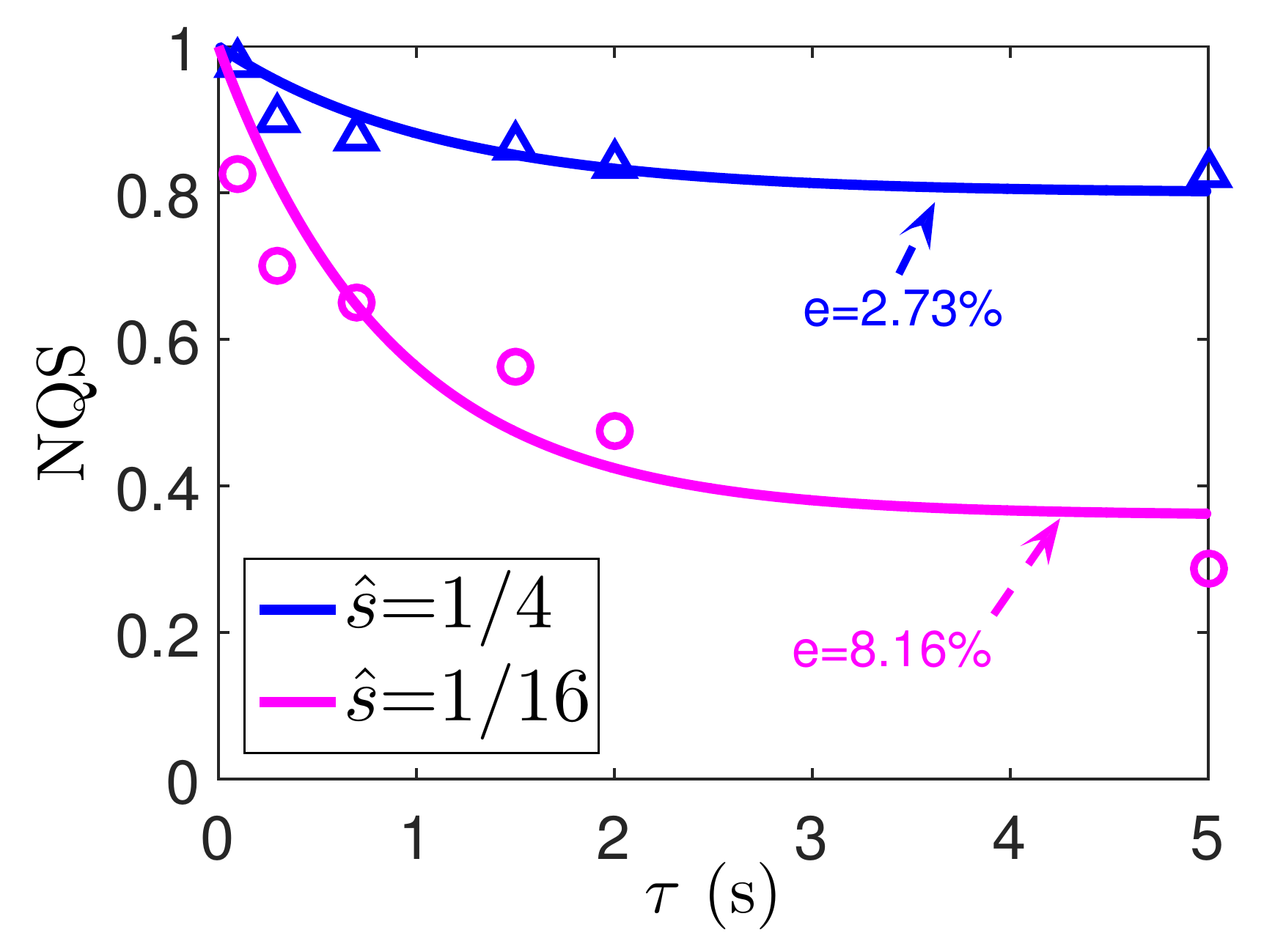}}
 \subfigure[Harbor*]{ \includegraphics[width=1.5in]{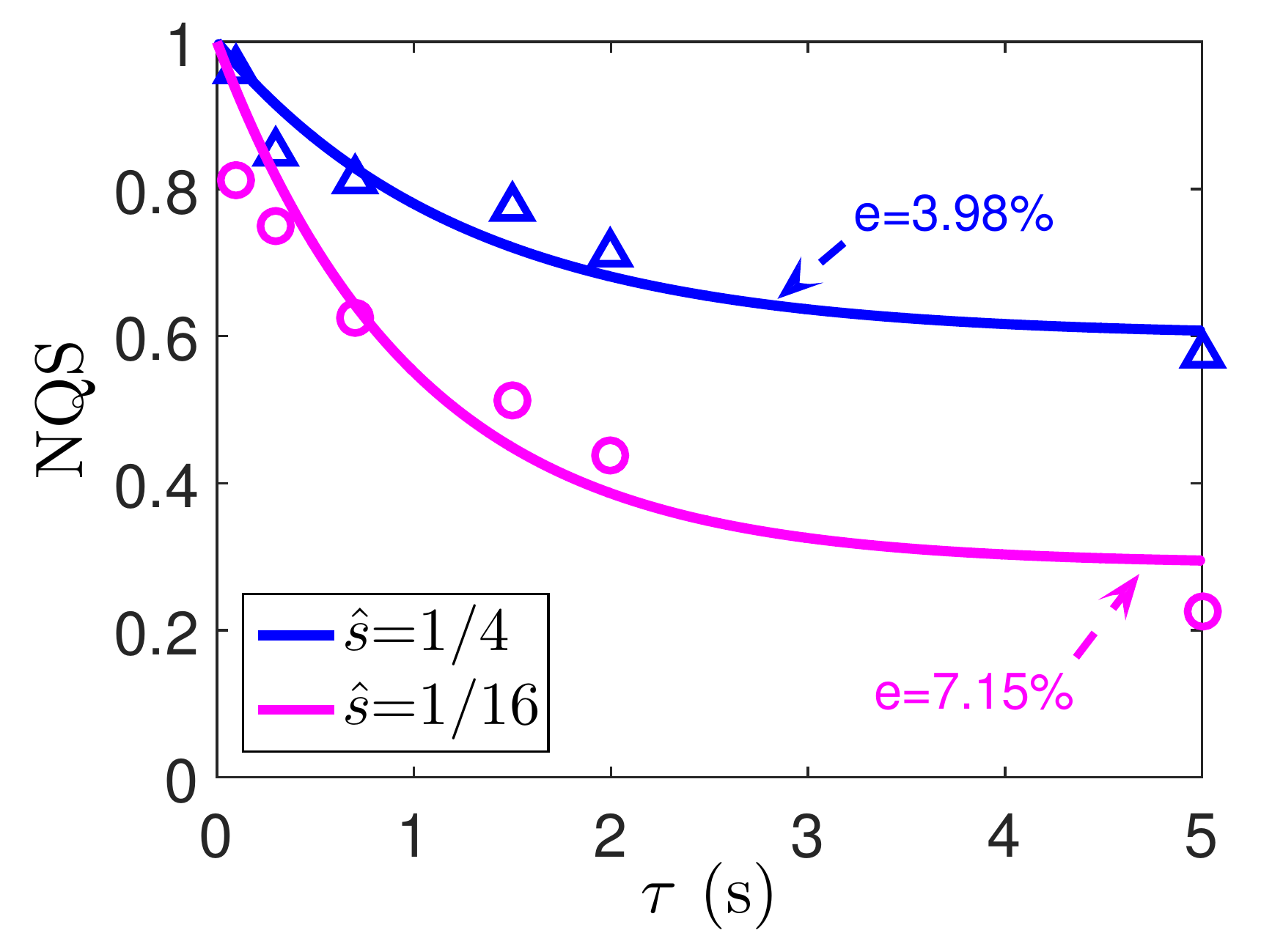}}
 \subfigure[Trolley*]{ \includegraphics[width=1.5in]{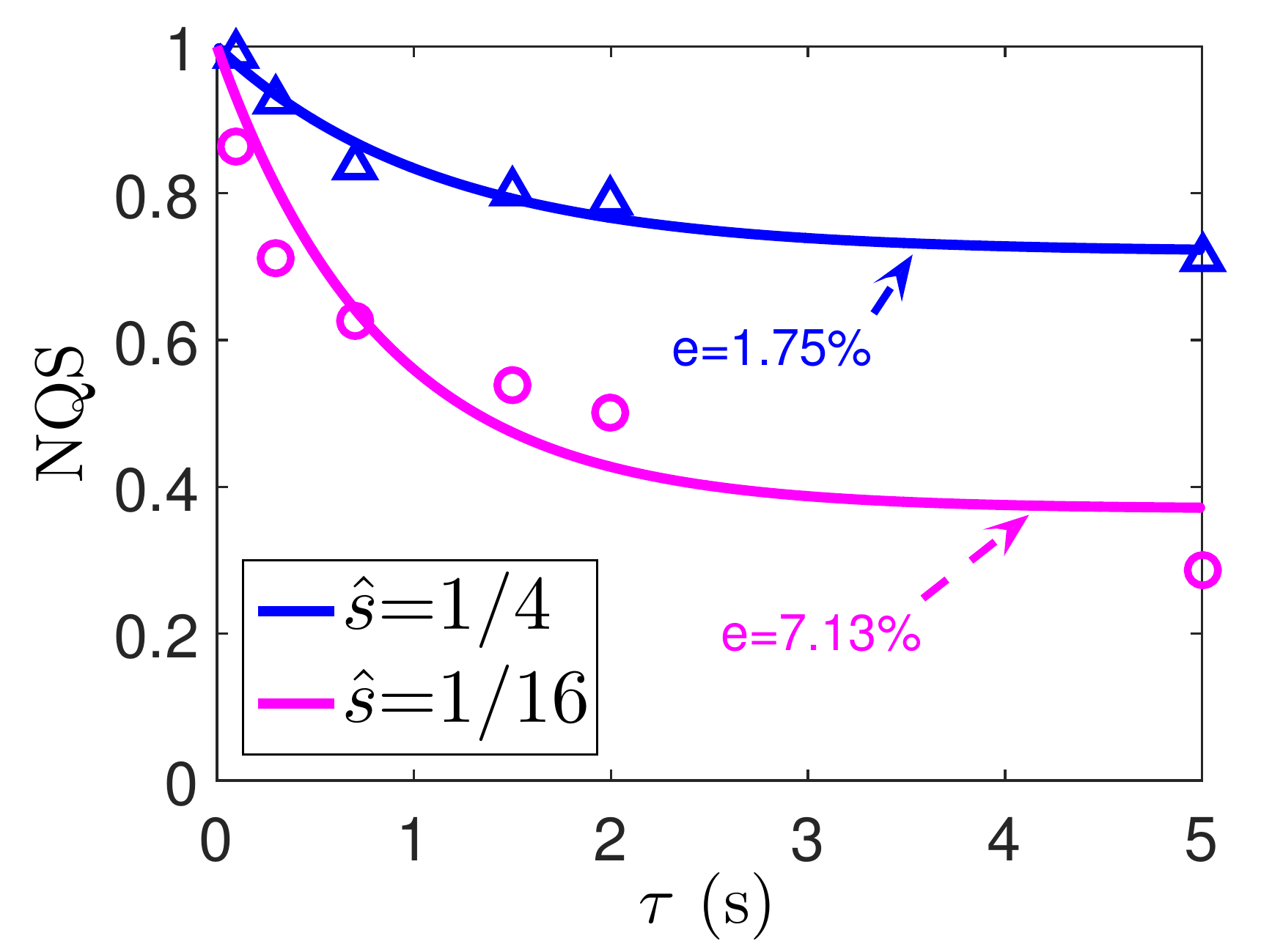}}
 \subfigure[Elephants]{ \includegraphics[width=1.5in]{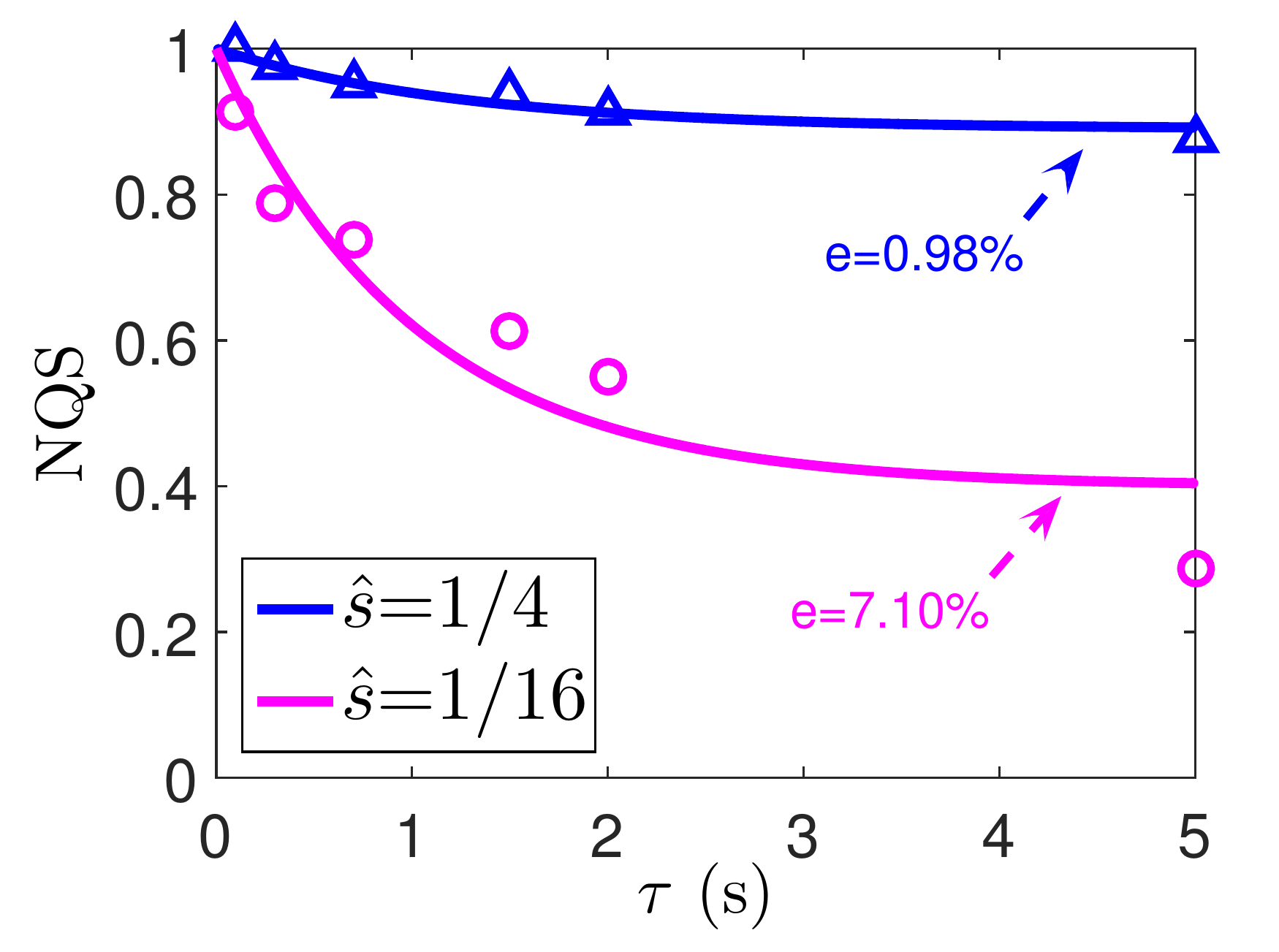}}
 \subfigure[Rhinos]{ \includegraphics[width=1.5in]{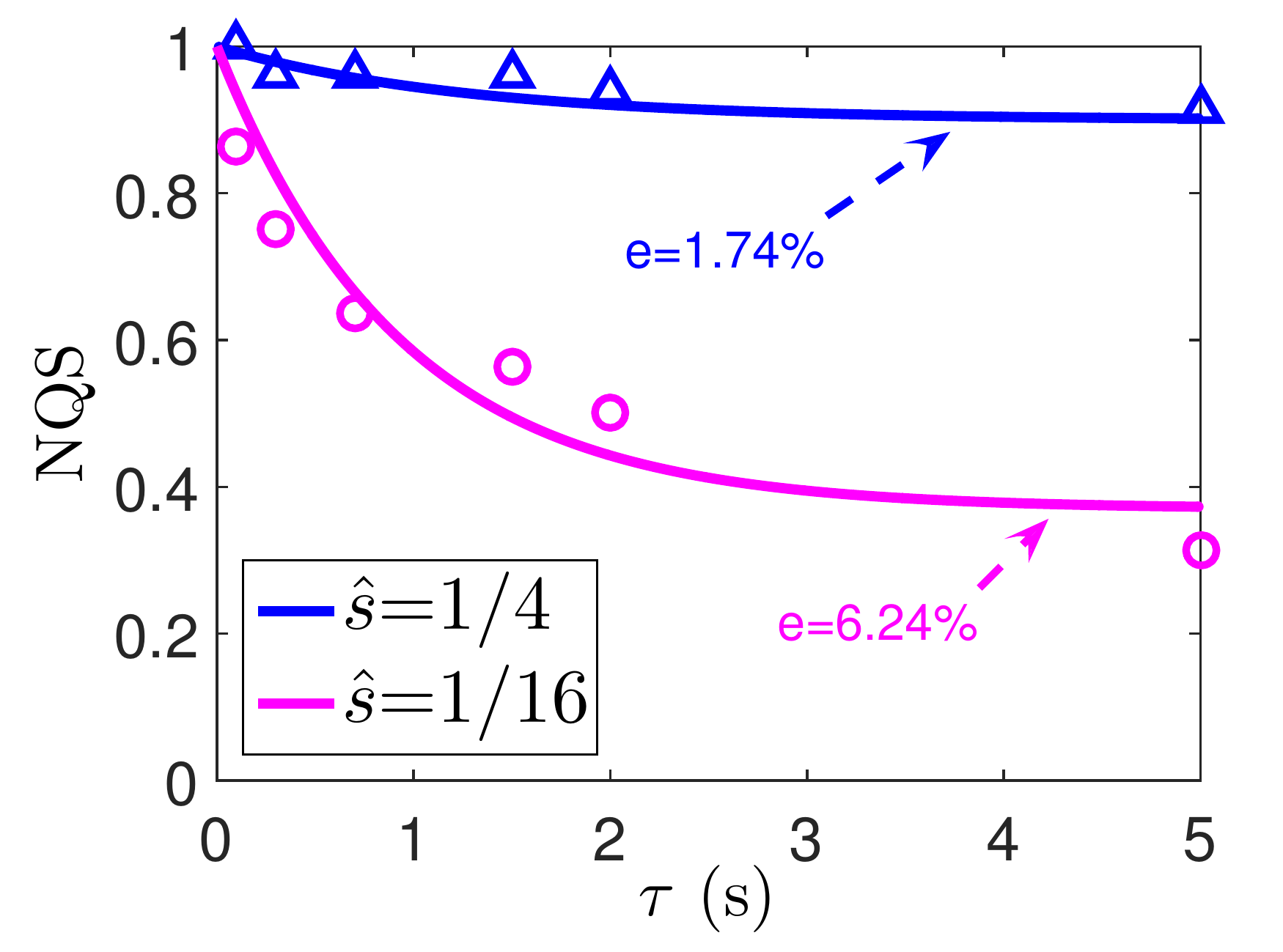}}
 \subfigure[Diving]{ \includegraphics[width=1.5in]{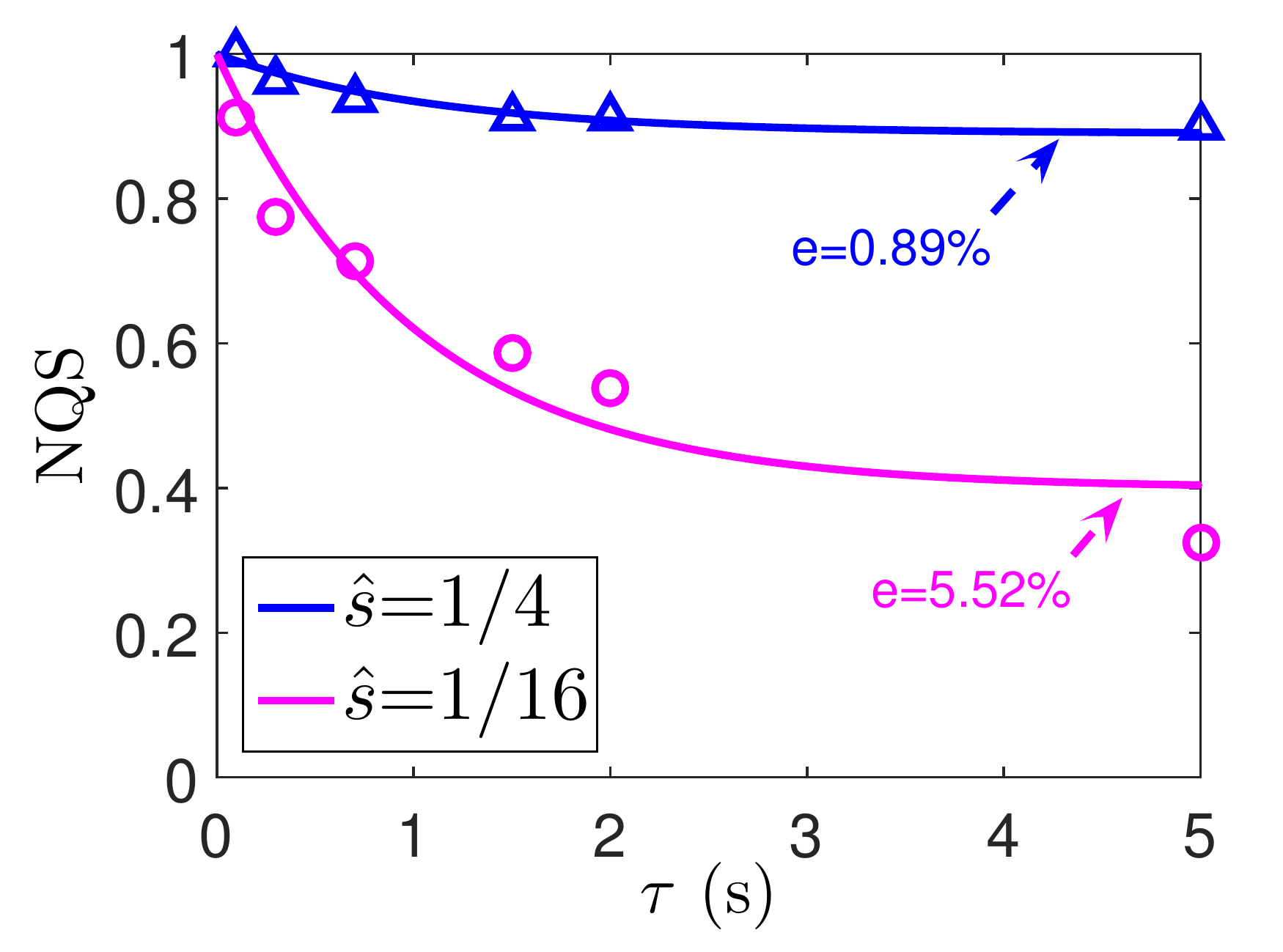}}
 \subfigure[Venice]{ \includegraphics[width=1.5in]{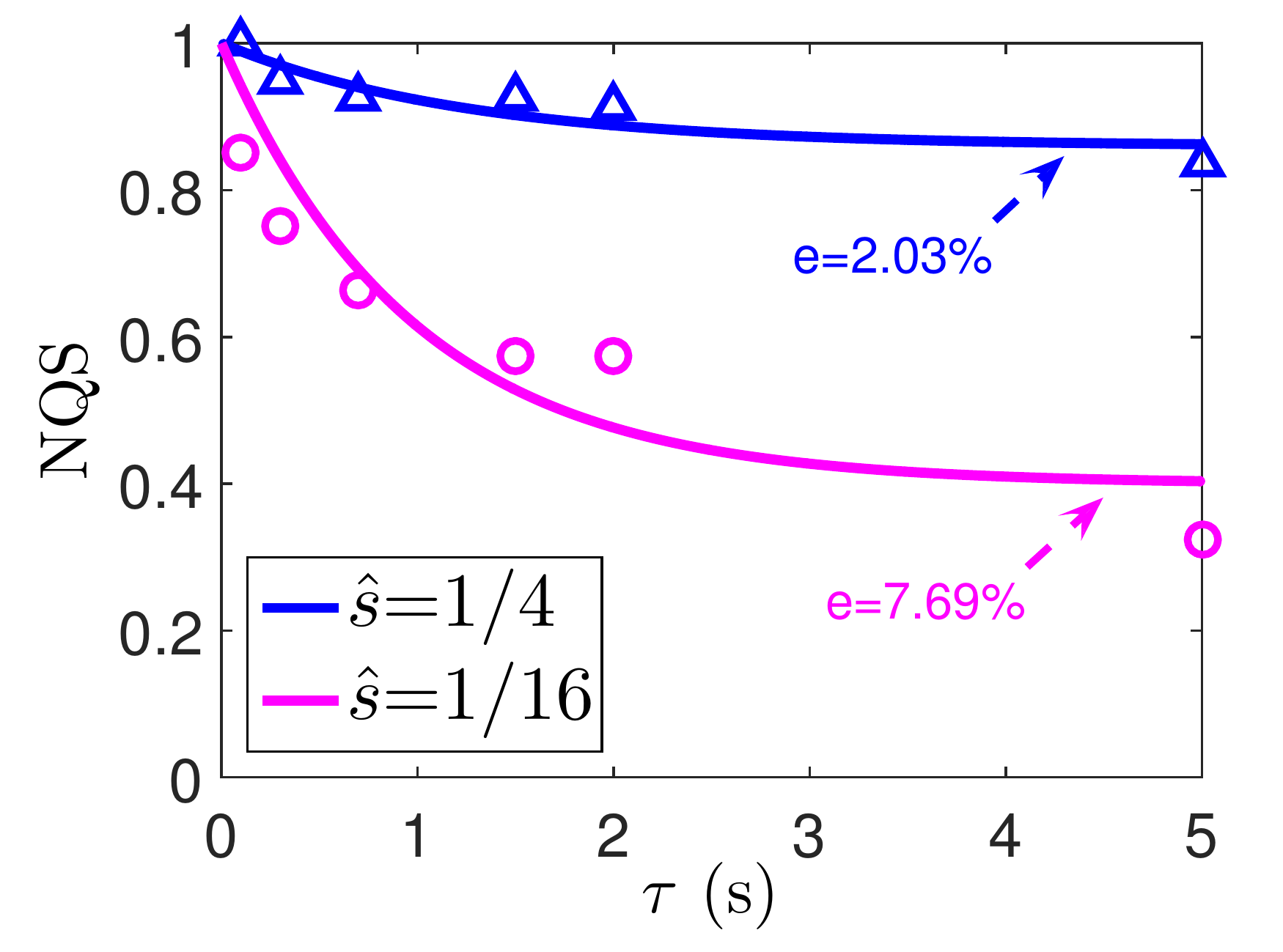}}
\caption{Normalized quality of $s$-impact (NQS) with respect to the refinement duration $\tau$: points are collected MOS and curves are fitted using the analytical model~\eqref{Qrate}}
\label{parametersFitting_SR}
\end{figure}

In our experiment, each processed video sample (PVS) consists of three consecutive parts, i.e., the  viewing period of FoV\#1, viewport adaptation period and the viewing period of FoV\#2, as shown in Fig.~\ref{ViewportAdaptation}. Users start at the FoV\#1, then navigate and focus their attention to the FoV\#2. Quality refinement happens when we stabilize our focus in FoV\#2. Herein, the first temporal segment of the FoV\#2 is a few seconds long and encoded at a reduced quality, followed by the original high-quality one after refinement. To cover sufficient scales of the quality variation and refinement duration, we set six distinct refinement durations ($\tau$ = $0.1$, $0.3$, $0.7$, $1.5$, $2$, $5$ second or s). Meanwhile, we apply five different quantization parameters (QP, or equivalent quantization stepsize $q$) (i.e., QP = $22$, $27$, $32$, $37$, $42$) and three spatial resolutions (i.e., native, $1/4$ and $1/16$ downscaled version)~\cite{pv_mobileQSTAR,Ma_RQModel,Xiaokai_TIP,ratecontrol}. Note that we keep the frame rate unchanged in this work. The total length of a PVS is $10$ seconds as recommended by the ITU-T BT. 500~\cite{BT500}.

We choose to use the popular HTC Vive with its associated HMD to perform the subject quality assessment. This is mainly because the HTC Vive system offers the state-of-the-art immersive experience. It enables the freedom of navigation when interacting with the content in the virtualized environment. Such freedom gives the user dramatical experience. However, it would introduce unexpected noise when we let the subject rate the content freely without imposing any viewing restriction with the HMD. We mainly crop and edit FoV sequences from the original immersive video to emulate the FoV adaptation (shown in Fig.~\ref{ViewportAdaptation}). With such setup, user are asked to stablize their body and head when performing the rating. (Eye movement is allowed with restriction to fit the real scenario that our eyes are often unconsciously move around when focusing our fixation.)
This generally reduces the uncertain rating noise. Note that HTC Vive HMD offers the 110$^\circ$ viewing range horizontally. Thus, we set 120$^\circ$ FoV when cropping it from the original content to fully overlay the HMD.

Intuitively, the perceptual impact introduced by the quality variations where both $q$ and $s$ are applied is inseparable~\cite{Xiaokai_TIP, pv_mobileQSTAR}.
However, in order to simplify the model complexity, we still assume the separable response of the $q$-impact and $s$-impact on the perceptual quality with respect to the refinement duration  in this work. Hence, we collect the MOSs for various $q$s and $s$s independently, resulting in 24 and 12 ten-second-long PVSs for each test video in separated tests, respectively. For either $q$-impact or $s$-impact, all PVSs are placed randomly for every test video.
In the meantime, each subject is asked to rate all PVSs corresponding to a subgroup of the test videos (i.e., three videos for considering the $q$ artifacts and six for $s$). We manually enforce that every subgroup selected for subjective assessment covers the sufficient spatio-temporal activities.
Participants are asked to give their MOSs (i.e., from 1 to 5 shown in Fig.~\ref{AssessmentProcedure}(c)) when finishing each ten-second-long PVS clip, within a 3 seconds short pause. Users rest another 20 seconds  when completing all PVSs for the same video content and moving to the next one. With aforementioned setup, each subject will complete his/her assessments within 30-minutes without feeling dizziness and tiredness.

Among these test sequences shown in Fig.~\ref{ImmersiveImages}, we have selected the ``KiteFlite", to train the participants to familiarize with test protocol and have the correct sensation of the quality variations. The rest eight videos are used for perceptual tests. All videos are rendered using the HTC Vive systems with its HMD. The rating MOS score ranged from 1 (refinement is very annoying) to 5 (refinement is imperceptible) is utilized, and the corresponding measurement is explained in Fig.~\ref{AssessmentProcedure}, along with the assessment procedure aforementioned. The human subjects are from widely diverse academic majors in universities. All of the viewers are found to have normal visual (after correction) and color perception.

After data collection, we firstly convert the raw ratings to Z-scores~\cite{num_scaling_SPIE} for normalizing all the scores of each viewer, i.e.,
\begin{equation}\label{Z-score}
z_{mij}=\frac{x_{mij}-\mu(X_i)}{\sigma(X_i)},
\end{equation}
where $x_{mij}$ and $z_{mij}$ are the raw rating and the Z-score of $i$-th PVS in $m$-th test video, from $j$-th viewer, respectively. $X_i$ denotes all ratings of $i$-th PVSs from all subjects. $\mu(\cdot)$ and $\sigma(\cdot)$ represent the mean and the standard deviation taking operator, respectively.

Based on Z-scores, we carry out the post screening method introduced in BT. 500~\cite{BT500} to remove all ratings by certain viewers, whose ratings are distant from most of the viewers. On average, one subject is excluded for each test video. We then perform further method to process the remaining ratings in the raw score domain, which is referred to \cite{spatial_tempral} and aims at removing any ``obvious" errors in raw ratings to keep the consistency. Particularly, we make use of the fact that a PVS coded at a higher $q$ (or/and a lower $s$) would not have a higher rating than the one coded at a lower $q$ (or/and a higher $s$) under the same $\tau$, and the rating can be higher while $\tau$ is shorter under the same coding parameters, if the viewer's judgement is consistent. Thus, we analyze the ratings from each viewer, and remove all the ratings of a very subject whose ratings contain more than 1/8 inconsistent outcomes (i.e., 3 out of 24 in $q$-impact test, while 1 out of 12 in $s$-impact test) for the same test video. For the remaining outliers, we replace them with the average value of those consistent ratings for adjacent $q$ (or/and $s$) and $\tau$. The MOS score for a particular PVS is derived by averaging the common scores after all the steps.

\subsection{Analytical Models}

Fig.~\ref{MOS_QP} and~\ref{MOS_SR} plot the MOS (with standard deviation) with respect to the normalized $q$ and $s$, i.e., $\hat{q} = q_{\min}/q$, and
$\hat{s} = s/s_{\max}$, respectively. Here, $q_{\min}$ = 8 (at corresponding QP 22) and $s_{\max}$ is the native spatial resolution
of the selected immersive videos which may be sampled at 3840$\times$1920, 3840$\times$2048, or 3840$\times$2160 in different dataset. In other words, we assume the high-quality content is prepared at its native spatial resolution and QP 22.
Clearly, MOS degrades consistently as $\tau$ increases, for any fixed $\hat{q}$ or $\hat{s}$. It can also be found that the degradation was faster when the quality difference is larger before and after refinement (i.e., lower $\hat{q}$ or $\hat{s}$), especially when $\tau$ is less than 2 seconds. All these discoveries motivated us to seek a simple but effective parametric model to explain the impacts of the compression factor $\hat{q}$, and spatial resolution $\hat{s}$ on the perceptual quality, with respect to the refinement duration $\tau$, in the applications of FoV adaptation.

Then, we plot the normalized quality of $q$-impact with respect to the refinement duration $\tau$ (NQQ) in Fig.~\ref{parametersFitting_QP} (i.e., discrete points), and so as the normalized quality of $s$-impact (NQS) in Fig.~\ref{parametersFitting_SR}. In theory, both the NQQ and NQS should be 1 (i.e., with highest MOS $Q$ = $Q_{\max}$) if the quality refinement duration $\tau$ is zero (i.e., does not take time to refine the quality from the reduced-quality version to the original high-quality one). Therefore, we propose to use the exponential function to describe the NQQ and NQS in terms of the $\tau$, i.e.,
\begin{align}\label{Qrate}
\hat{Q}=\frac{Q}{Q_{\max}}=a\cdot e^{-b\cdot\tau}+c,
\end{align}
where $a$, $b$, and $c$ are model parameters. Since $Q$ = $Q_{\max}$ when $\tau = 0$, we have $c = 1 - a$.

\begin{table}[t]
\centering
\caption{Fixed parameters for $a$ and $b$}
\label{tab:a_b_para}
\begin{tabular}{|c|c|c|c|}
\hline
& $k_1$ & $k_2$ & $k_3$\\
\hline
$a(\hat{q})$ & 0.8 & 39.55 & 2.73 \\
$b(\hat{q})$ & 1.45 & 47.14 & 3.29 \\
\hline
\hline
$a(\hat{s})$ & 0.8 & 4.65 & 0 \\
$b(\hat{s})$ & 4.53 & 0.3 & -3.37 \\
\hline
\end{tabular}
\end{table}

\begin{figure}[b]
\centering
 \subfigure[]{\includegraphics[width=1.6in]{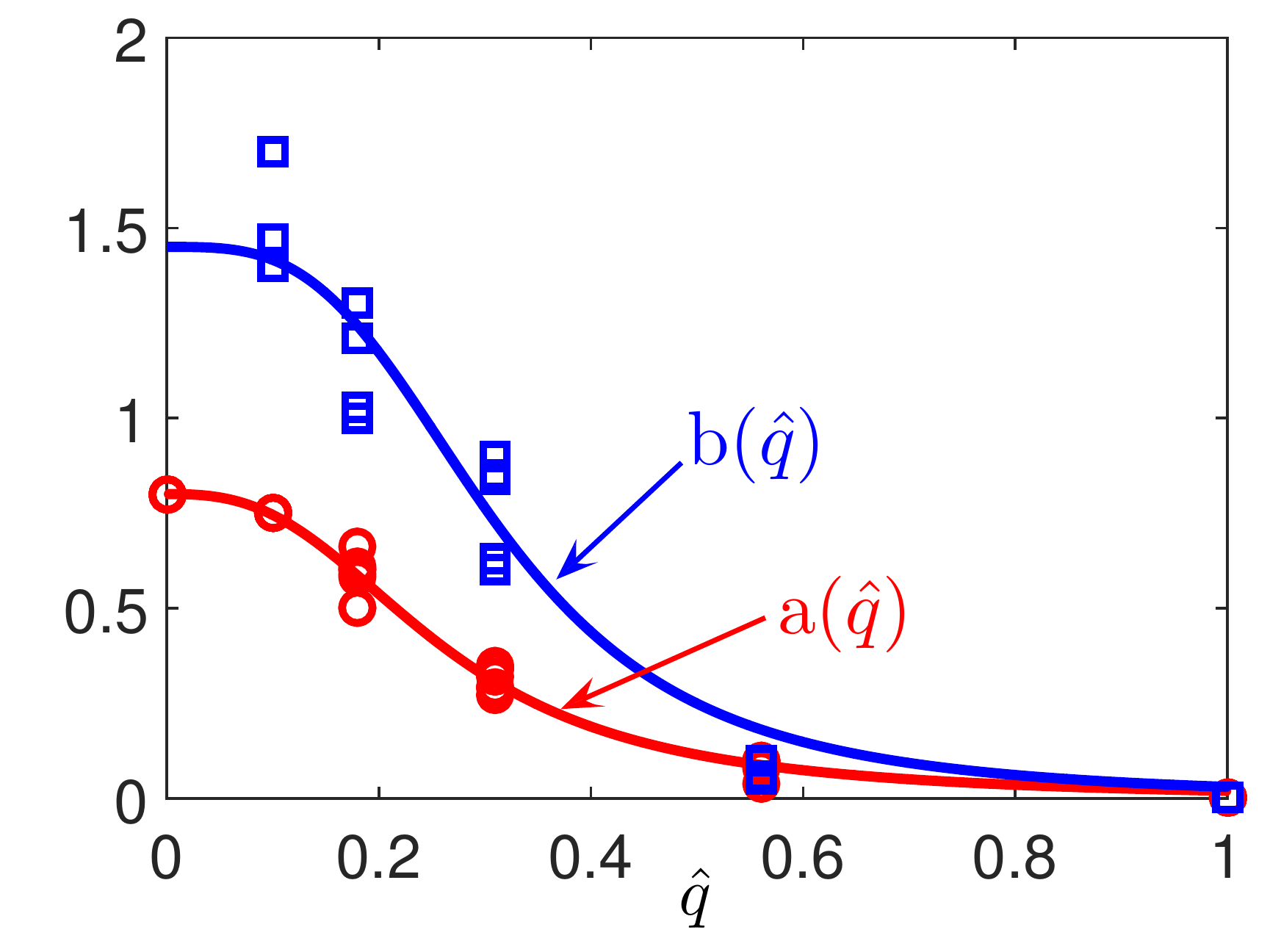}}
 \subfigure[]{\includegraphics[width=1.6in]{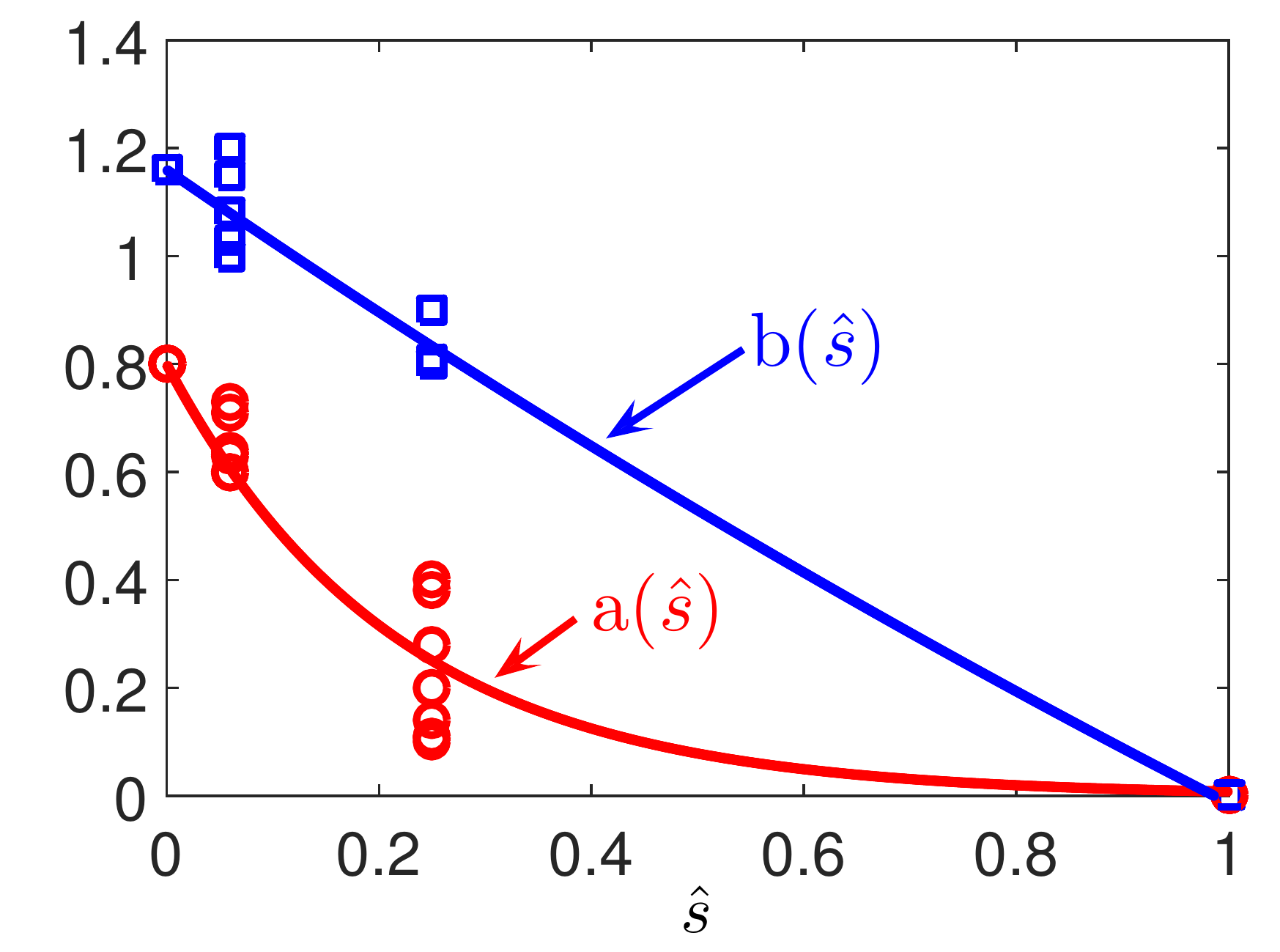}}
\caption{Illustration of parameters $a$ and $b$ with respect to the $\hat{q}$ (a) and $\hat{s}$ (b). }
\label{Parameters}
\end{figure}

\subsubsection{Parameter Prediction}

We derive the optimal values of parameters $a$ and $b$ following the least squared error (LSE) fitting criteria~\cite{num_scaling_SPIE}, and plot the model curves in Fig.~\ref{parametersFitting_QP} and~\ref{parametersFitting_SR}. Results indicate that Eq.~\eqref{Qrate} could describe the trend of NQQ and NQS very well with very small root mean squared error (RMSE) $e$.

It is shown that parameters $a$ and $b$ are $q$ and $s$ dependent, but since we model the perceptual response by adapting the quality through varying the $q$ and $s$ against the $q_{\min}$ and $s_{\max}$ independently, Figure~\ref{Parameters} illustrates $a$ and $b$ with respect to the $\hat{q}$ and $\hat{s}$ respectively. We have found that a Butterworth function could explain the $a(\hat{q})$ and $b(\hat{q})$, i.e.,
\begin{equation}\label{a_q}
\frac{k_1}{1+k_2\cdot{\hat{q}}^{k_3}},
\end{equation}
while a exponential function for $a(\hat{s})$ and $b(\hat{s})$,  i.e.,
\begin{equation}\label{a_s}
k_1\cdot e^{-k_2\cdot{\hat{s}}}+k_3,
\end{equation} with all fixed parameters shown in Table~\ref{tab:a_b_para}.

\begin{figure}[t]
\centering
 \subfigure[Balboa*]{ \includegraphics[scale=0.1]{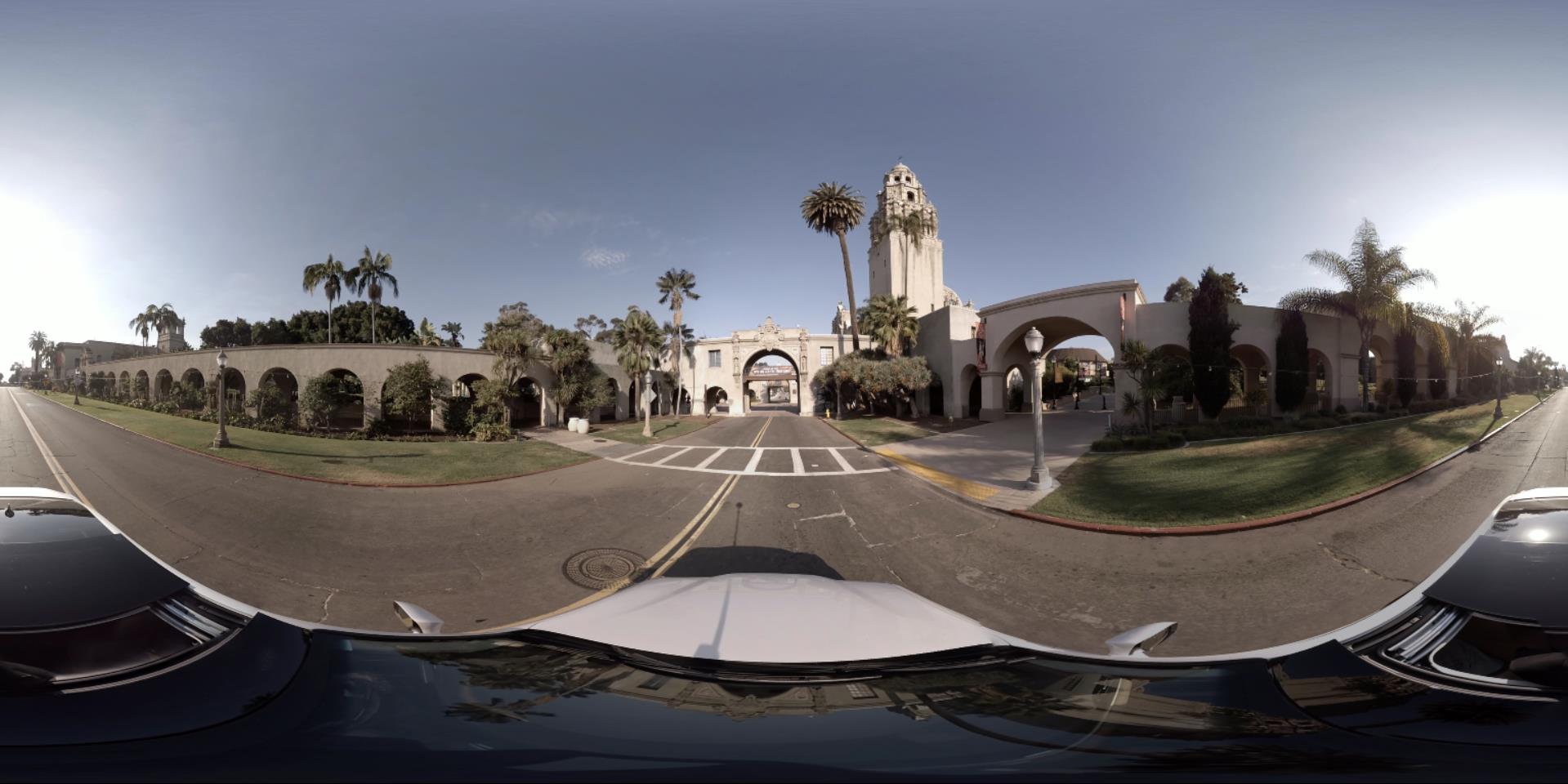}}
 \subfigure[PoleVault*]{ \includegraphics[scale=0.1]{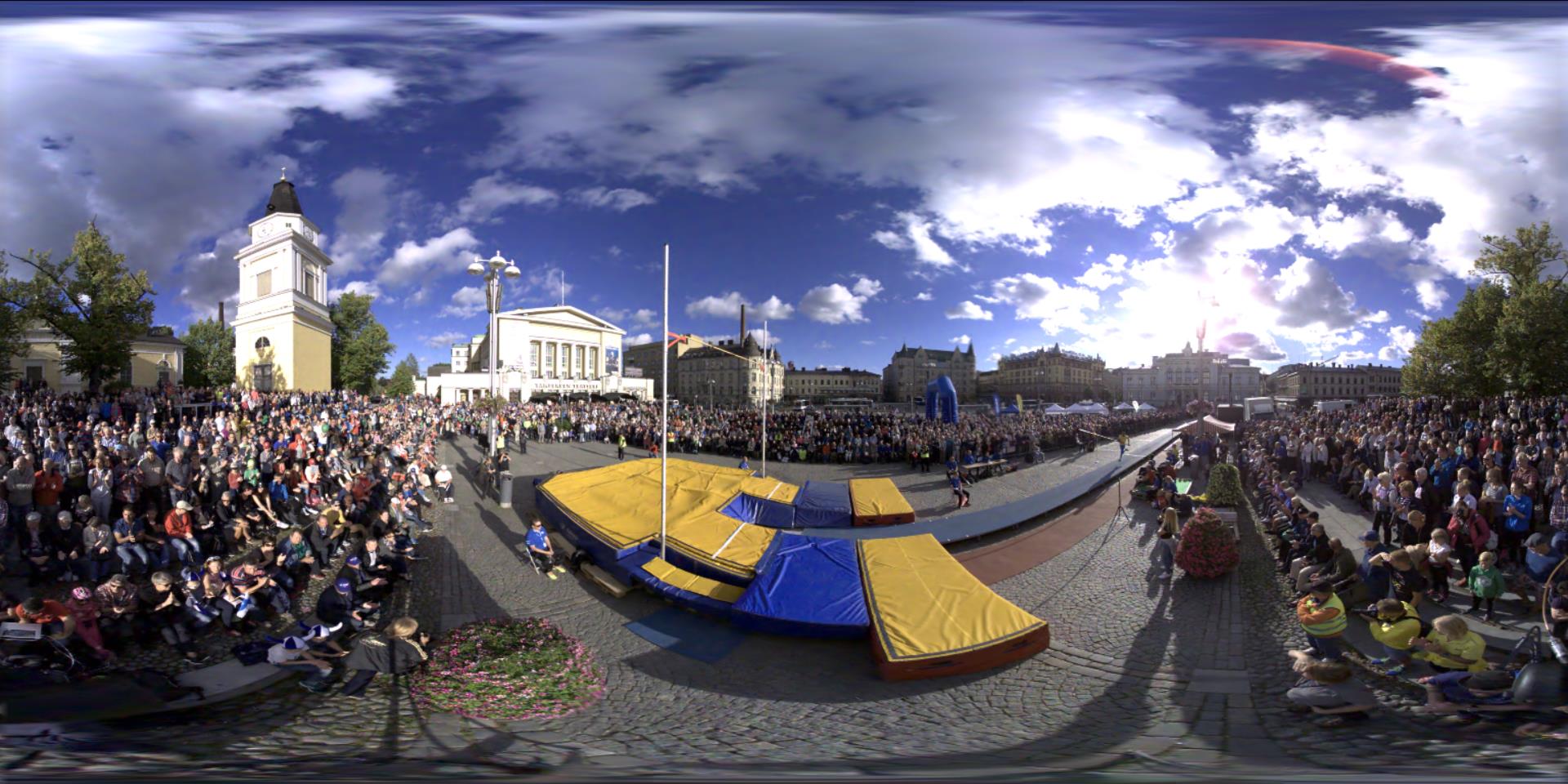}}
 \subfigure[Hangpai2\dag]{ \includegraphics[scale=0.08]{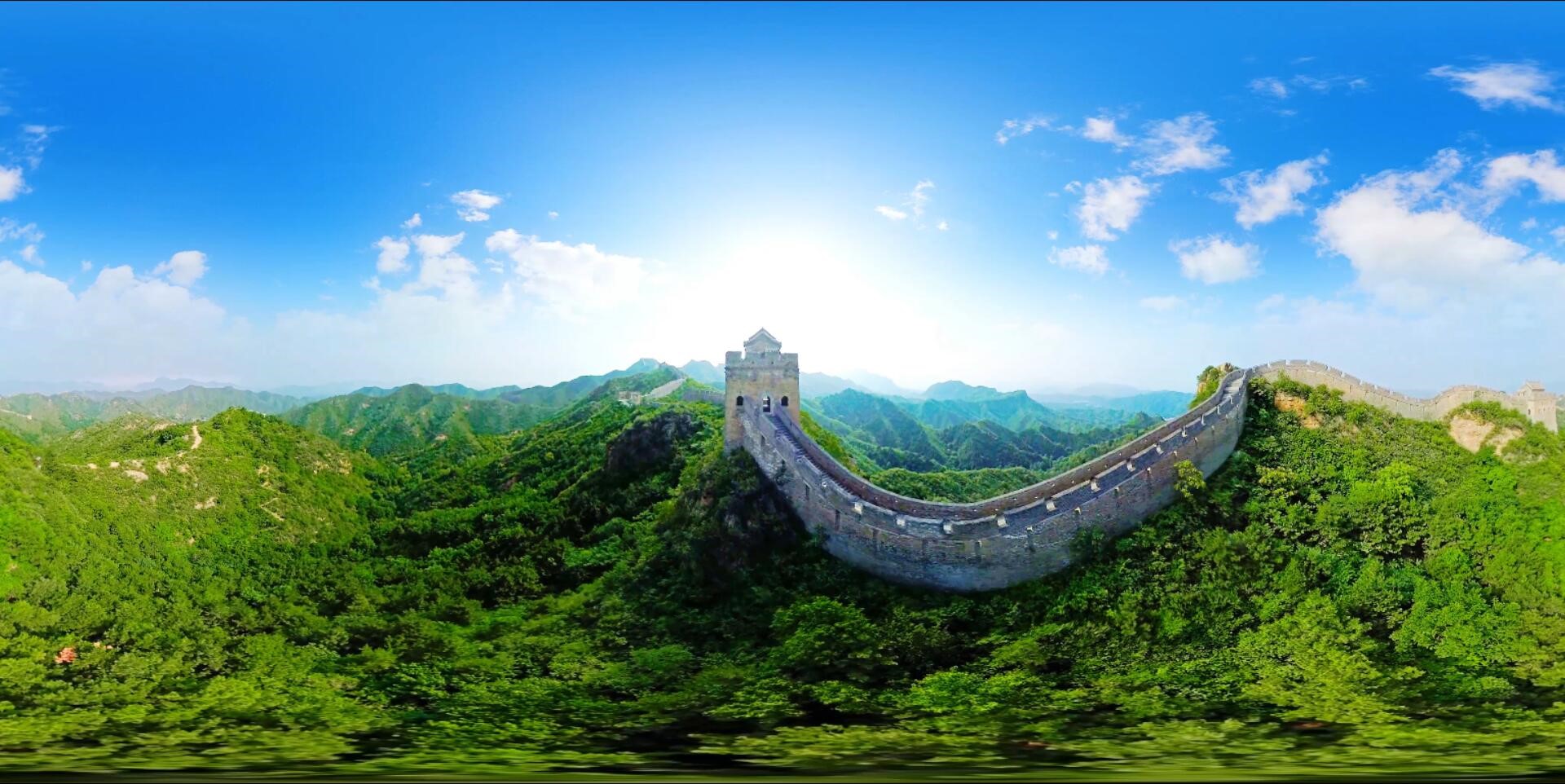}}
 \subfigure[Hangpai3\dag]{  \includegraphics[scale=0.08]{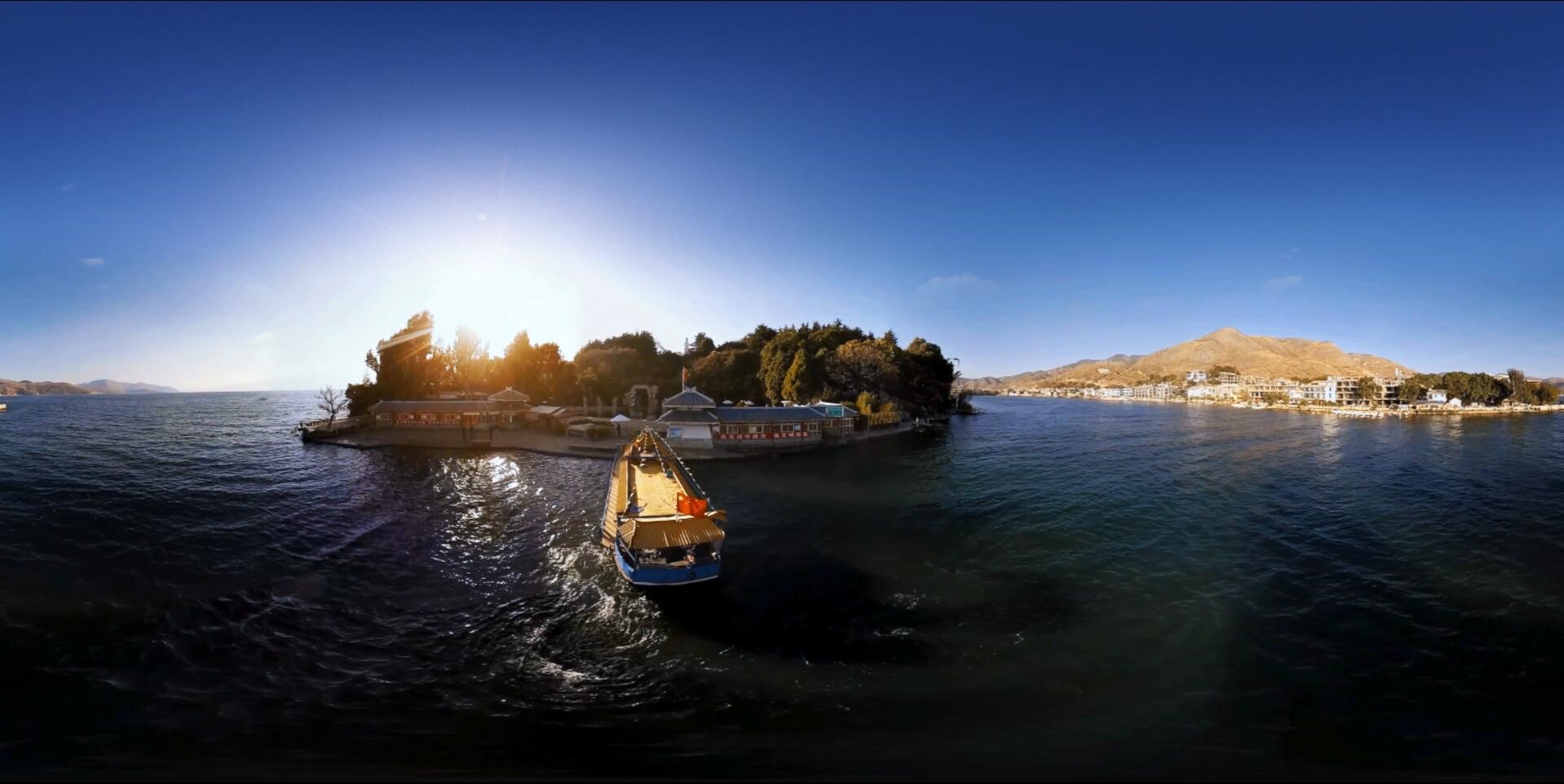}}
 \subfigure[Elephants2]{ \includegraphics[scale=0.076]{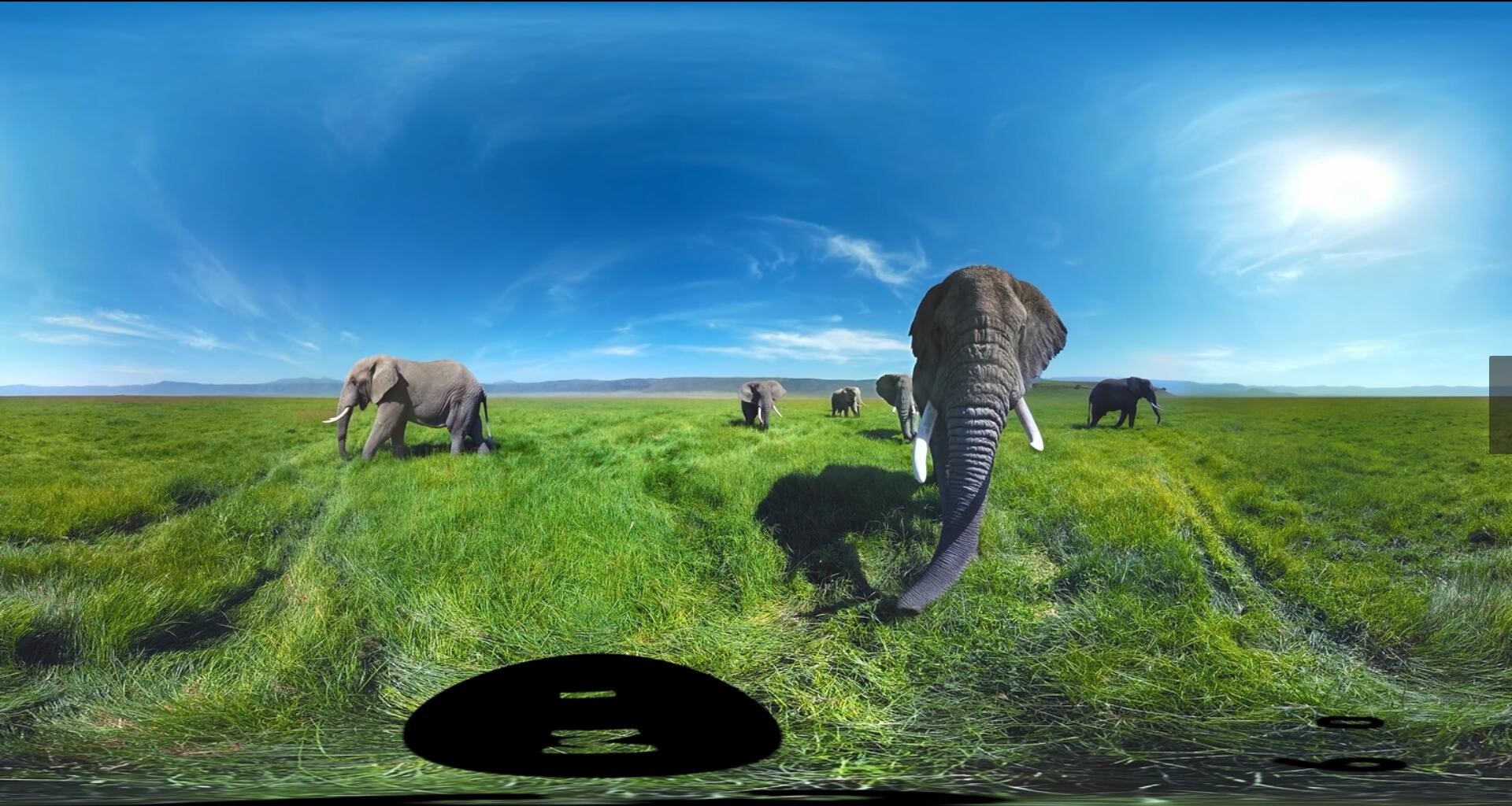}}
 \subfigure[NewYork]{ \includegraphics[scale=0.1]{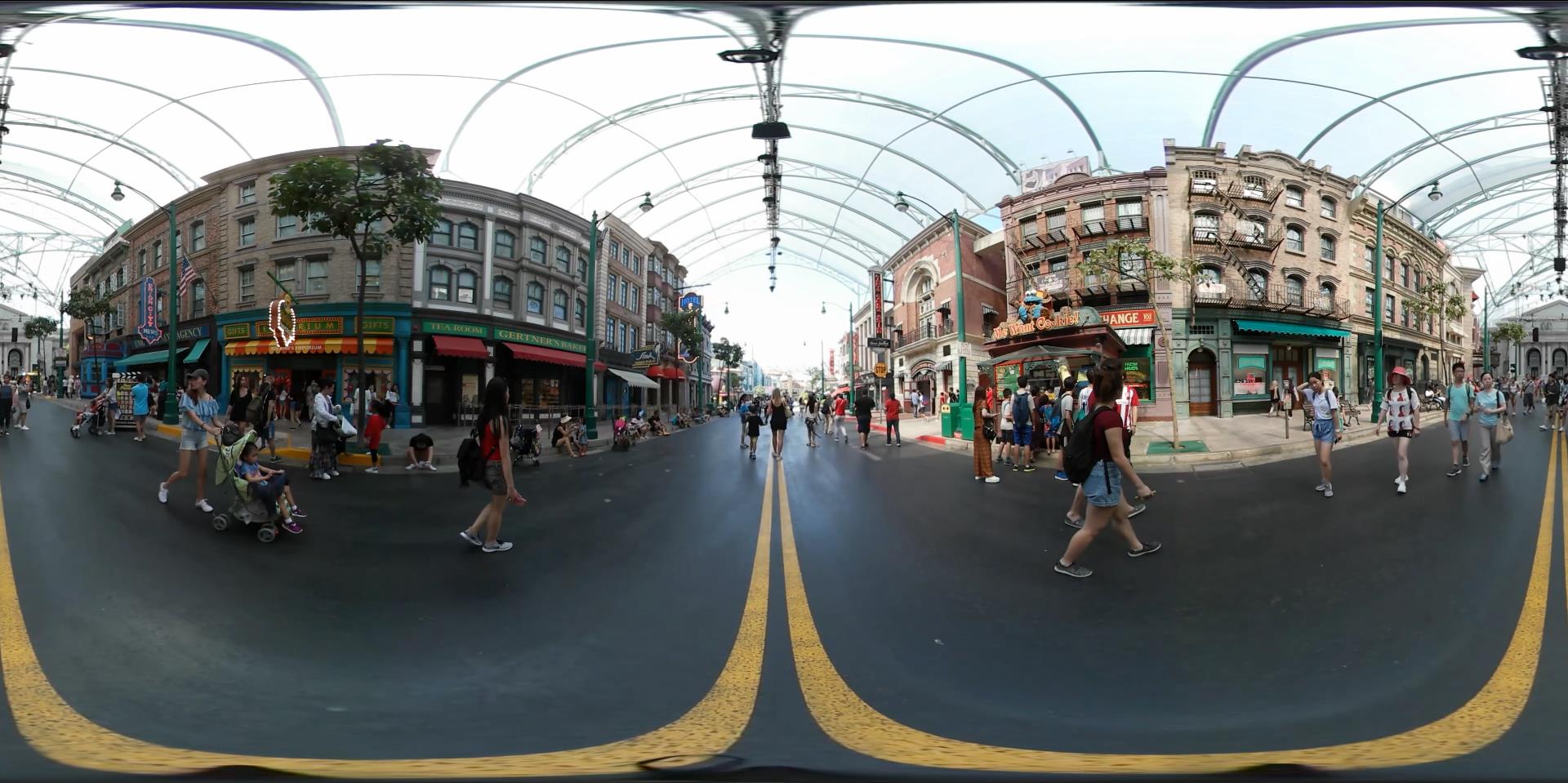}}
 \subfigure[Snowberg]{ \includegraphics[scale=0.1]{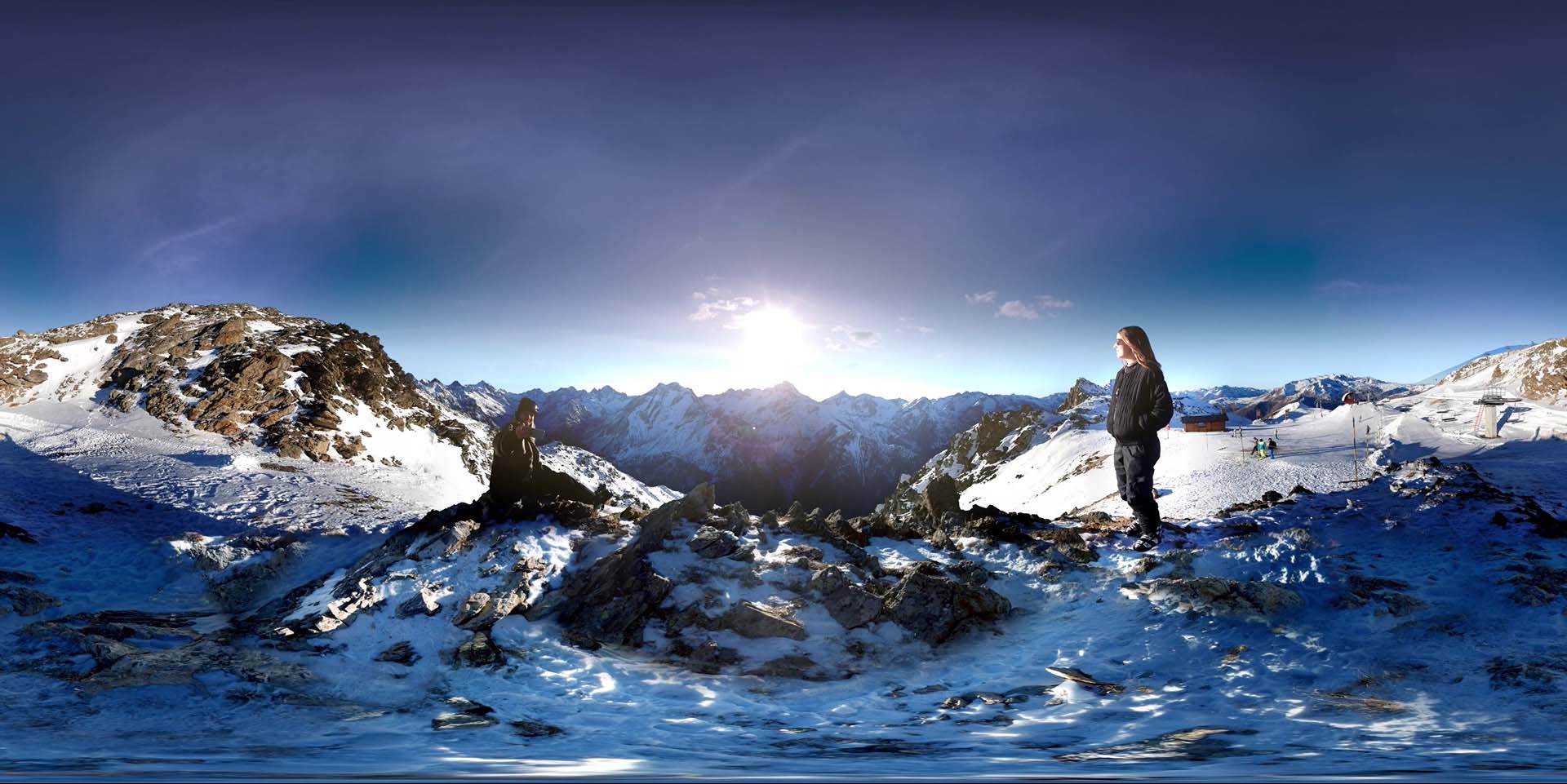}}
 \subfigure[Street2]{ \includegraphics[scale=0.075]{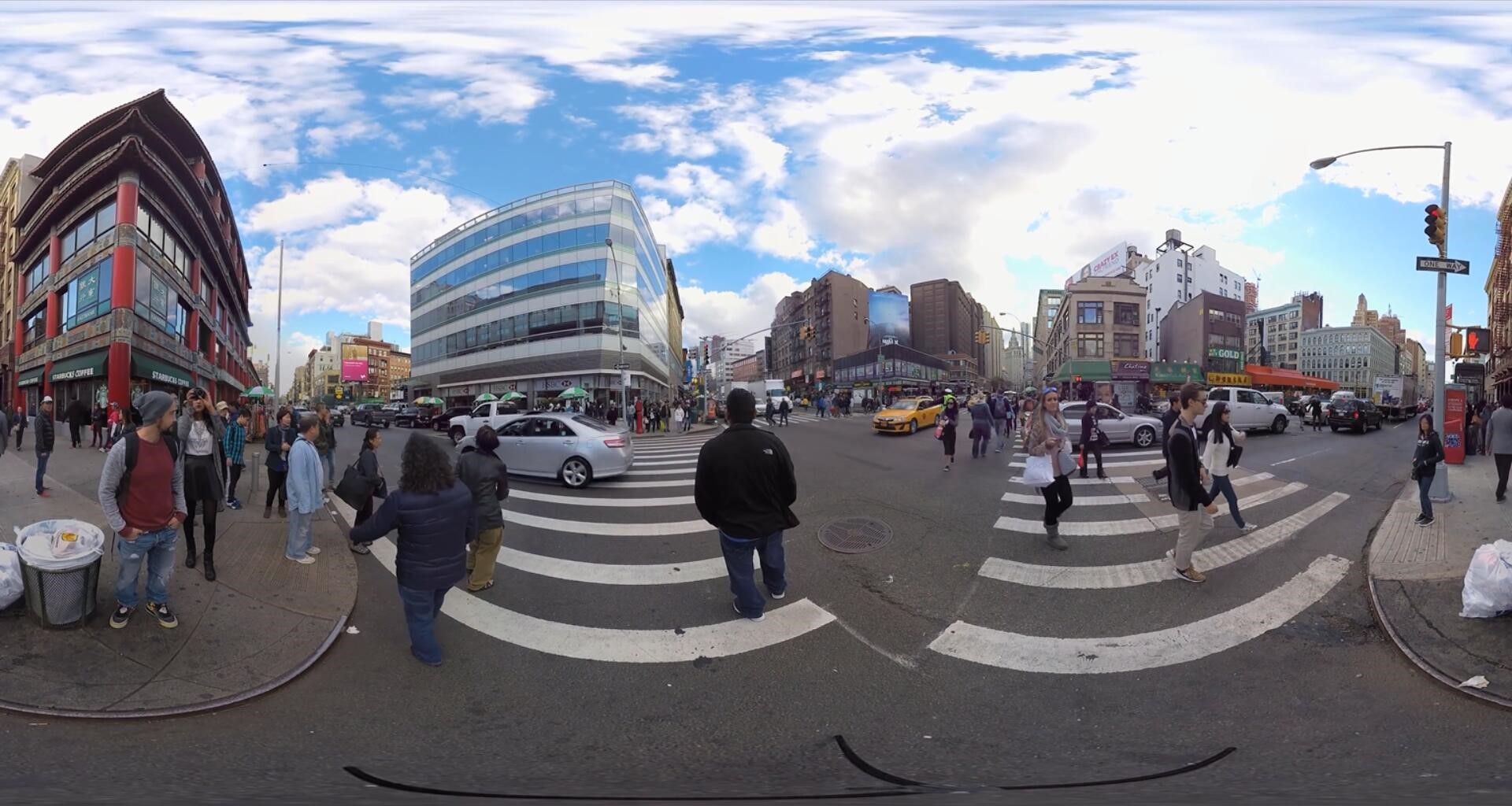}}
\caption{Illustration of Samples of the Immersive Videos Used for Cross Validation}
\label{ValidationImages}
\end{figure}

\begin{figure}[t]
\centering
 \subfigure[Balboa*]{ \includegraphics[scale=0.25]{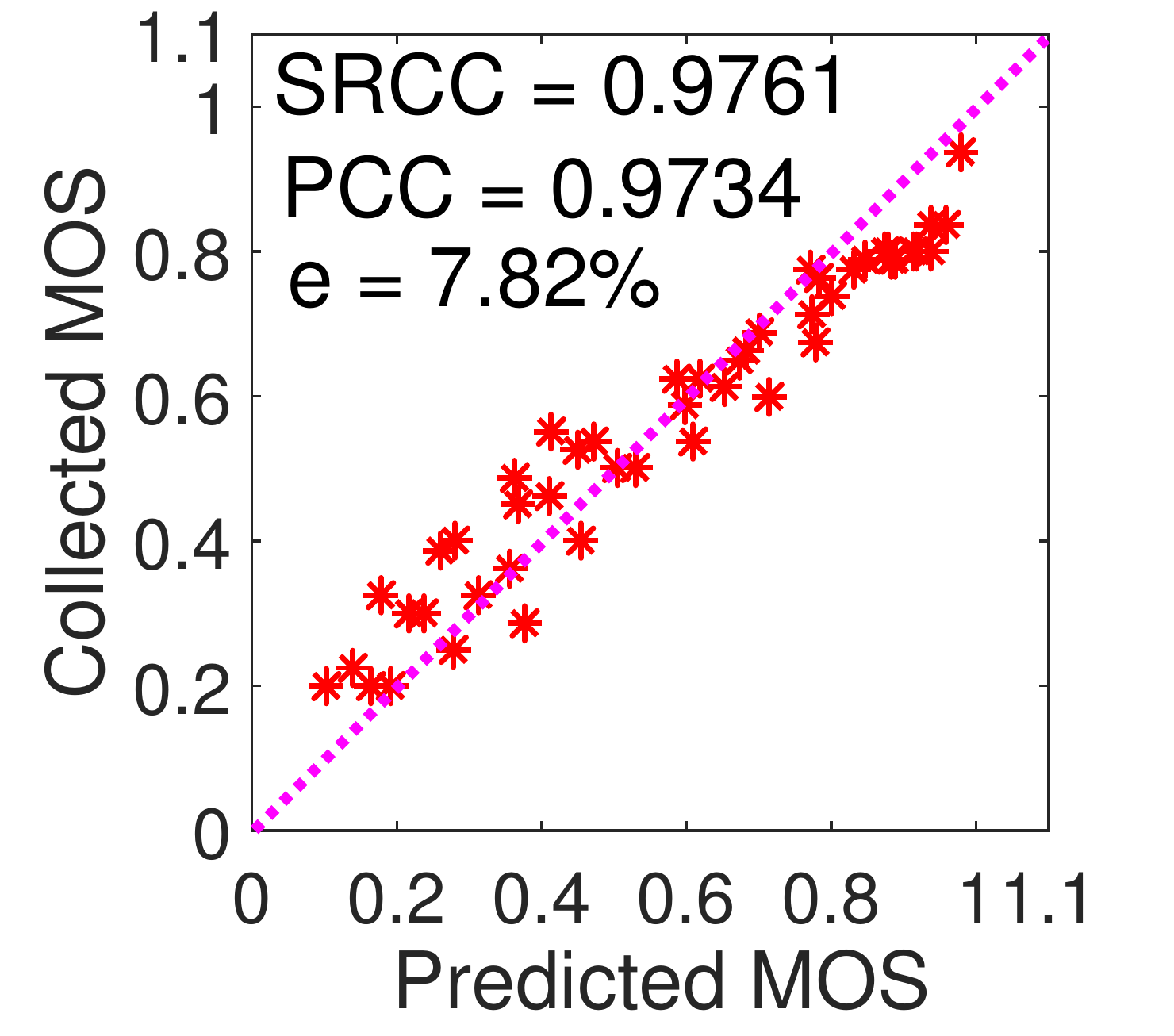}}
 \subfigure[PoleVault*]{ \includegraphics[scale=0.25]{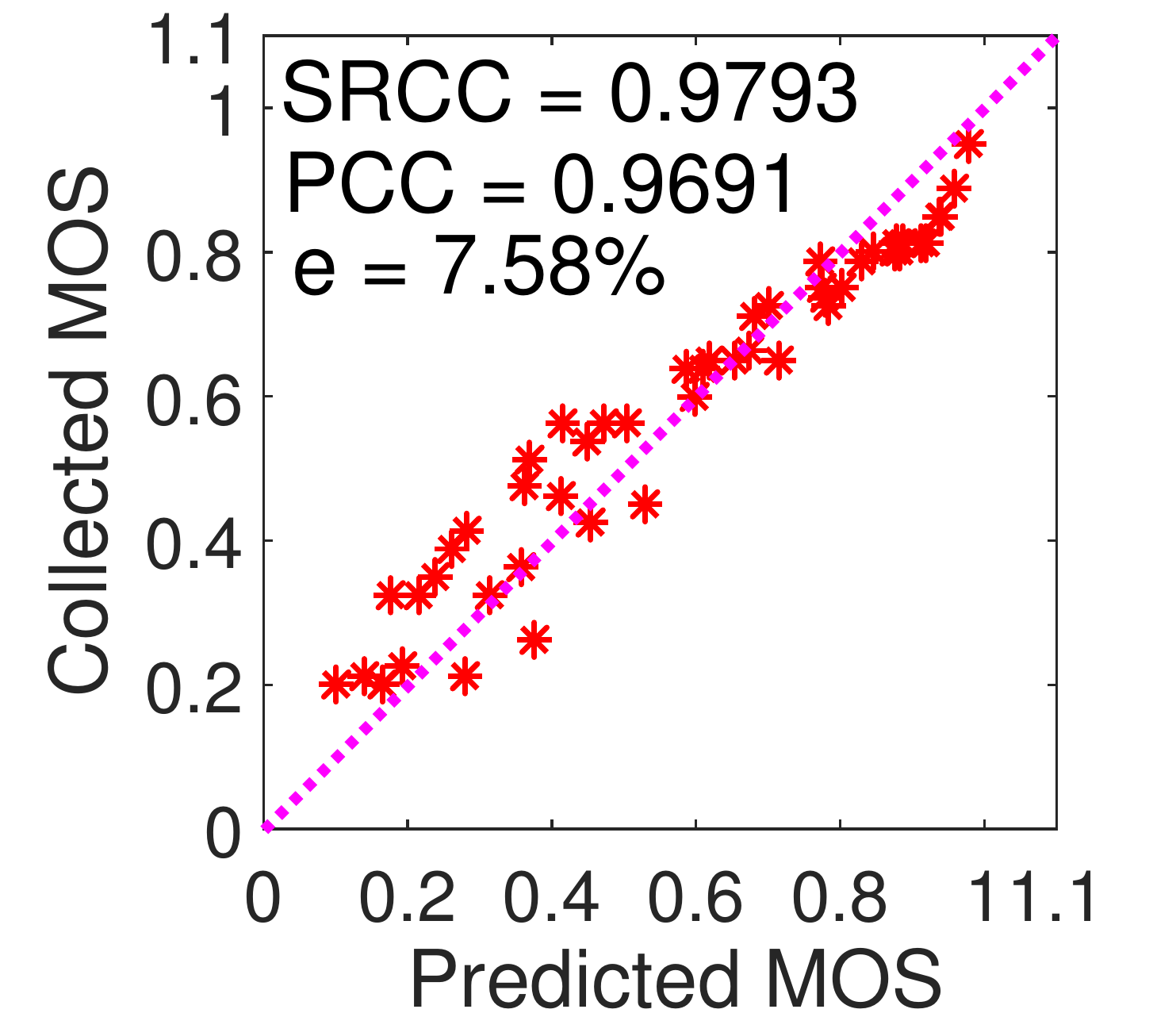}}
 \subfigure[Hangpai2\dag]{ \includegraphics[scale=0.25]{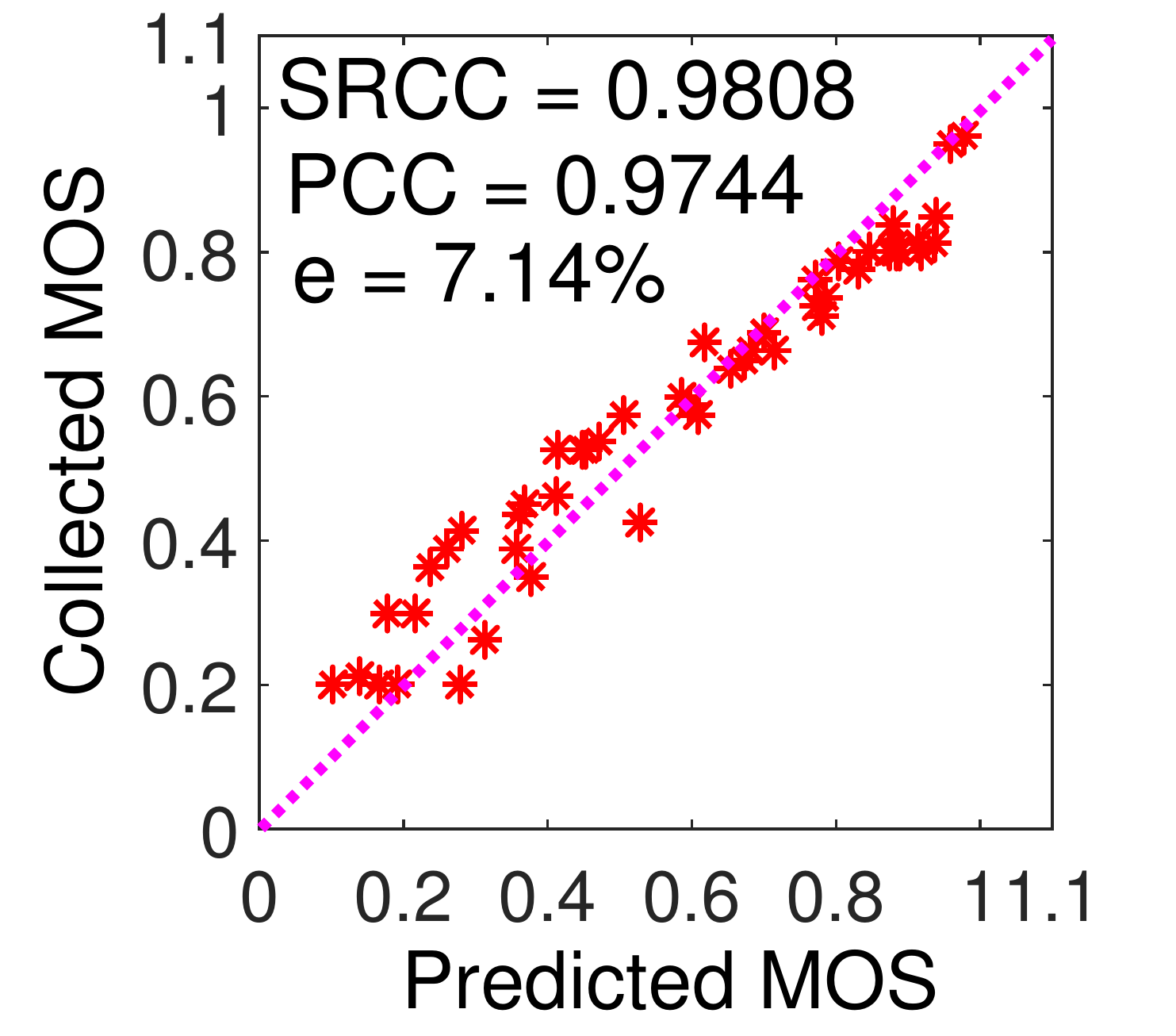}}
 \subfigure[Hangpai3\dag]{ \includegraphics[scale=0.25]{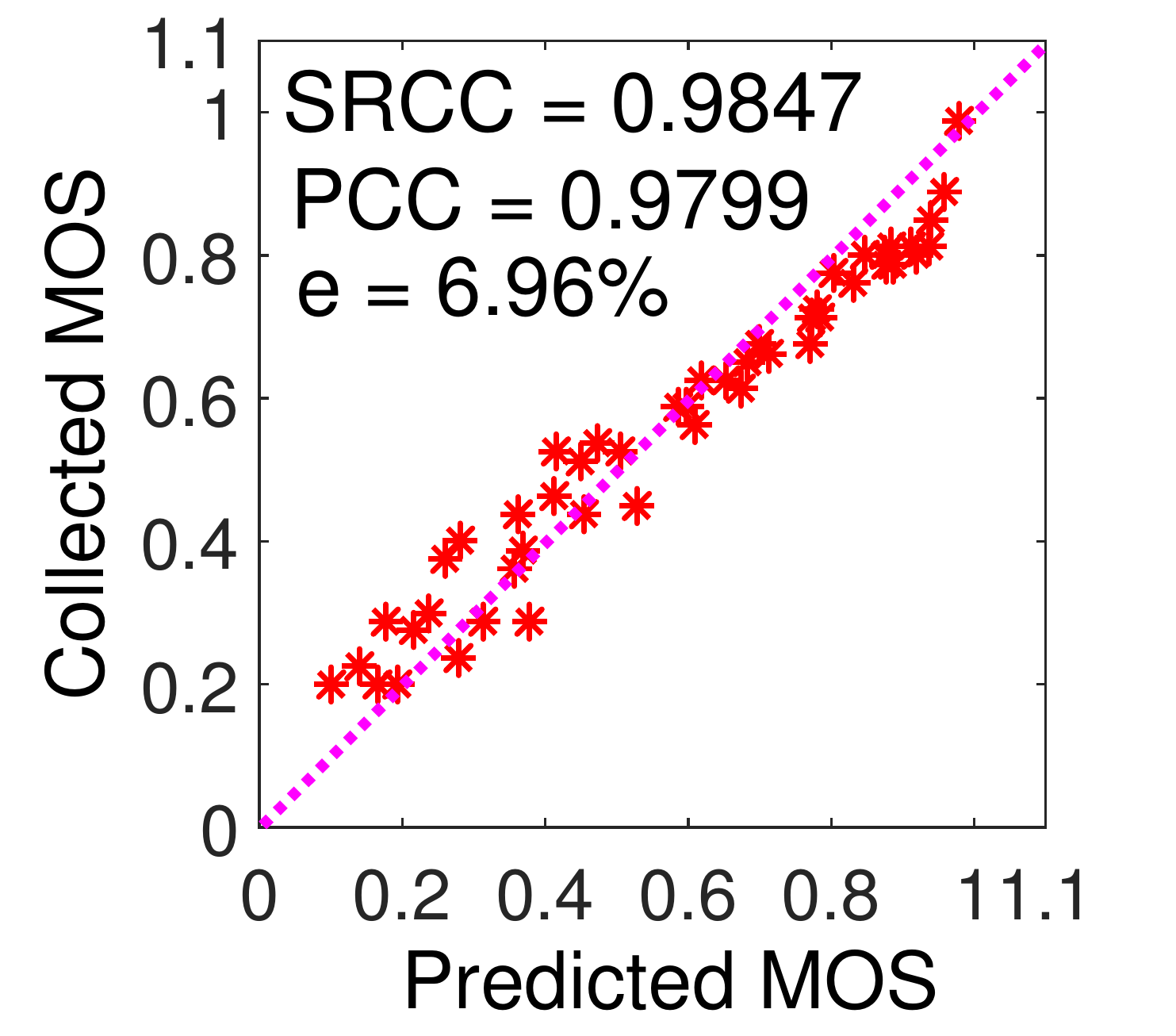}}
 \subfigure[Elephants2]{ \includegraphics[scale=0.25]{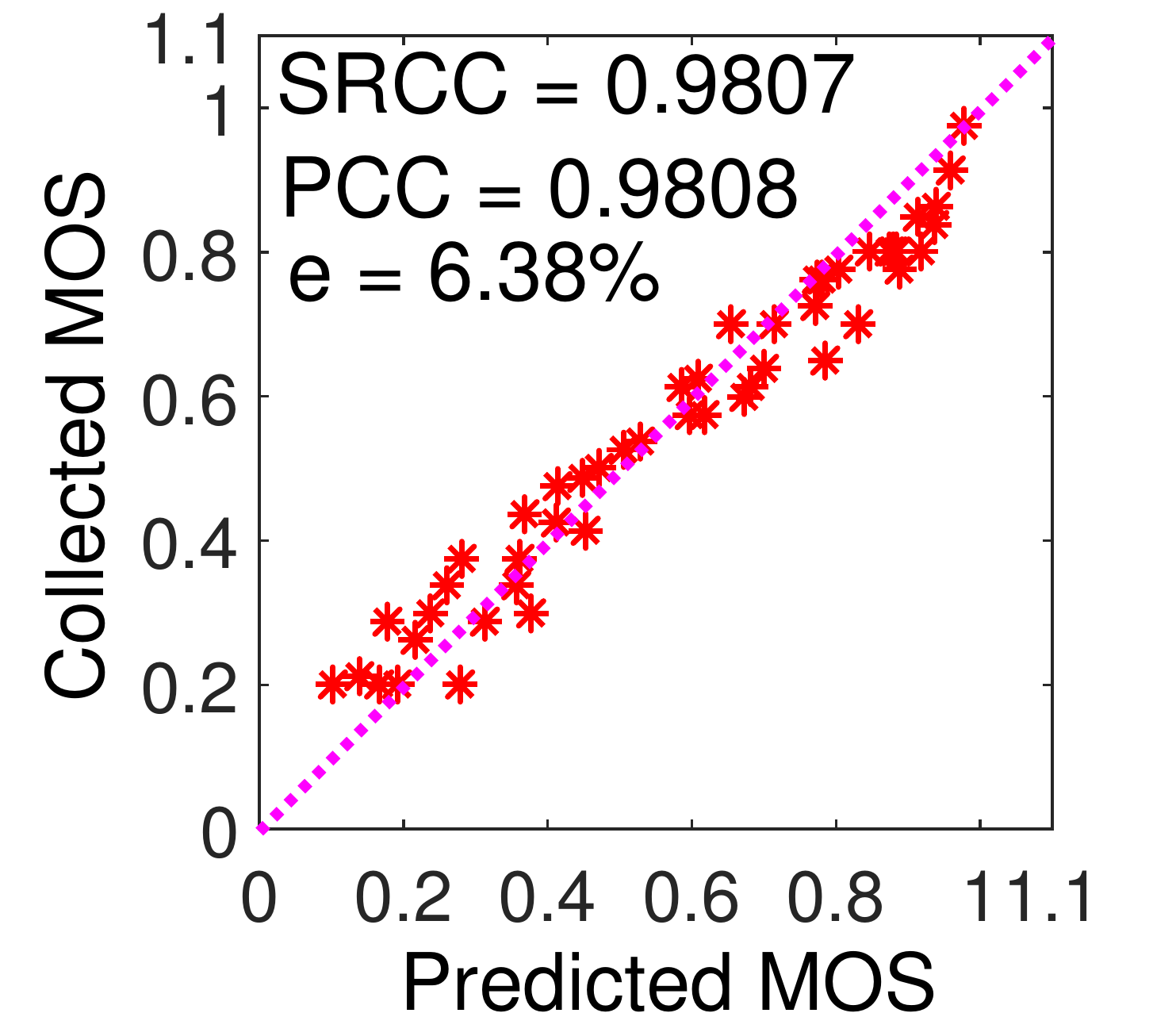}}
 \subfigure[NewYork]{ \includegraphics[scale=0.25]{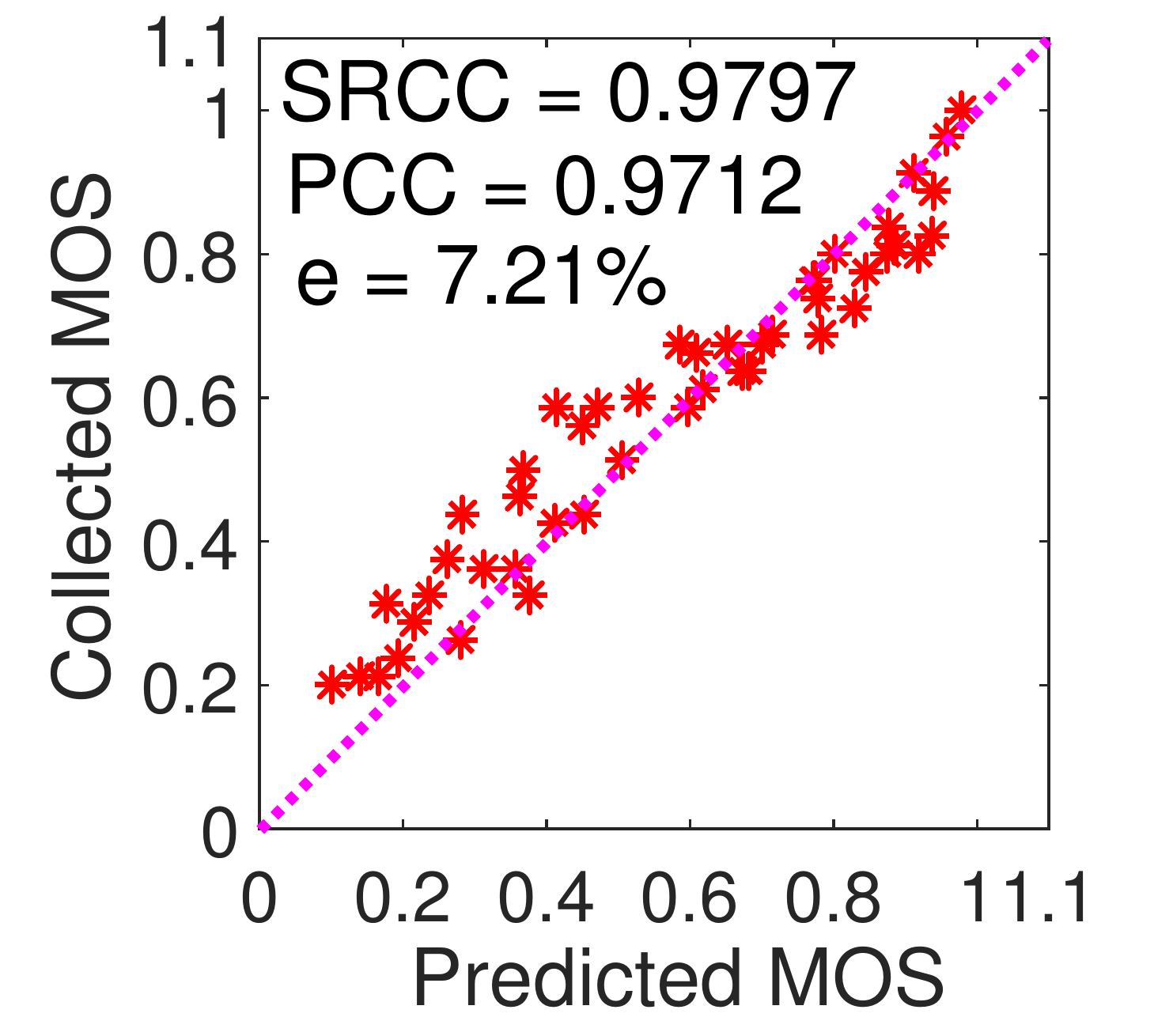}}
 \subfigure[Snowberg]{ \includegraphics[scale=0.25]{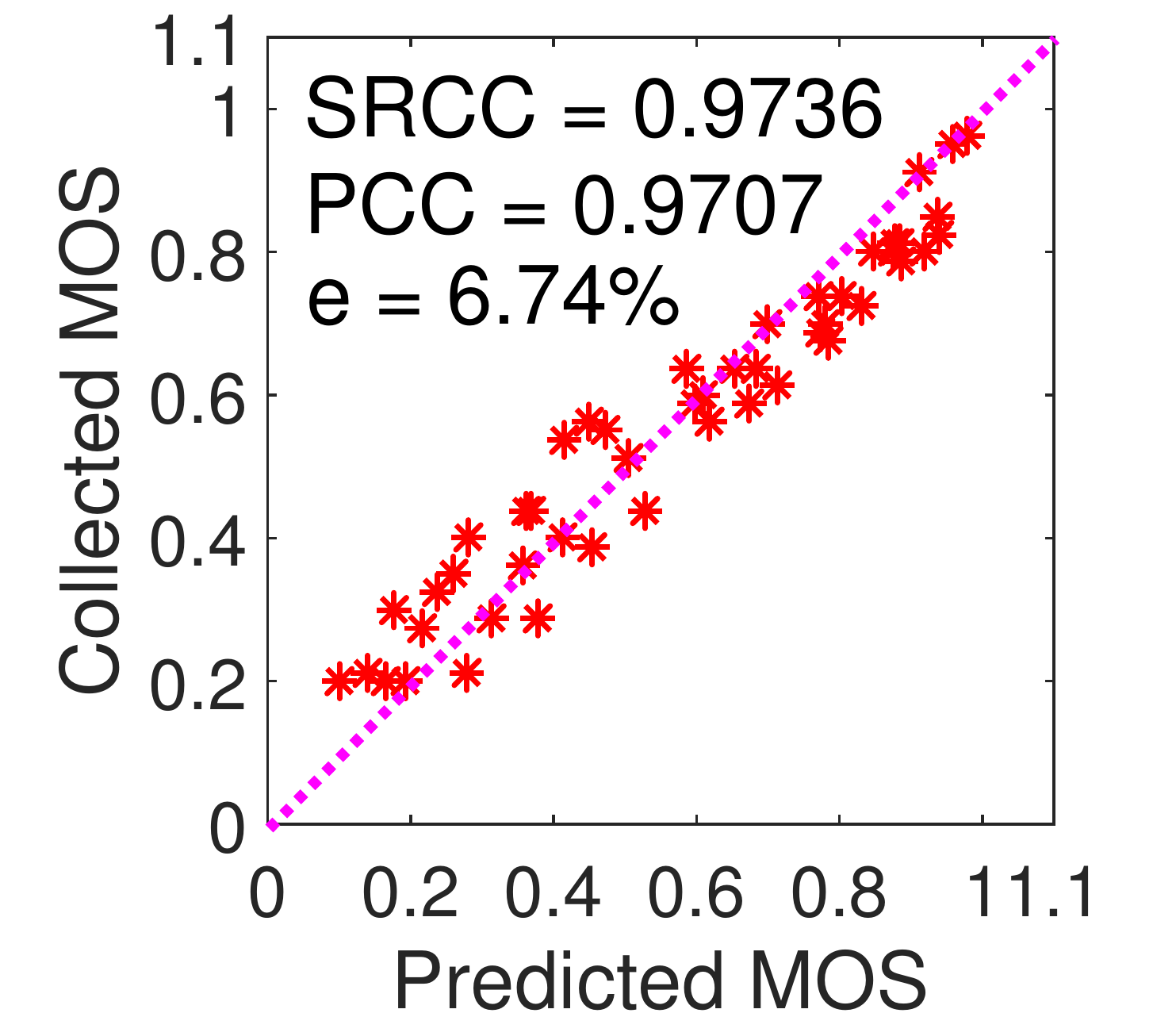}}
 \subfigure[Street2]{ \includegraphics[scale=0.25]{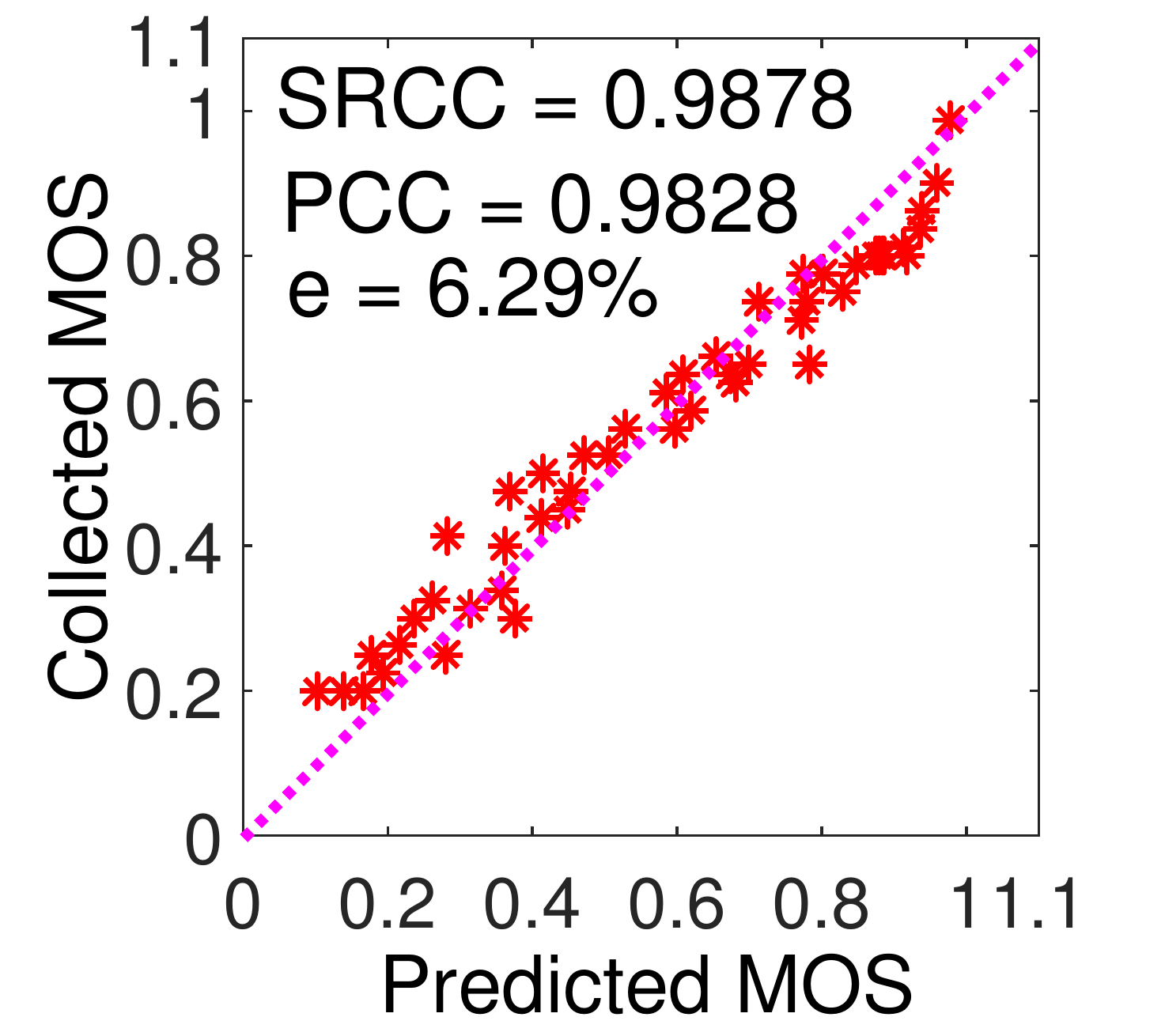}}
\caption{Illustration of the model accuracy: collected MOS are from the subjective
assessments and predicted MOS are derived using model \eqref{eq:overall_quality_model} with parameters derived in \eqref{a_q} and \eqref{a_s}.}
\label{ScatterDiagrams}
\end{figure}

\subsubsection{The Overall Analytical Model}
Towards the goal of developing an analytic model $Q(\tau,\hat{q},\hat{s})$ that can be used for assessing the perceptual quality when adapting FoV in immersive video applications, we could finally reach at model \eqref{eq:overall_quality_model}, following the aforementioned derivations, i.e.,
\begin{align}
Q(\tau,\hat{q},\hat{s}) = Q_{\max}\cdot\hat{Q}_{\rm NQQ}(\tau,\hat{q})\cdot\hat{Q}_{\rm NQS}(\tau,\hat{s}), \label{eq:overall_quality_model}
\end{align}
where
\begin{align}
\hat{Q}_{\rm NQQ}(\tau,\hat{q}) &= a(\hat{q})\cdot e^{-b(\hat{q})\cdot{\tau}}+(1-a(\hat{q})), \\
\hat{Q}_{\rm NQS}(\tau,\hat{s}) &= a(\hat{s})\cdot e^{-b(\hat{s})\cdot{\tau}}+(1-a(\hat{s})).
\end{align}

\subsubsection{Model Cross-Validation}
To ensure our model~\eqref{eq:overall_quality_model} is generally applicable, we invite another 79 subjects to participate the
cross-validation assessment. Each participant assesses all PVSs associated with one or two test videos.
Another two JVET test sequences (marked with *), two VRU (Virtual Reality Unity organization in China) test sequences (marked with \dag)~\cite{VRU}, and four YouTube 360$^\circ$ videos, as shown in Fig.~\ref{ValidationImages}, are chosen to produce the PVSs for validation. All PVSs are prepared with  two spatial resolutions (in addition to the native resolution) and four QPs jointly, as well as six $\tau$s, resulting in $2\times4\times6 = 48$ test samples for each video content. Note that this is different from
the aforementioned PVSs in Section~\ref{sec:subjective_assessment} where either $q$ or $s$ is fixed when adapting another factor.
We directly evaluate the joint impacts of the $q$ and $s$ on the perceptual quality with respect to the $\tau$ to validate the
accuracy of the \eqref{eq:overall_quality_model}.

As presented in Fig.~\ref{ScatterDiagrams}, we have found that model \eqref{eq:overall_quality_model} could predict the actual MOS very well (i.e., with both Pearson correlation coefficient (PCC)~\cite{Pearson_coeff} and Spearman's rank correlation coefficient (SRCC)~\cite{Xiaokai_TIP} close to 0.98), even with all fixed parameters.

\section{Quality-Bandwidth Optimized Streaming} \label{sec:application}

\begin{figure*}[t]
 \centering
 \subfigure[]{\includegraphics[width=0.9\linewidth]{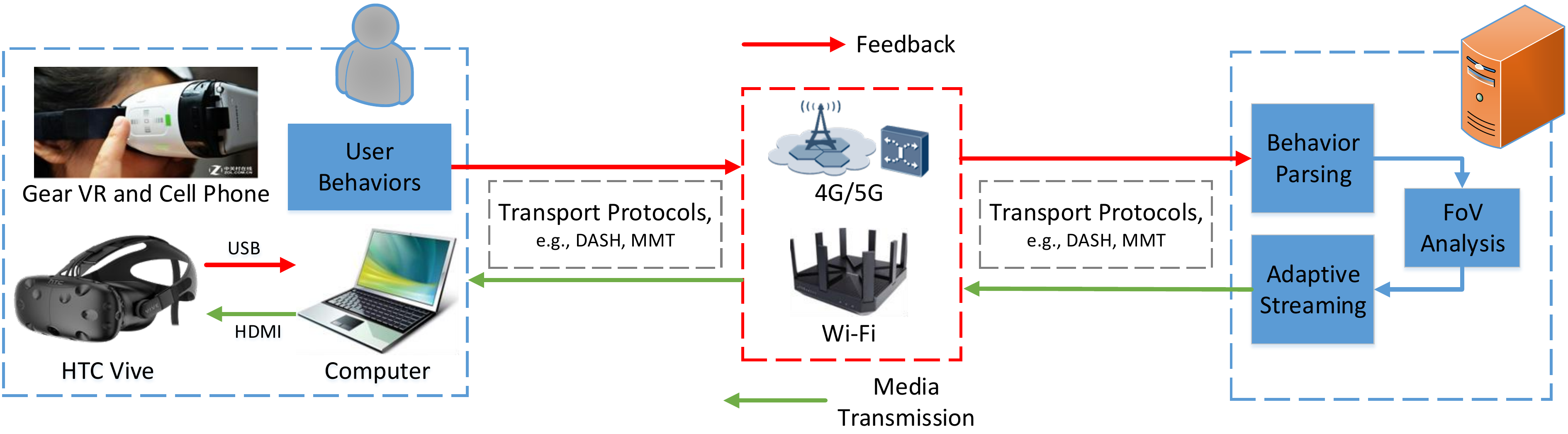} \label{FoV_Transmission_System}}
 \subfigure[]{\includegraphics[scale=0.25]{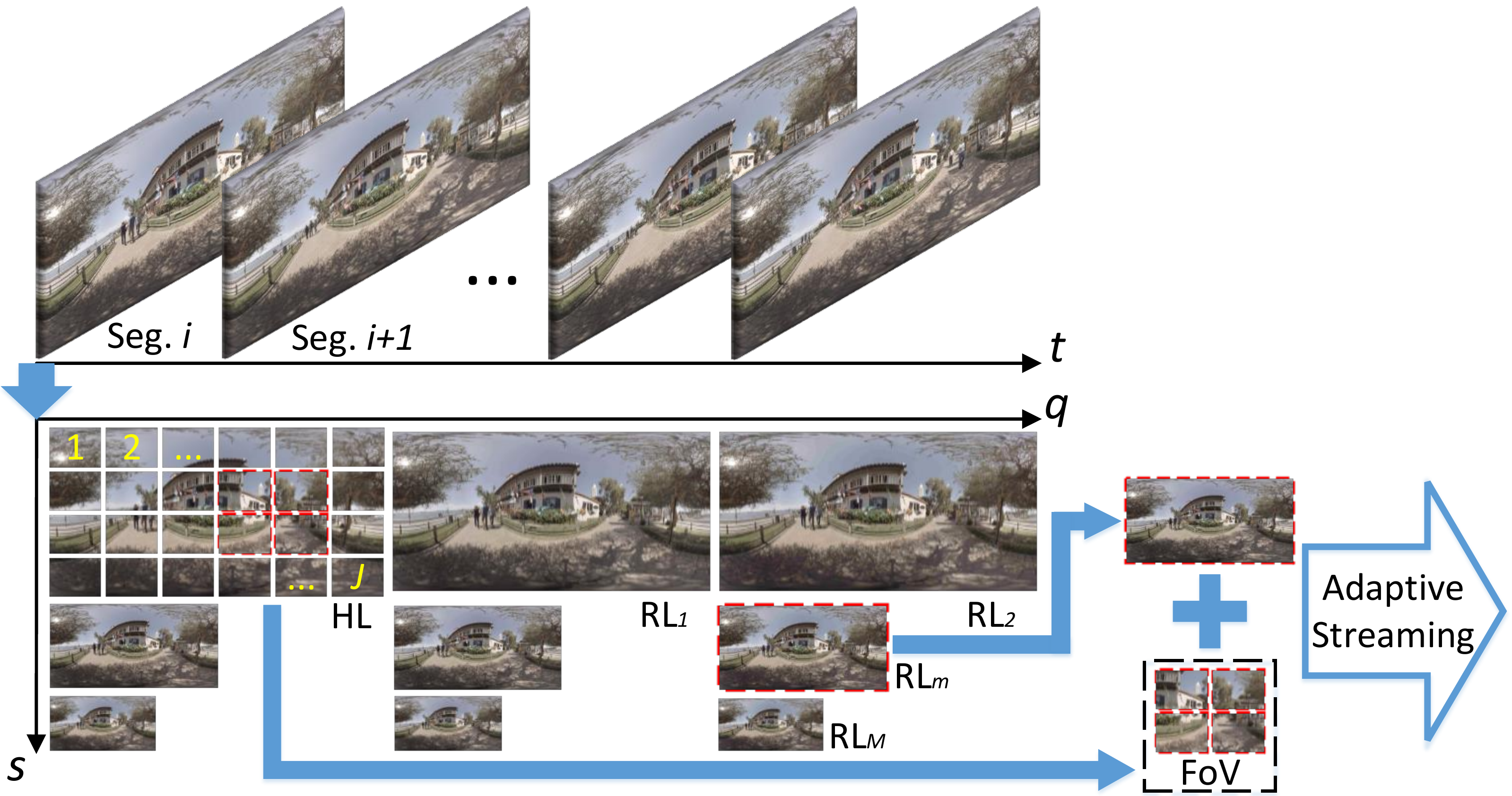}\label{QualityBandwidthOptimization}}
 \subfigure[]{\includegraphics[width=0.45\linewidth]{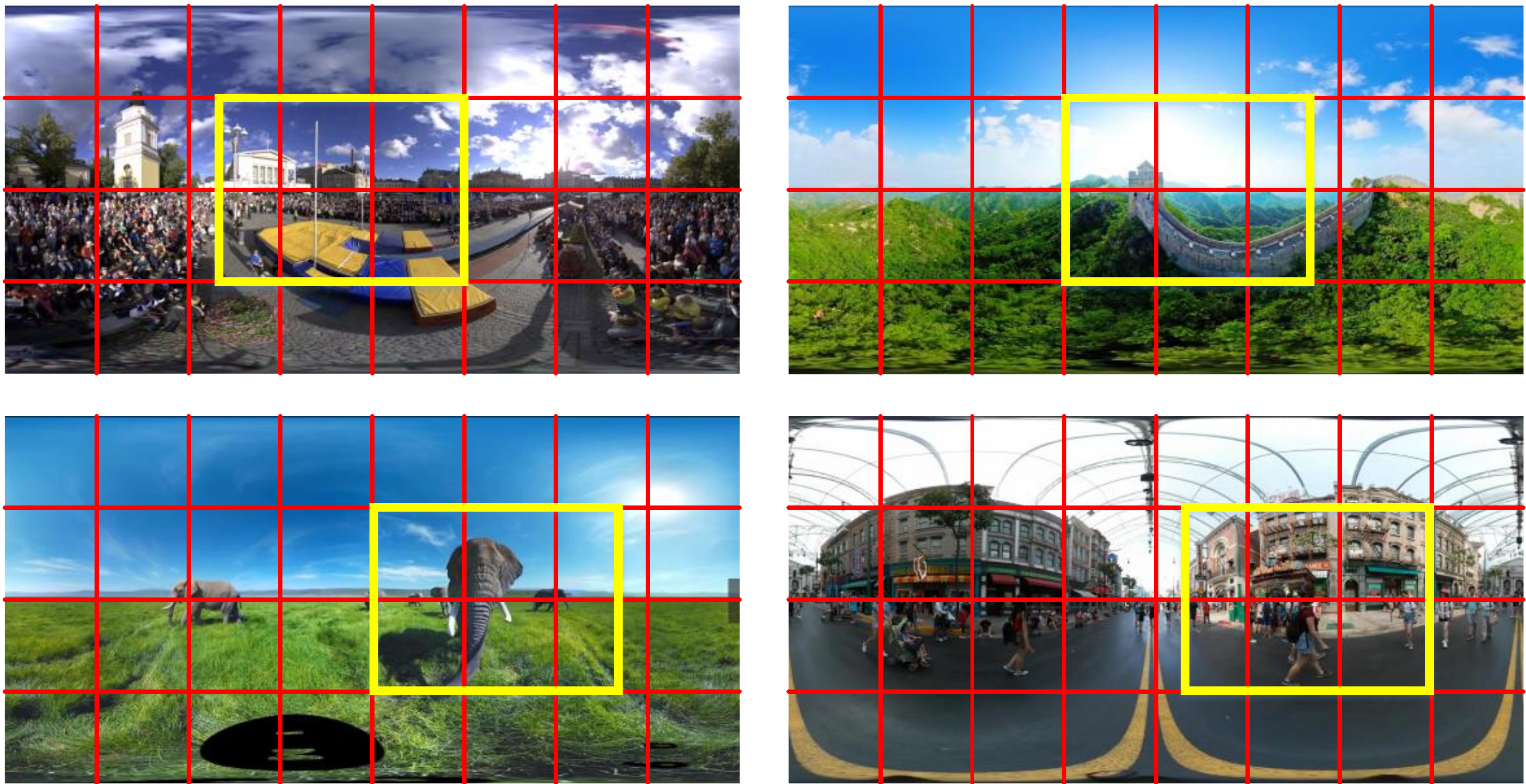} \label{SaliencyArea}}
 \caption{Illustration of the FoV adaptive streaming system for immersive video: (a) end-to-end architectural overview (b) multiscale adaptive FoV streaming (c) exemplified a $8\times4$ tile partition (red grid) and FoVs (highlighted in yellow). Each FoV contains meaningful saliency area.}

\end{figure*}

In this section, we explore the opportunity to apply our proposed model to guide the adaptive FoV streaming of the immersive video application under the bandwidth (or bit rate) constraint. Fig.~\ref{FoV_Transmission_System} exemplifies an architectural overview of the end-to-end system.  In such setup, user's behavior and content saliency are captured using the HMD, hand/eye tracker, etc. These information travel across
the network and parsed at edge servers to exact the appropriate video/image tiles (or patches) accordingly.  Here, to ensure the massive requests
from diverse users, we propose to prepare a multiscale repository where
video tiles are cached at various quality scales (i.e., via different combination of the quantization and spatial resolution) and users could retrieve
their interests adaptively without interfering others. Such multiscale mechanism has been used in practical transport protocols, such as the DASH (Dynamic Adaptive Streaming over HTTP)~\cite{DASH} and MMT (MPEG Media Transport)~\cite{MMT}.

\subsection{Problem Formulation}
As discussed previously, immersive video streaming can be realized cost-efficiently via the FoV or viewport adaptation~\cite{OptimizingCell,TwoTierSystem}. Often times, we will set current FoV with original high quality, but with reduced-quality elsewhere.  This would involve the quality refinement when we focus our attention to a new FoV. Apparently, our model can be of great benefit to reach an optimal solution under a limited bandwidth supply. In this work, we exemplify our model application in a practical adaptive FoV streaming system shown in Fig.~\ref{QualityBandwidthOptimization}.

More specifically, user will receive one or multiple high-quality tiles corresponding to current FoV (referred to as the ``high-quality layer" HL with $q_{\min}$ and $s_{\max}$), as well as another selective reduced-quality video covering full-size panoramic scene (i.e., referred to as the ``reduced-quality layer" RL with $q>q_{\min}$ and $s<s_{\max}$). This is to ensure the subject to perceive the content momentarily when adapting his/her focus.  In order to
support a great amount of potential users heterogeneously,  we propose to produce all quality scales immediately at server side, through scaling
the spatial resolution and applying various quantization stepsize. Note that for simplifying the implementation, we only allow the multiple tiles at HL and enforce a single tile for all RLs.

Ideally, the optimization can be formulated as
\begin{align}
 \max \limits_{\tau,\hat{q},\hat{s}}\quad
 &Q, \label{eq:OptimalProblem} \\
 s.t.\quad
 &R_{i}^{\rm FoV}+R_{mi}^{\rm RL}\leq B,\\
 &0<\hat{q},\hat{s}\leq1.
\end{align}
Thereinto,
\begin{align}
&\tau=\frac{R_{i}^{\rm FoV}+R_{mi}^{\rm RL}}{B}\cdot T, \\
&R_{i}^{\rm FoV}=\sum\nolimits_{j=1}^{n}R_{ij}^{\rm HL}, \\
&R_{mi}^{\rm RL}=R(\hat{q},\hat{s}).
\end{align}
$R_{i}^{\rm FoV}$ is the total bitrate of $n$ tiles covering the current FoV and $R_{\rm ij}^{\rm HL}$ is the HL bitrate of the $j$-th spatial tile, for the $i$-th segment in time dimension. $R_{mi}^{\rm RL}$ is the RL bit rate corresponding to the $i$-th segment coded at the $m$-th quality level shown in Fig.~\ref{QualityBandwidthOptimization}. In addition, $B$ and $T$ denote the constrained streaming bandwidth and the minimum duration of a segment that is sufficient for decoding and rendering, respectively.

In practice, the total bitrate consumption cannot be greater than $B$ for smooth playback. On the other hand, our model also suggests that
the user's subjective experience would be better if the quality gap is smaller and/or $\tau$ is shorter. However, smaller quality gap means the higher RL bitrate; and the more data exchange, the longer $\tau$ lasts. Thus, a fundamental problem is to select an appropriate RL that balances the trade-off between
the bandwidth requirement and refinement duration. Fortunately, this can be resolved via our model analytically. More details on the analytical derivation in different scenarios are unfolded as follows.

\subsection{Optimal Solution Under Continuous q}
We first solve the rate-constrained optimization problem in~\eqref{eq:OptimalProblem}, assuming the spatial resolution $s$ is at its native resolution (i.e., $\hat{s}=1$) and the quantization stepsize $q$ can be any value in the range of $q\in[q_{\min}, 160]$ ($q_{\min} = 8$). Here, $q$ = 160 corresponds to the QP 51 in either H.264/AVC~\cite{H264} and HEVC standard~\cite{HEVC}.

As demonstrated in our previous work~\cite{R-STAR}, bitrate consumption of any video can be represented by its spatial resolution, frame rate and quantization stepsize. Since we keep the frame rate unchanged and assume the native spatial resolution, the bitrate of a RL video is
\begin{align}
R_{mi}^{\rm RL}(\hat{q}) = R_{\max}\cdot\hat{q}^\alpha, \label{eq:RateModel}
\end{align}
where $R_{\max}$ is the actual rate when coding a video with $q_{\min}$ at its native spatial and temporal resolution. The model parameter $\alpha$ is content dependent~\cite{Ma2012Rate}, as listed in Table~\ref{Rmax-BDrate} for each cross-validation video sample.

Regarding the total bitrate of the current FoV, i.e., $R_i^{\rm FoV}$, it is closely related to number of tiles included.
In Fig.~\ref{SaliencyArea}, four cross-validation videos are presented as the example, where they are partitioned into $8\times4$ tiles with the FoVs highlighted.  Note that a FoV typically belongs to a meaningful salient region.

\begin{figure}[t]
\centering
 \subfigure[Balboa*]{\includegraphics[scale=0.23]{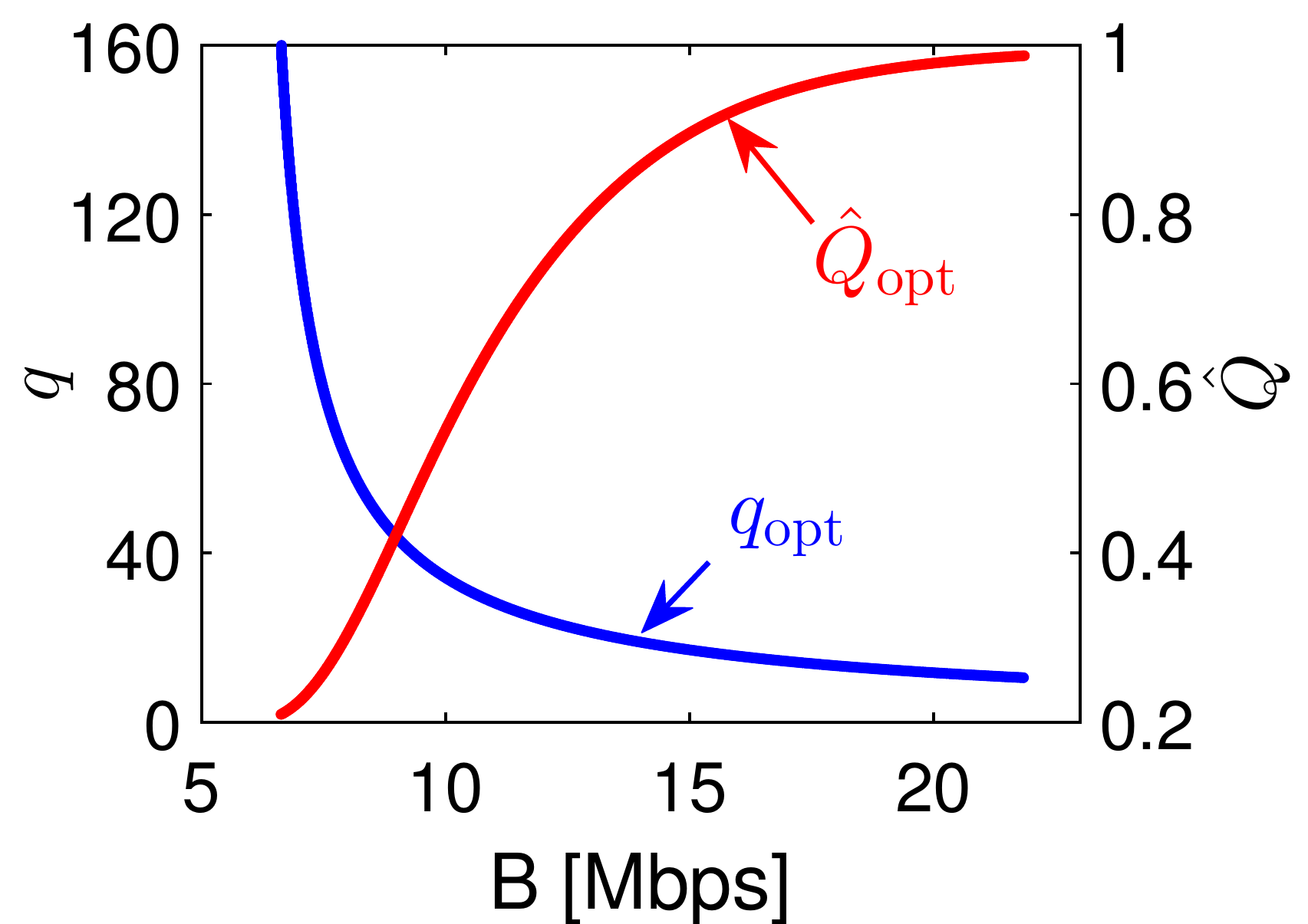}}
 \subfigure[PoleVault*]{\includegraphics[scale=0.23]{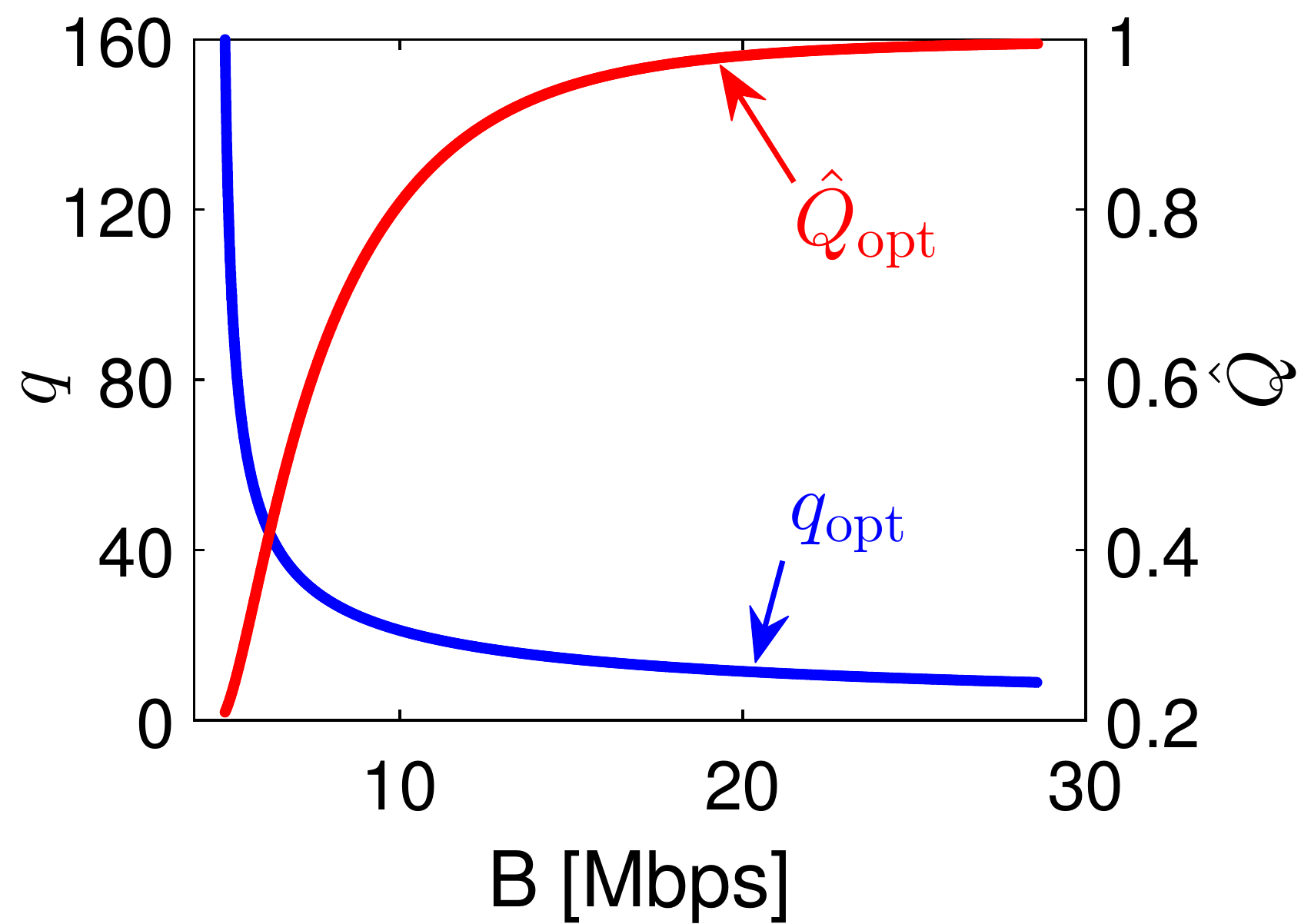}}
 \subfigure[Hangpai2\dag]{\includegraphics[scale=0.23]{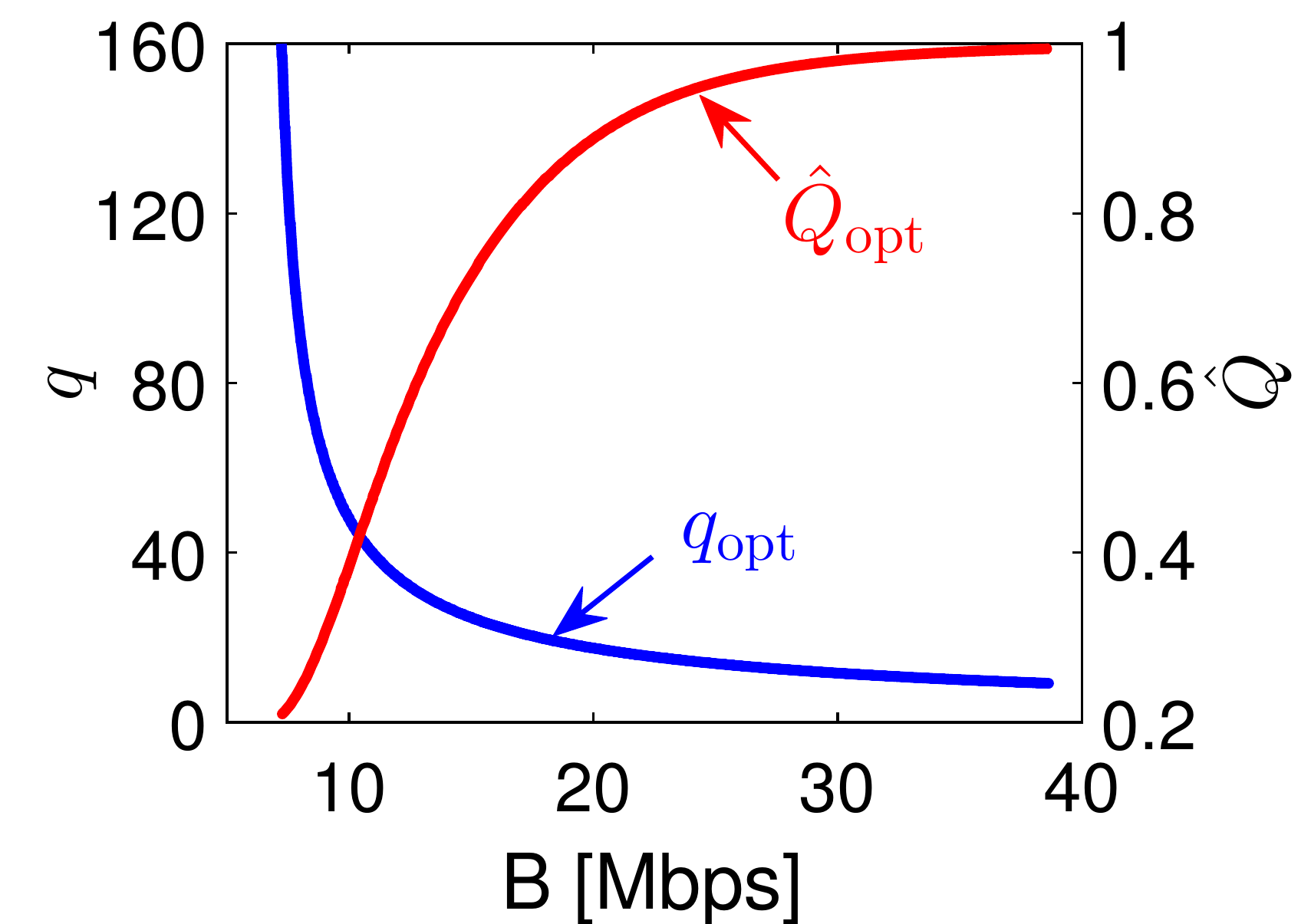}}
 \subfigure[Hangpai3\dag]{\includegraphics[scale=0.23]{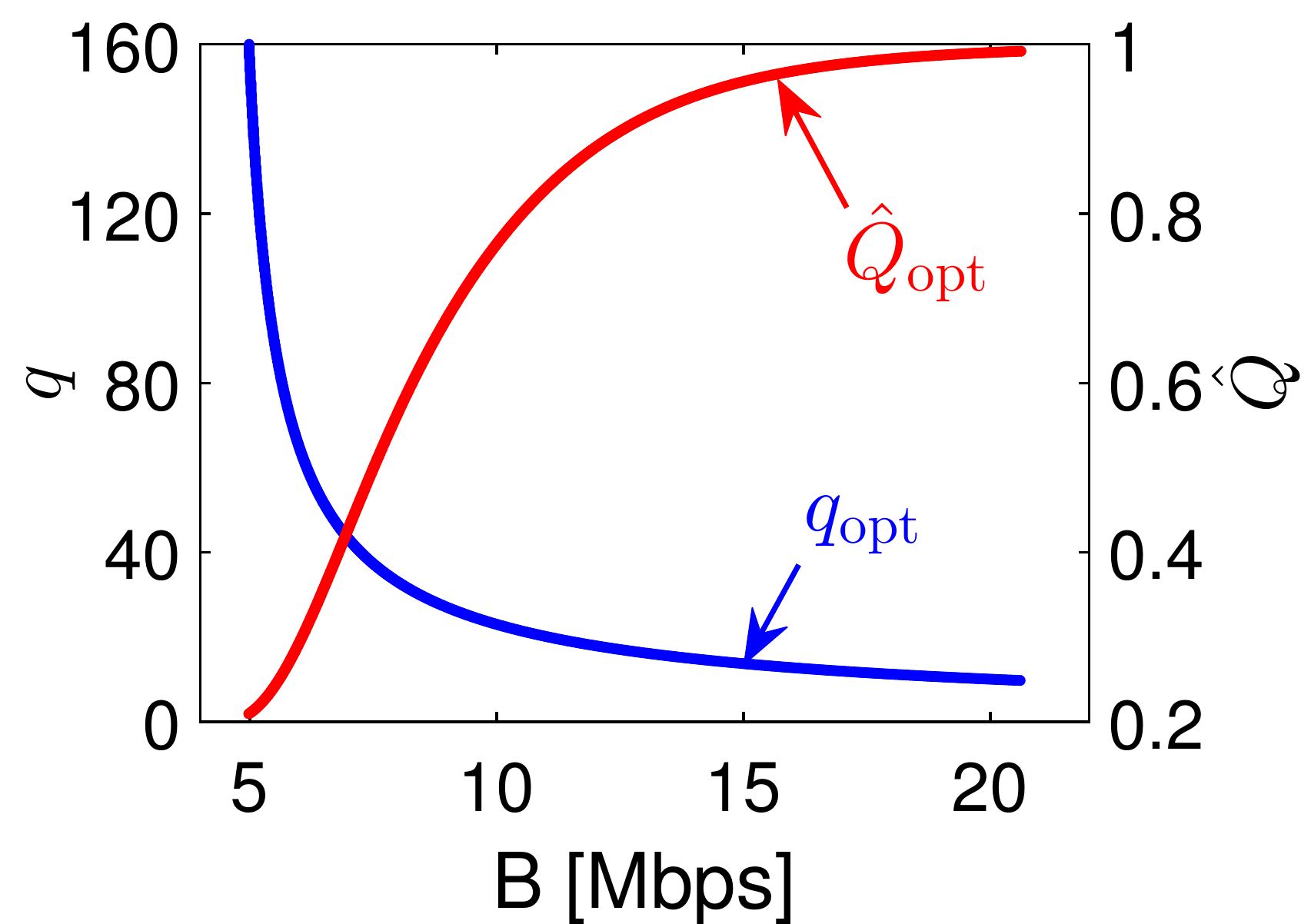}}
 \subfigure[Elephants2]{\includegraphics[scale=0.23]{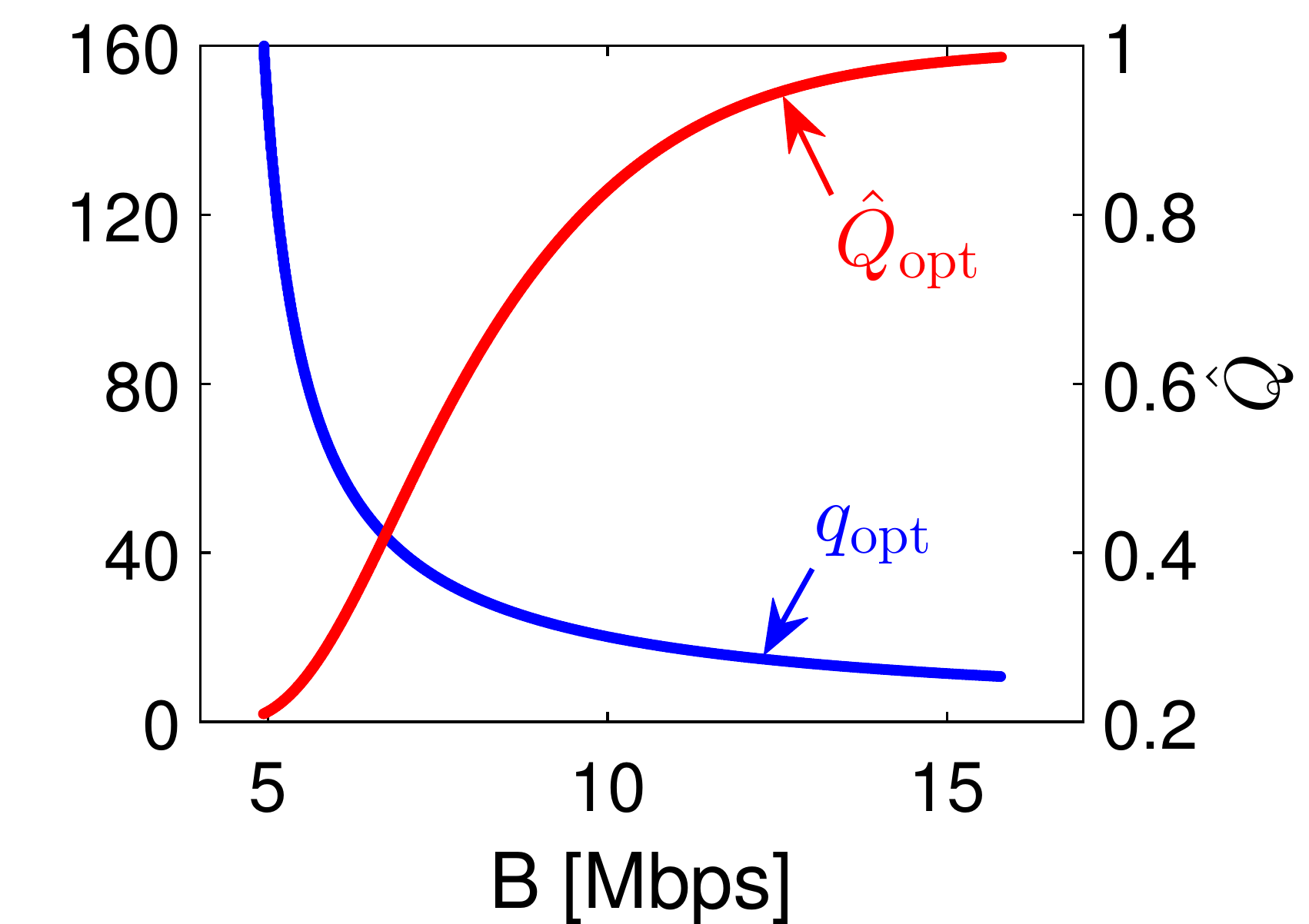}}
 \subfigure[NewYork]{\includegraphics[scale=0.23]{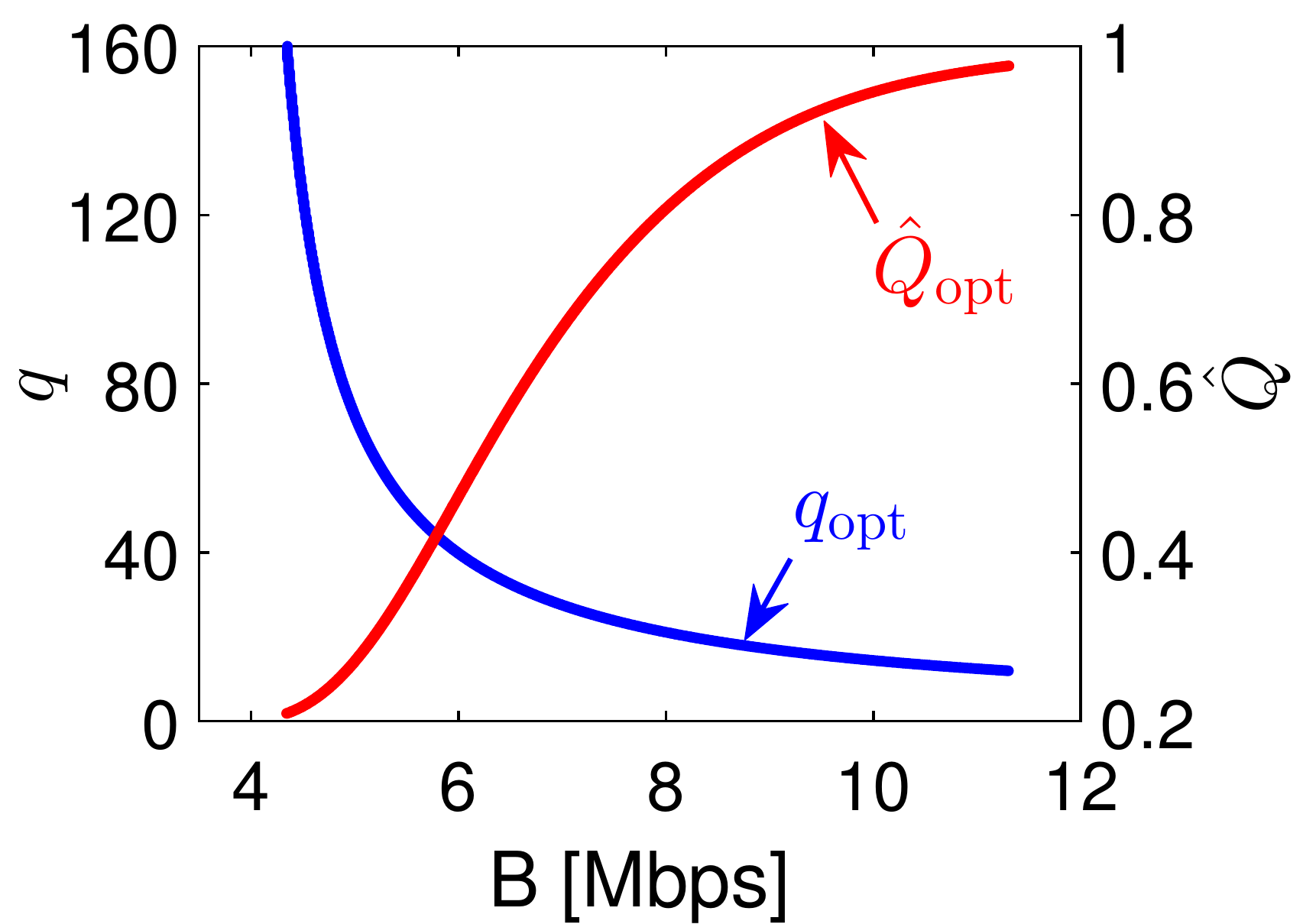}}
 \subfigure[Snowberg]{\includegraphics[scale=0.23]{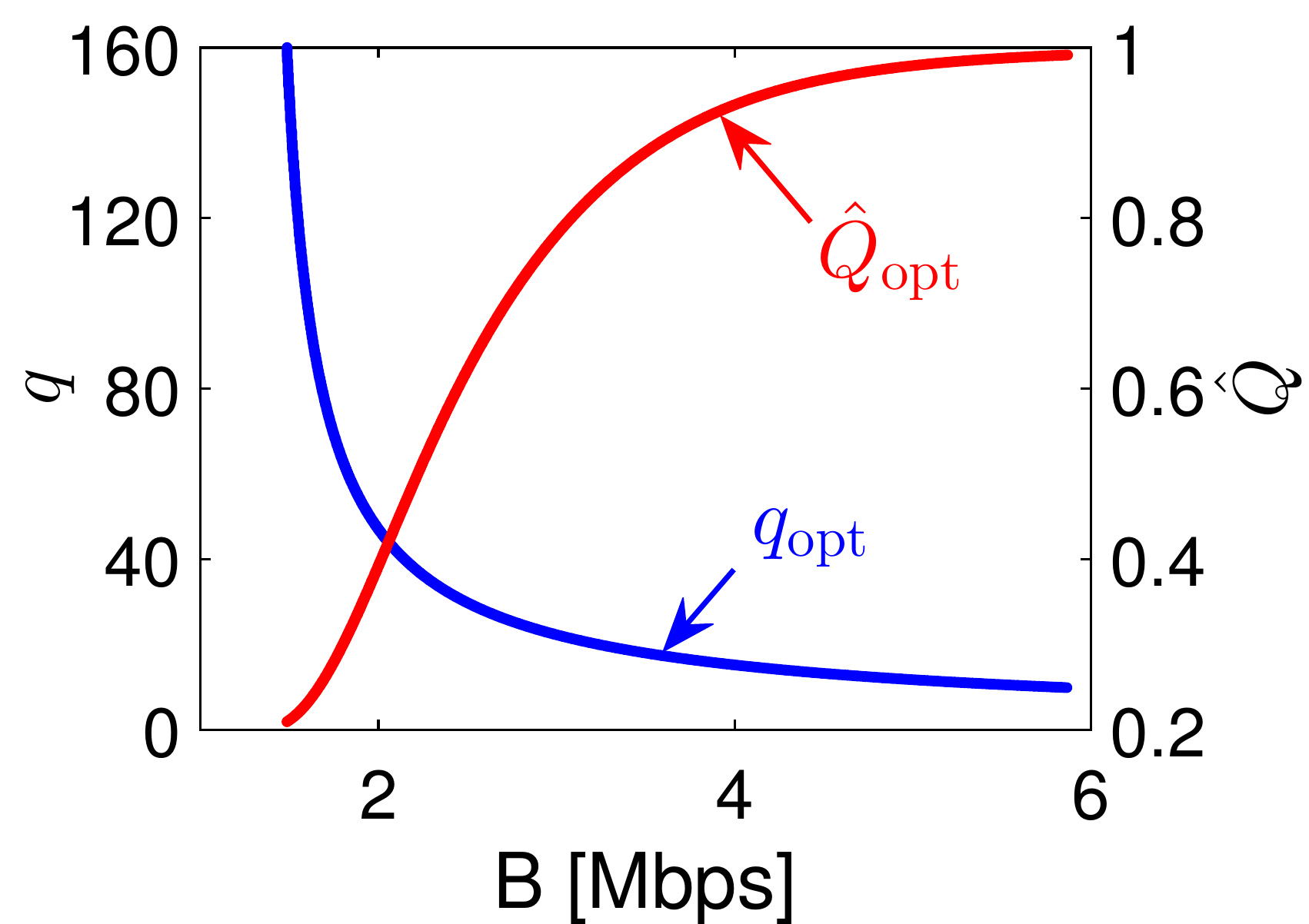}}
 \subfigure[Street2]{ \includegraphics[scale=0.23]{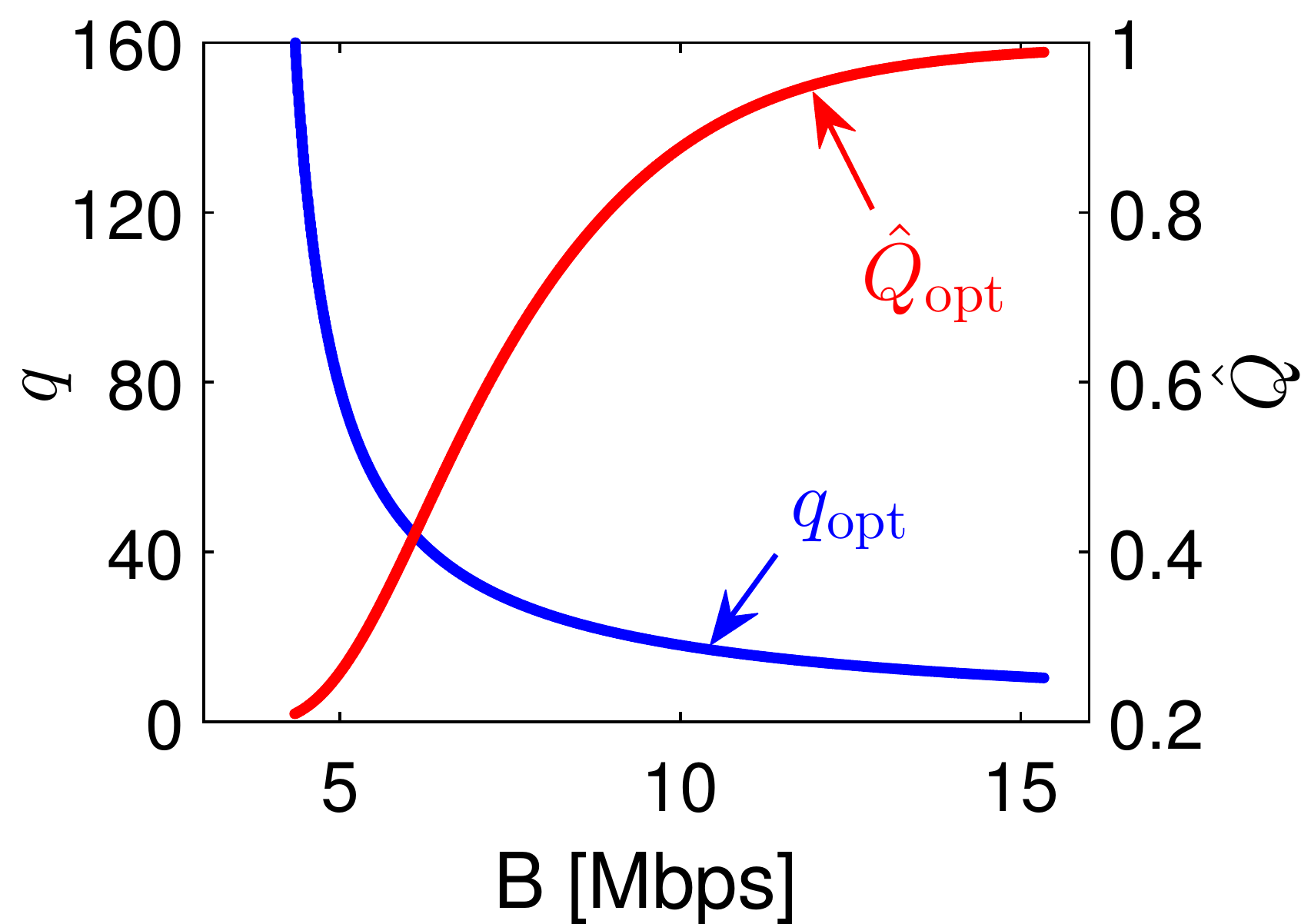}}
\caption{Optimal quantization stepsize $q_{\rm opt}$ and the corresponding normalized quality $\hat{Q}_{\rm opt}$ versus the constrained streaming bandwidth $B$ by assuming $q$ can take any continuous value within its range at $s_{\max}$. Left y-axis: $q$, right y-axis: $\hat{Q}$. The initialization duration $T$ of a segment is set to 5 seconds.}
\label{application_qQRopt}
\end{figure}

Thus, the optimization problem~\eqref{eq:OptimalProblem} can be transformed into
\begin{align}
 \max \limits_{\hat{q}}\quad
 &\hat{Q}=a(\hat{q})\cdot e^{-b(\hat{q})\cdot\frac{(R_i^{\rm FoV}+R_{\max}\cdot\hat{q}^\alpha)\cdot T}{B}}+1-a(\hat{q}),\label{eq:TransformedOptimization} \\
 s.t.\quad
 &R_i^{\rm FoV}+R_{\max}\cdot\hat{q}^{\alpha}\leq  B,\\
 &0.05\leq\hat{q}\leq1.
\end{align}

Mathematically, this function has an unique maximum $\hat{Q}$ (i.e., $\hat{Q}_{\rm opt}$) under a bandwidth $B$ target, by setting its derivative with respect to $\hat{q}$ to zero. Nevertheless, it is hard to solve this expression analytically. Thus, for any given $B$, we numerically determine the optimal quantization stepsize $q_{\rm opt}$ and the corresponding normalized maximum perceptual quality $\hat{Q}_{\rm opt}$ using~\eqref{eq:TransformedOptimization}. Fig.~\ref{application_qQRopt} shows $q_{\rm opt}$ and $\hat{Q}_{\rm opt}$ as functions of the rate constraint $B$. As expected, when $B$ increases, $q_{\rm opt}$ reduces and the achievable best quality $\hat{Q}_{\rm opt}$ increases continuously until the $q_{\rm opt}$ reaches to $q_{\min}$.

\begin{table*}[!tb]
\renewcommand\arraystretch{1.25}
\centering
\caption{The parameters $R_{\max}$, $\alpha$, $\beta$, the selected $R_i^{\rm FoV}$, and the BD-rate of each cross-validation videos} \label{Rmax-BDrate}%
 \begin{tabular}{|c|c|c|c|c|c|c|c|c|c|}
 \hline
                  & Balboa & PoleVault & Hangpai2 & Hangpai3 & Elephants2 & NewYork & Snowberg & Street2 & Ave.\\
 \hline
 $R_{\max}$ (Mbps) & 21.86  & 28.60     & 38.66    & 20.63    & 15.80      & 11.31   & 5.87     & 15.34 & 19.76 \\
 \hline
    $\alpha$      & 1.1621 & 1.7515    & 1.3522   & 1.2516   & 1.1220     & 1.0275  & 1.2349   & 1.1137 & 1.2519\\
 \hline
    $\beta$       & 0.8872 & 1.0864    & 0.9607   & 0.8880   & 1.0251     & 0.7898  & 1.0041   & 0.8537 & 0.9369\\
 \hline
 \hline
 $R_i^{\rm FoV}$ (Mbps) & 5.95 & 4.75     & 6.56     & 4.49     & 4.38       & 3.82    & 1.34     & 3.79 & 4.39\\
 \hline
    BD-rate       & -11.26\% & -3.32\%  & -6.60\%  & -8.19\%  & -14.57\%    & -16.30\%  & -3.10\%   & -11.54\% & -9.36\%\\
 \hline
 \end{tabular}
\end{table*}

\begin{figure}[b]
\centering
\includegraphics[width=0.8\linewidth]{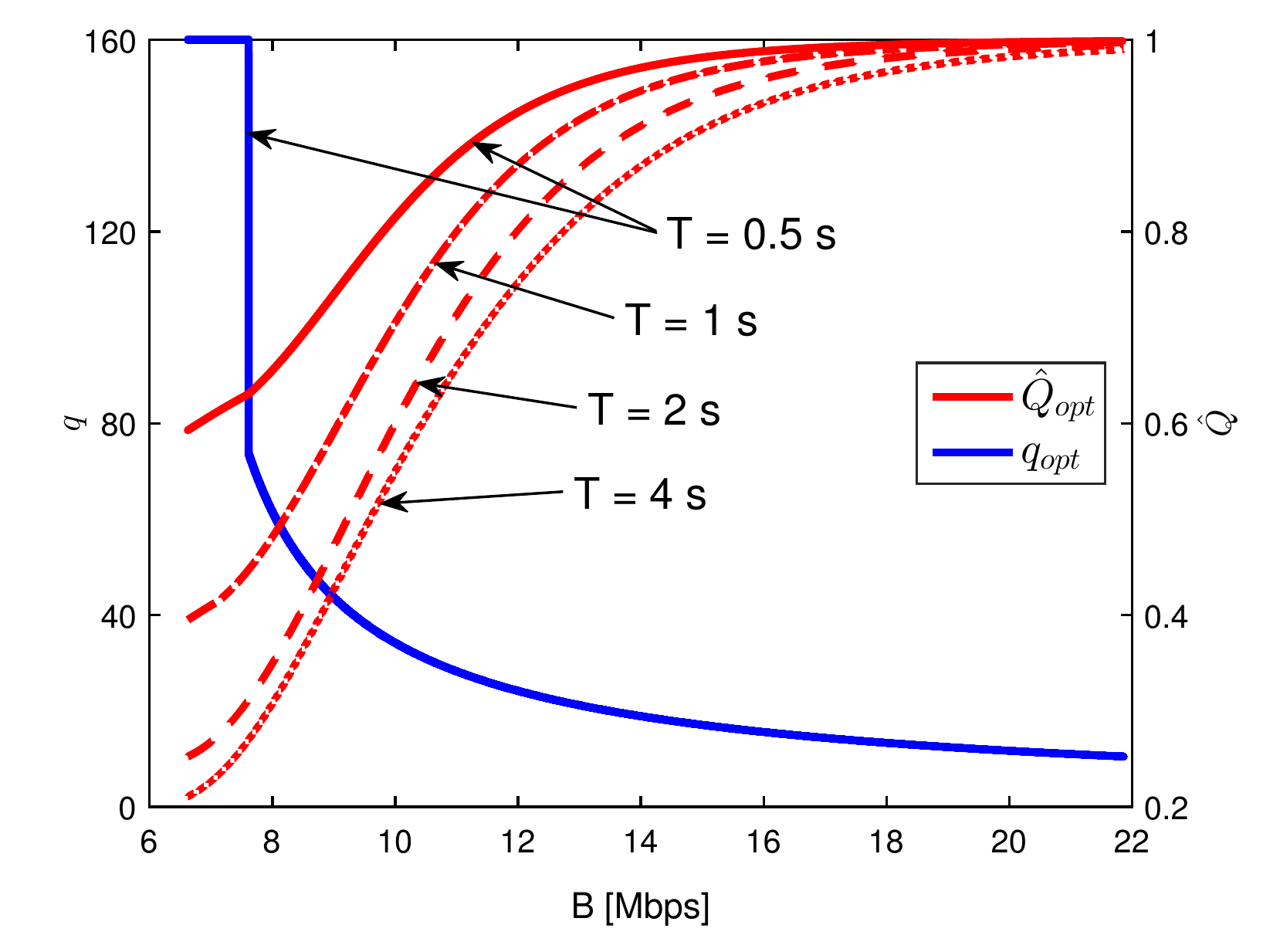}
\caption{Illustration of the effect of $T$ on bandwidth-quality optimization applying out model. Immersive video ``Balboa" is used as an example here.}
\label{application_Teffect}
\end{figure}

The initialization duration $T$ of a segment refers to the actual total amount of data that is required to be buffered
before rendering. It might take different values for different transport protocols (e.g., DASH versus MMT). $T = 5$ is exemplified
in Fig.~\ref{application_qQRopt}. To understand the impact of various $T$ on the bandwidth-quality optimization,
we use the ``Balboa" as an example, and plot the $\hat{Q}_{\rm opt}$ against the $B$ under different $T$, along with the corresponding $q_{\rm opt}$, as shown in Fig.~\ref{application_Teffect}. It can be found that as $T$ decreases, the optimal normalized
quality $\hat{Q}$ improves at same $B$, i.e., $\hat{Q}(T_1) > \hat{Q}(T_2)$ with $T_1 < T_2$.
Note that, if the $T$ is smaller enough, such as 0.5s, when $B$ is less than a threshold, the
optimal quantization stepsize $q_{\rm opt}$ stays at 160 (the highest value in practical codec), which means that
the best way in this case for adaptive FoV streaming is choosing the RL with worst quality to
minimize $\tau$, and the $\hat{Q}_{\rm opt}$ increases monotonically as the $B$ increases.

\begin{figure}[t]
\centering
 \subfigure[Balboa*]{ \includegraphics[scale=0.23]{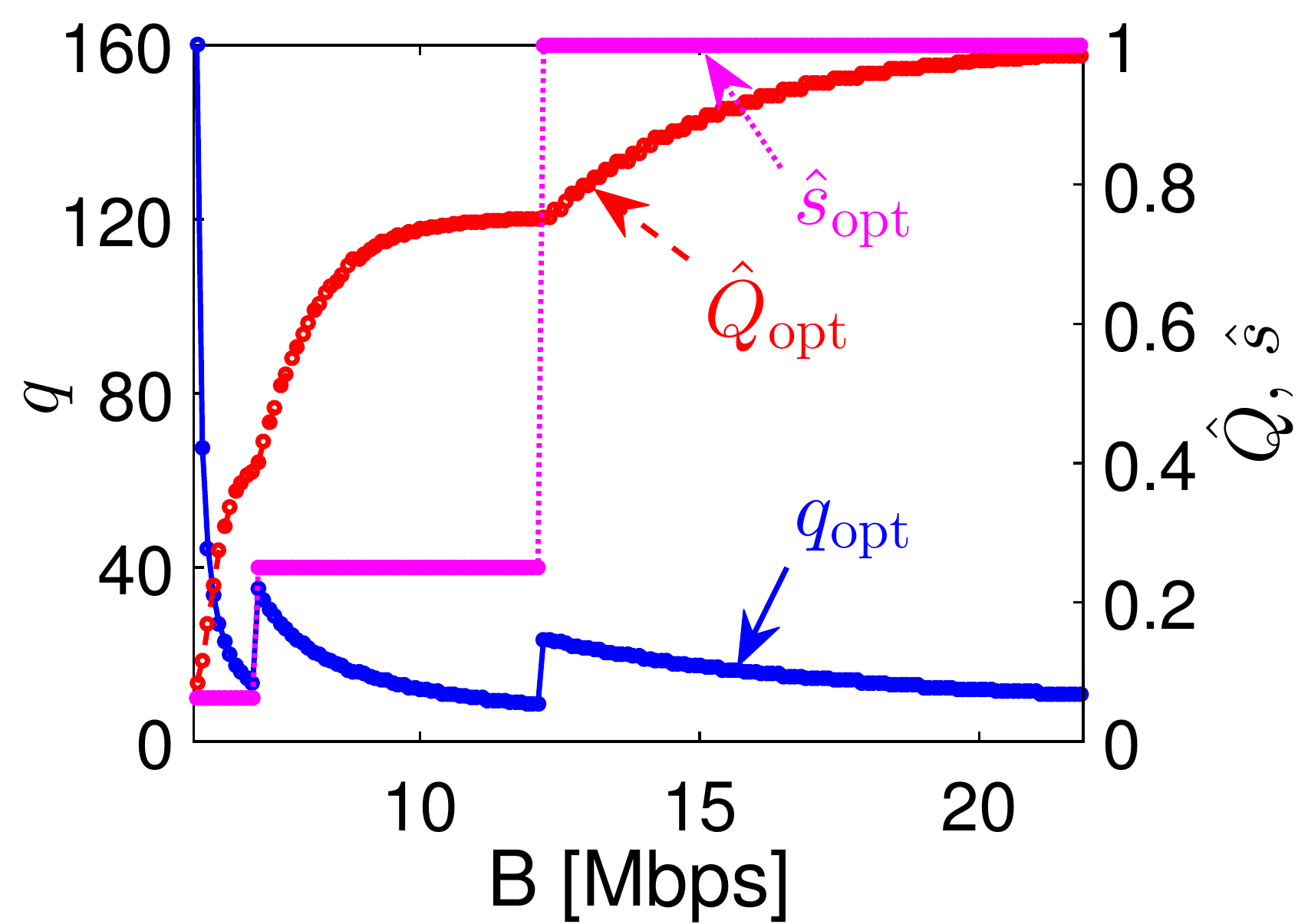}}
 \subfigure[PoleVault*]{ \includegraphics[scale=0.23]{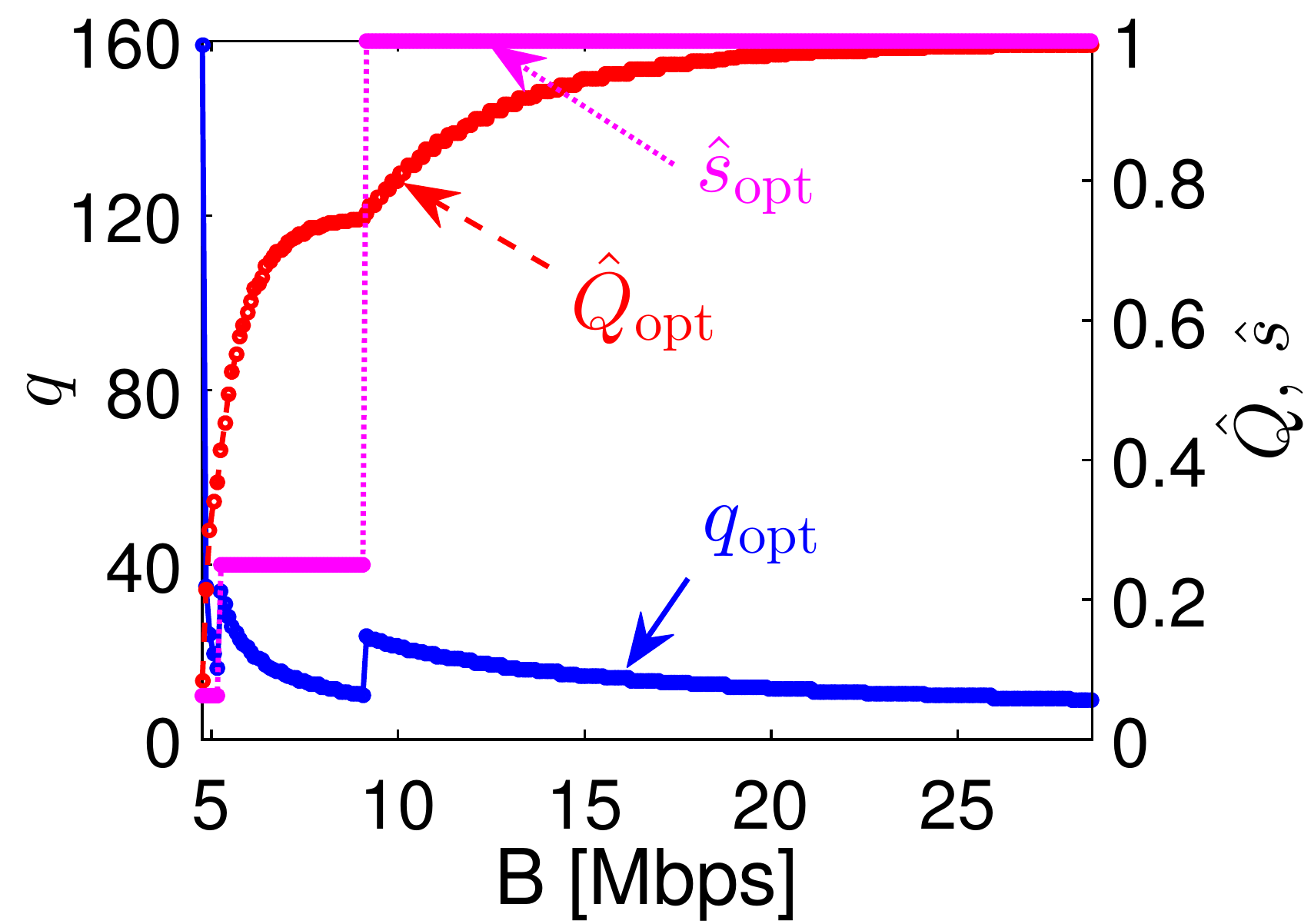}}
 \subfigure[Hangpai2\dag]{ \includegraphics[scale=0.23]{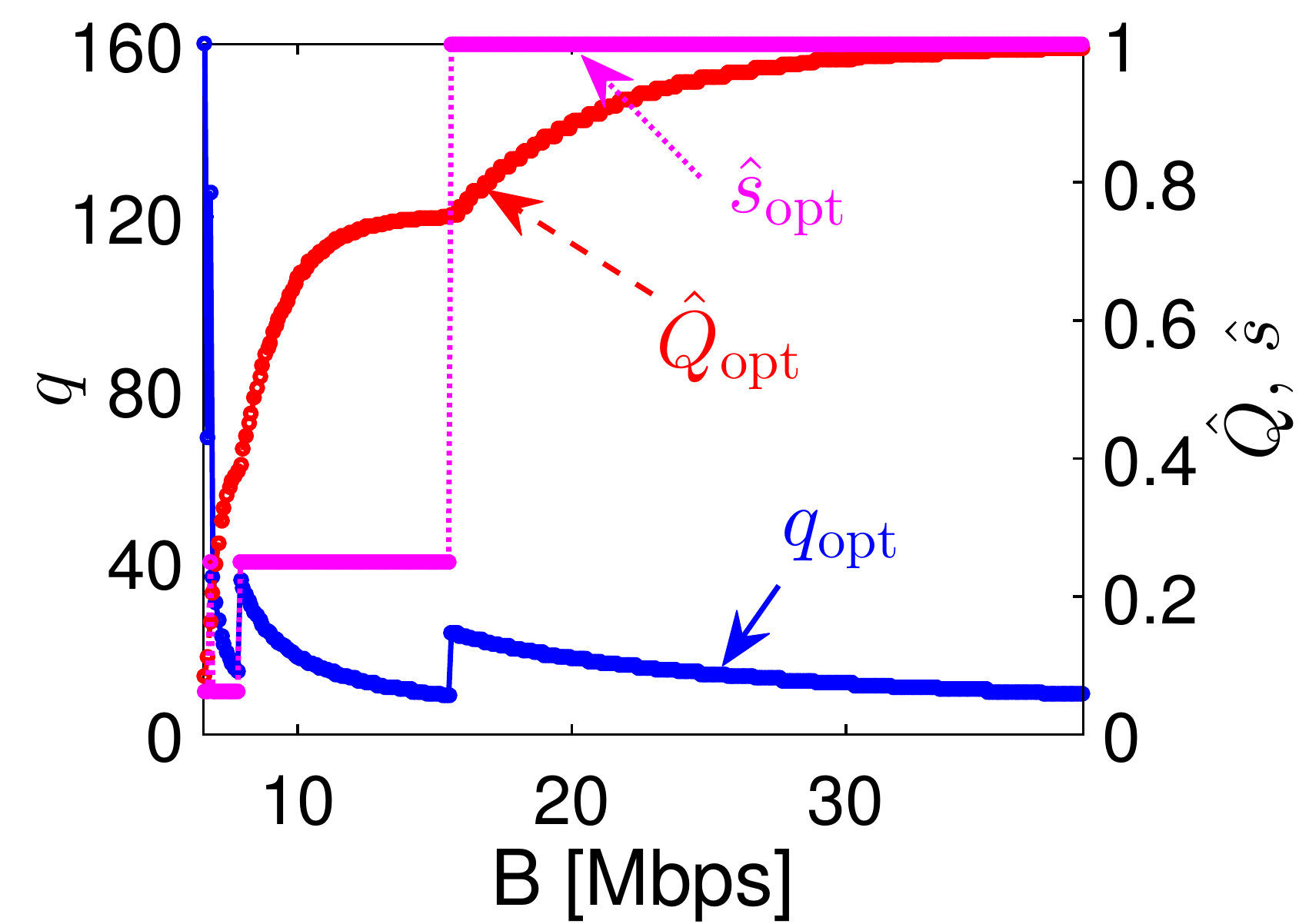}}
 \subfigure[Hangpai3\dag]{ \includegraphics[scale=0.23]{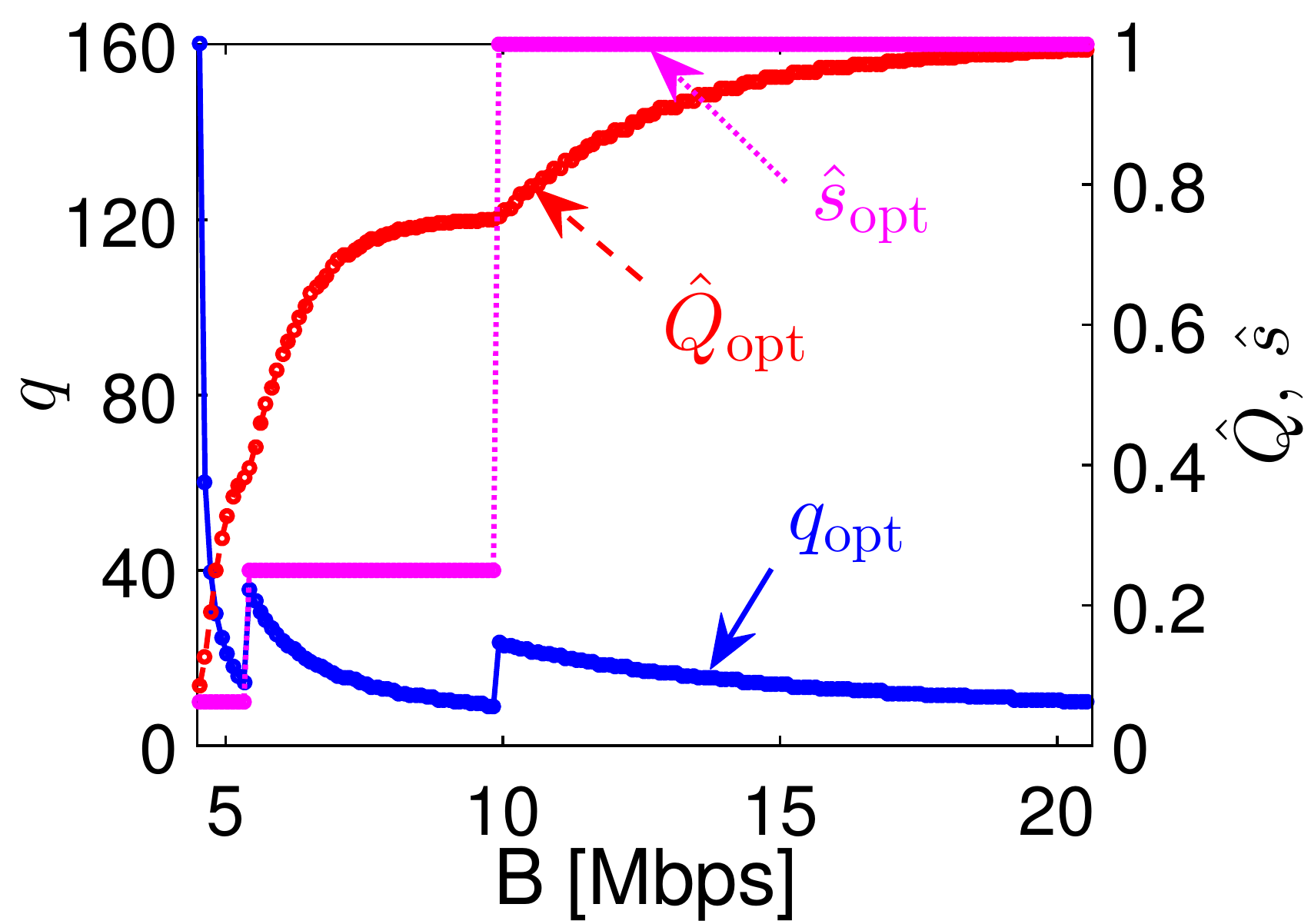}}
 \subfigure[Elephants2]{ \includegraphics[scale=0.23]{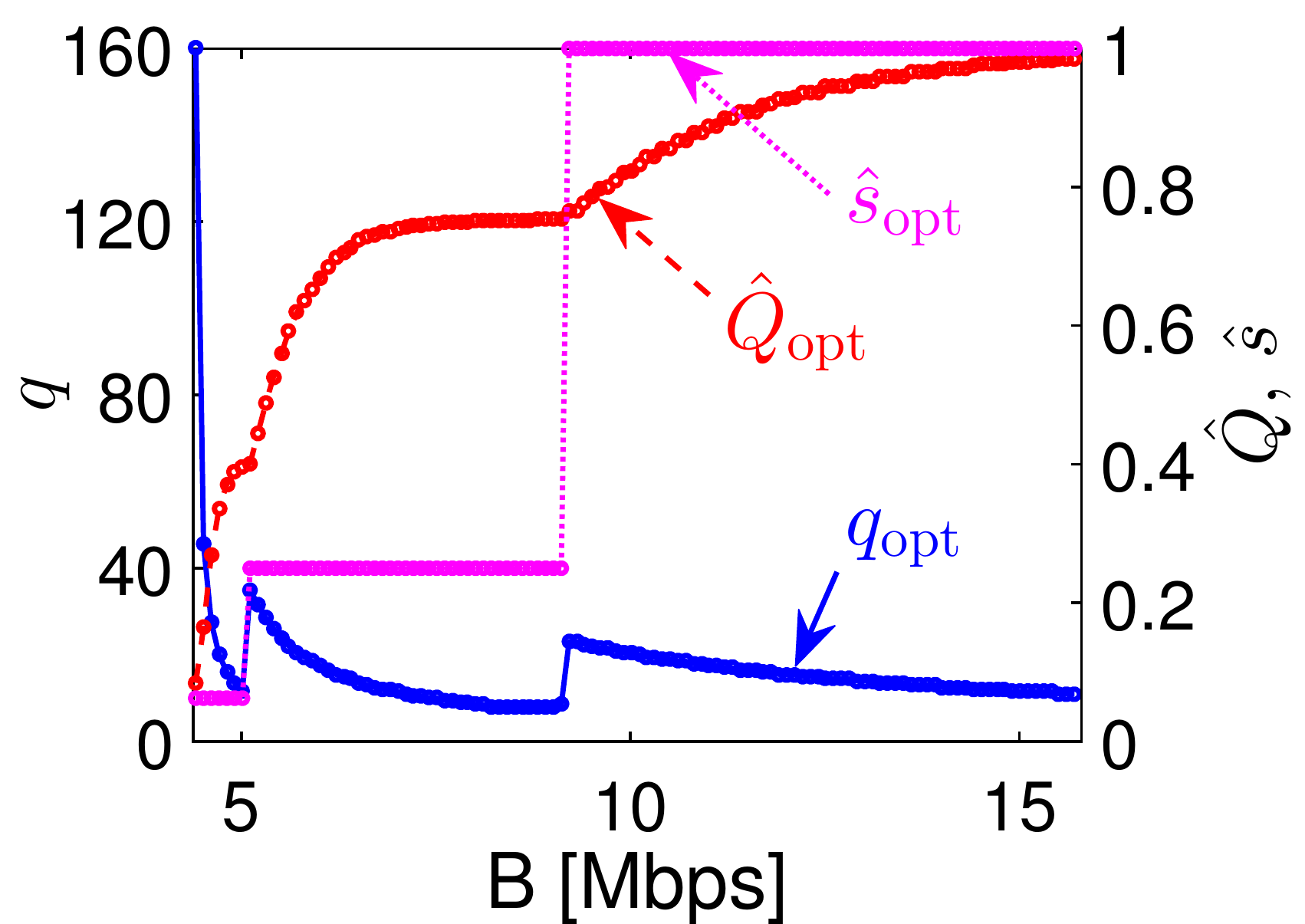}}
 \subfigure[NewYork]{ \includegraphics[scale=0.23]{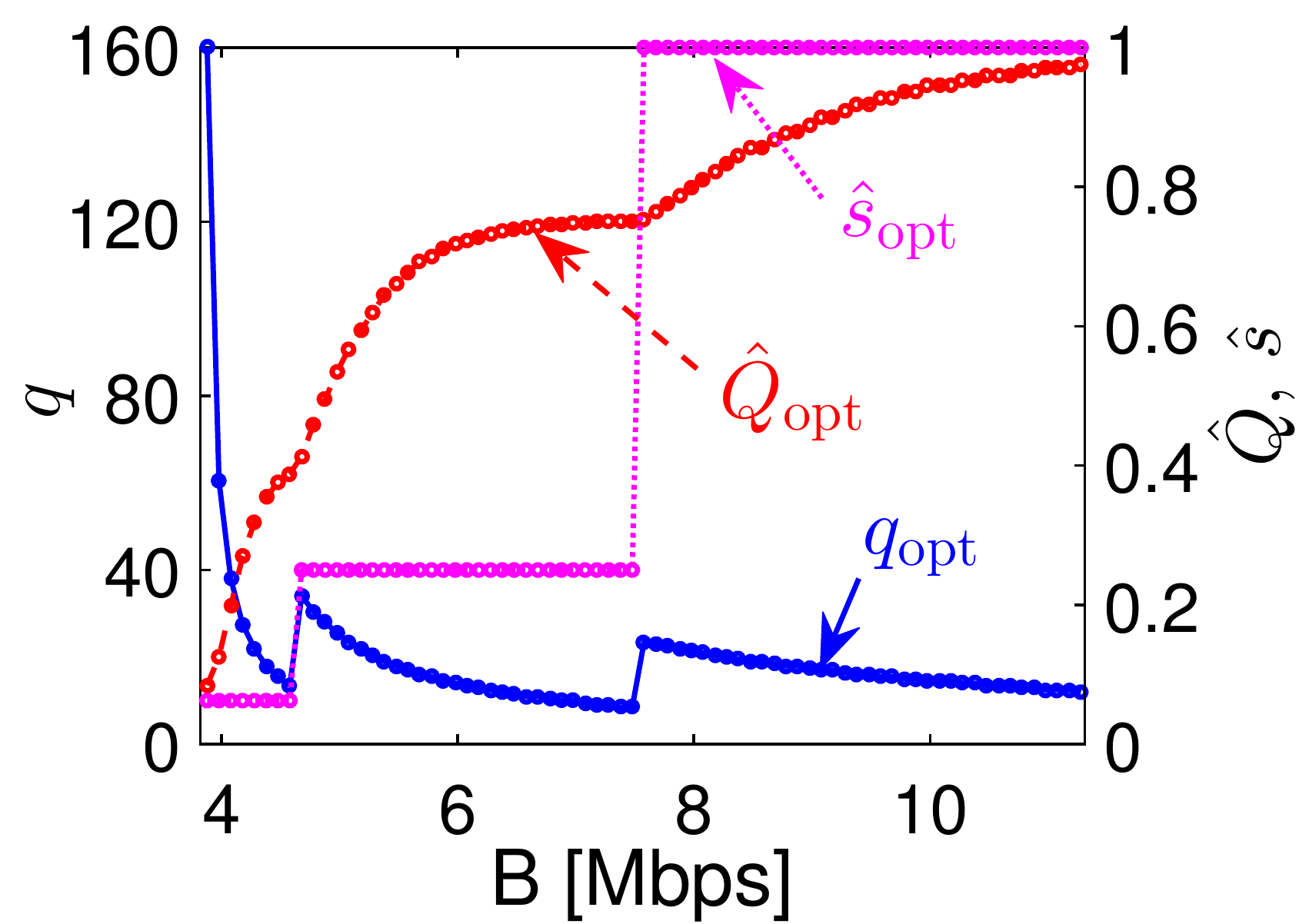}}
 \subfigure[Snowberg]{ \includegraphics[scale=0.23]{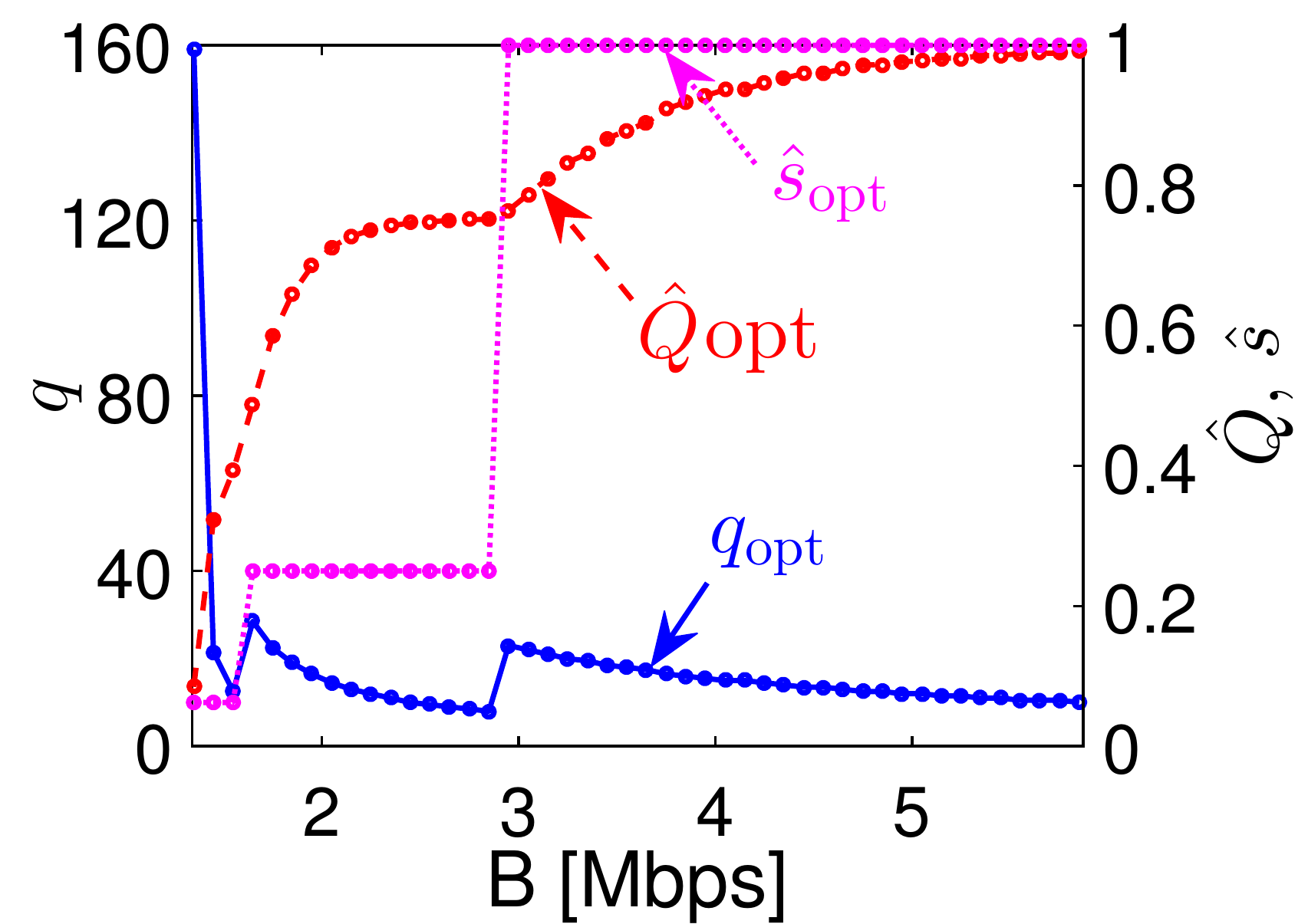}}
 \subfigure[Street2]{ \includegraphics[scale=0.23]{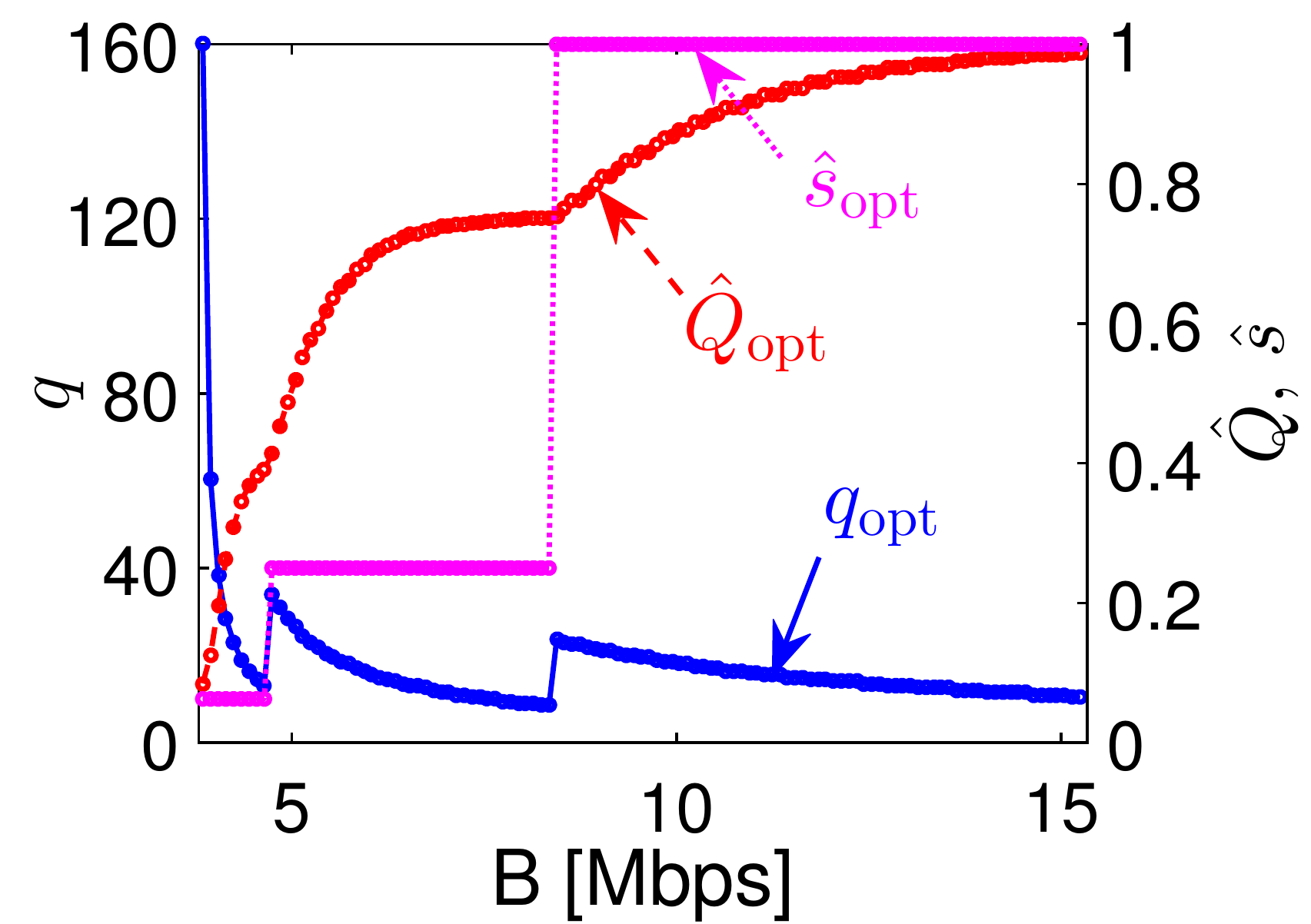}}
\caption{Optimal $q_{\rm opt}$, $\hat{s}_{\rm opt}$, and $\hat{Q}_{\rm opt}$ versus $B$ by assuming $\hat{s}$ can only take discrete values (i.e., 1/16, 1/4, 1), whereas $q$ can vary continuously. Left y-axis: $q$, right y-axis: $\hat{s}$, $\hat{Q}$. The initialization duration $T$ of a segment is set to 5 seconds.}
\label{application_qsQRopt}
\end{figure}

\begin{figure}[t]
\centering
 \subfigure[Balboa*]{ \includegraphics[scale=0.23]{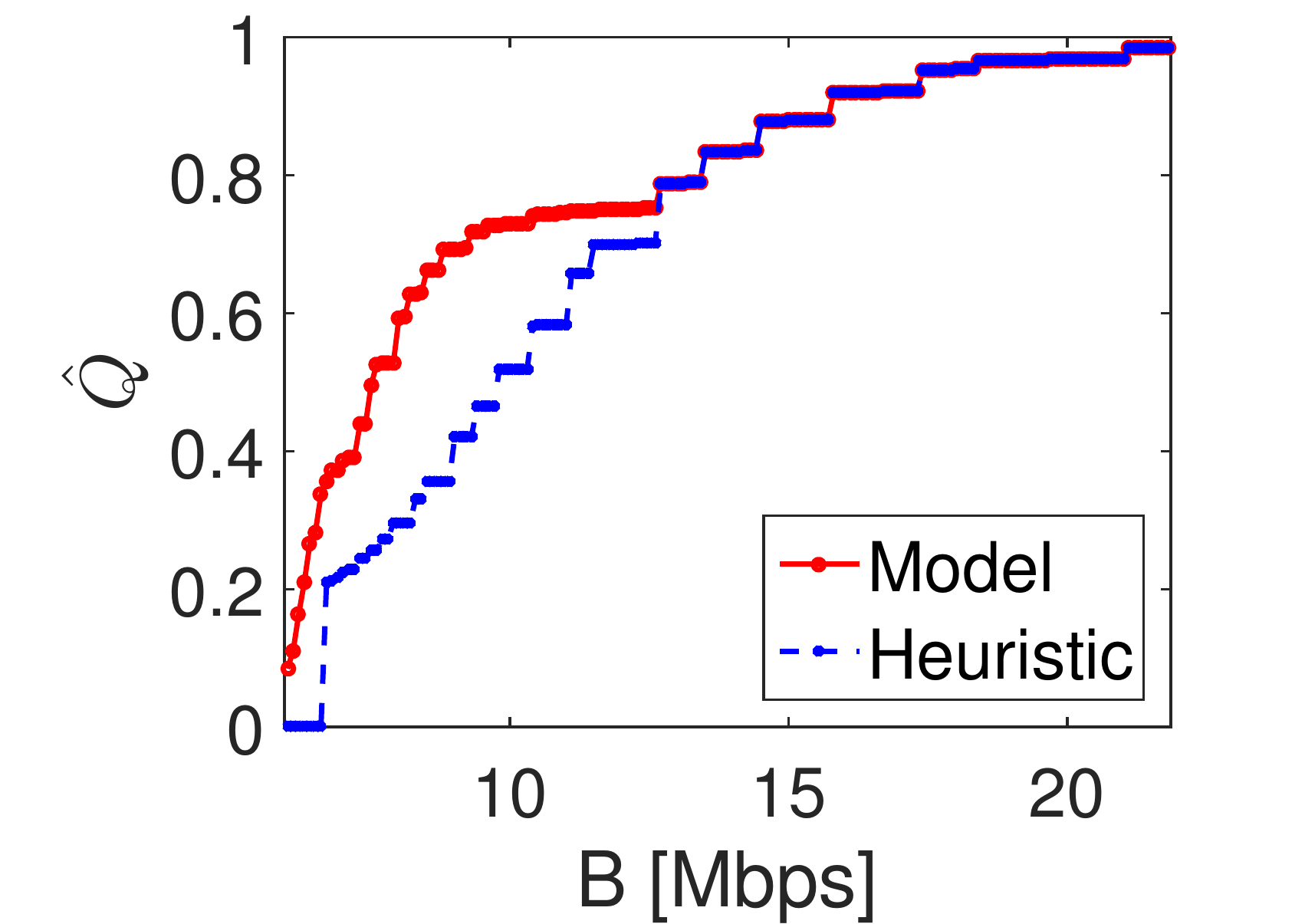}}
 \subfigure[PoleVault*]{ \includegraphics[scale=0.23]{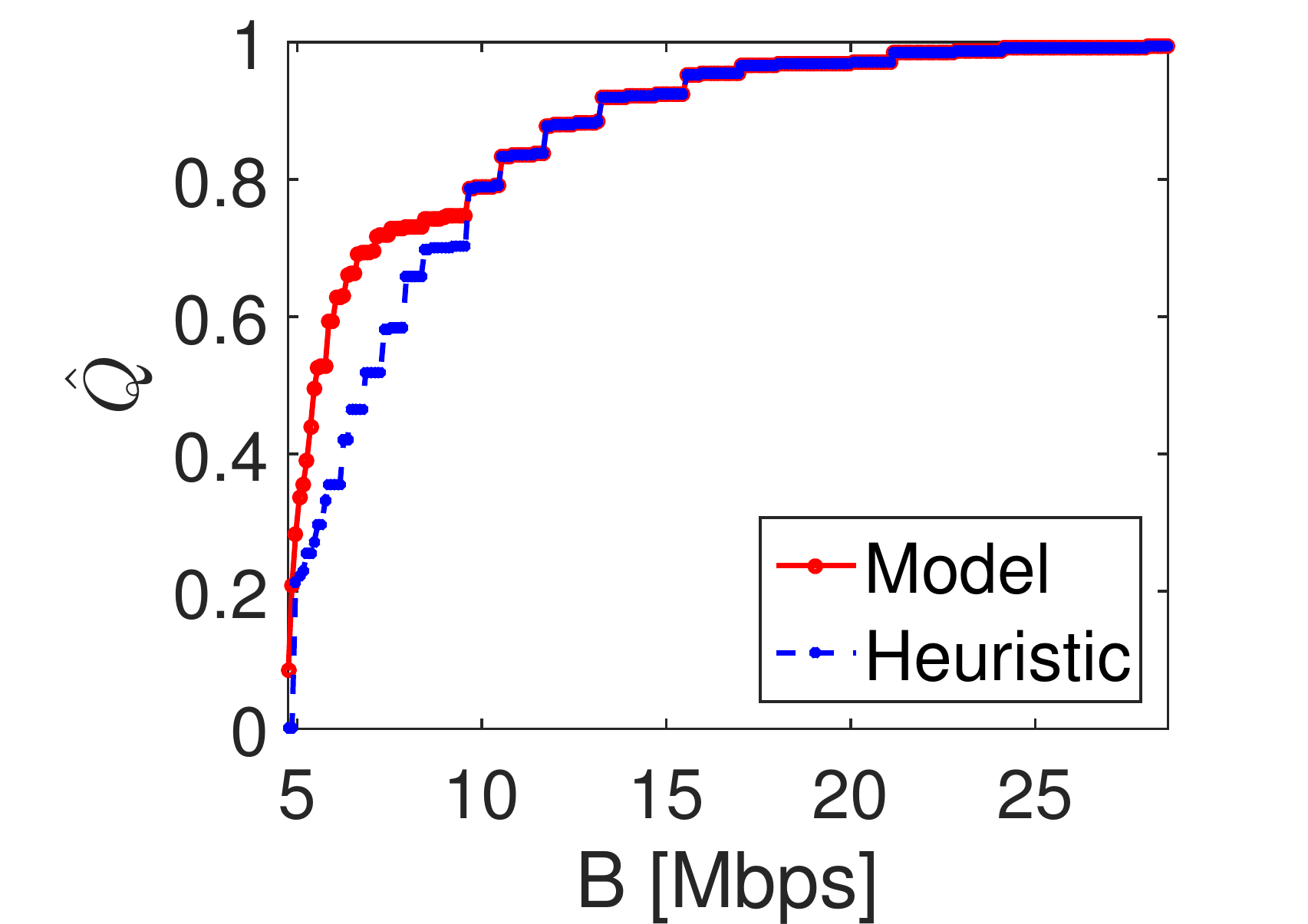}}
 \subfigure[Hangpai2\dag]{ \includegraphics[scale=0.23]{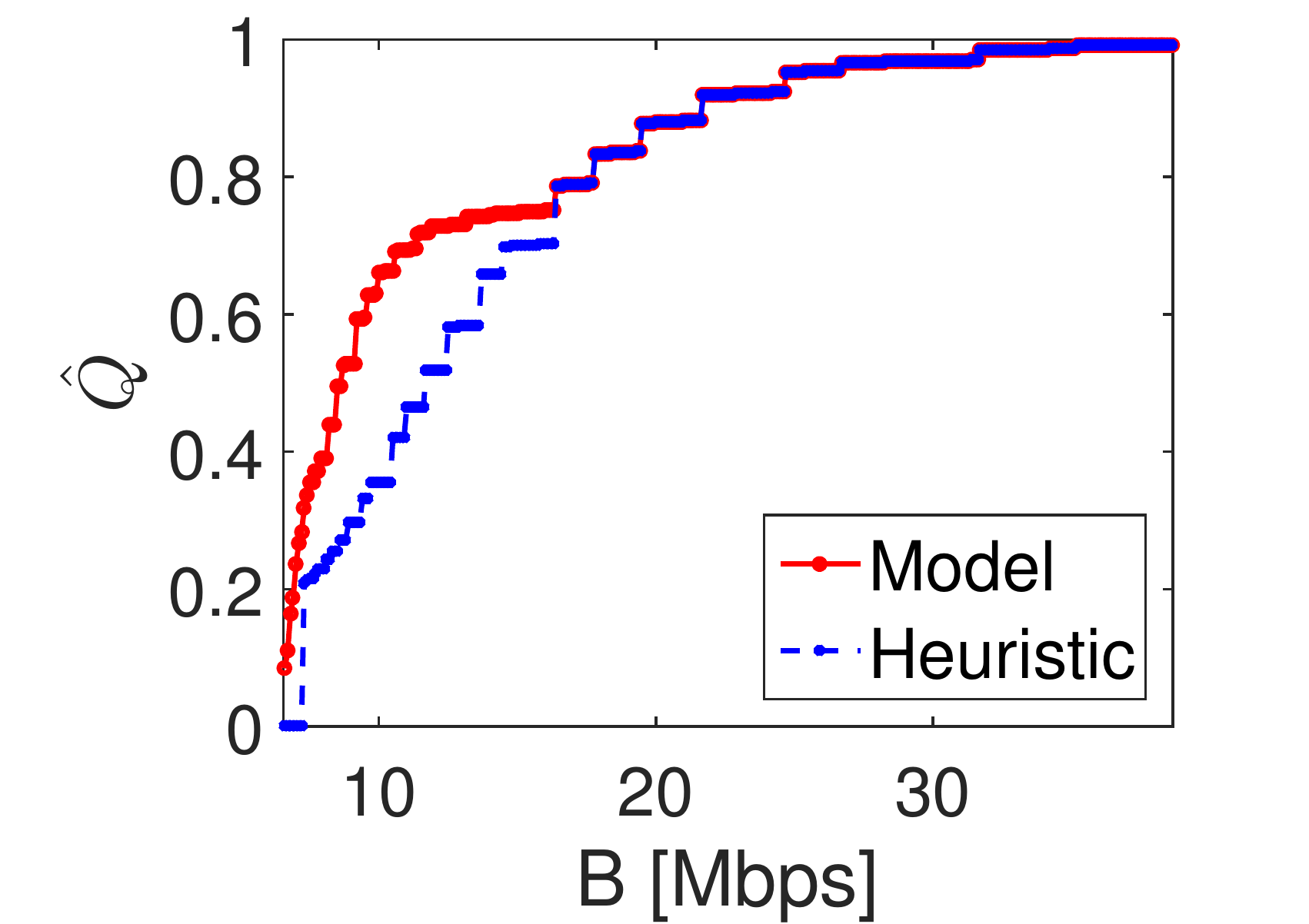}}
 \subfigure[Hangpai3\dag]{ \includegraphics[scale=0.23]{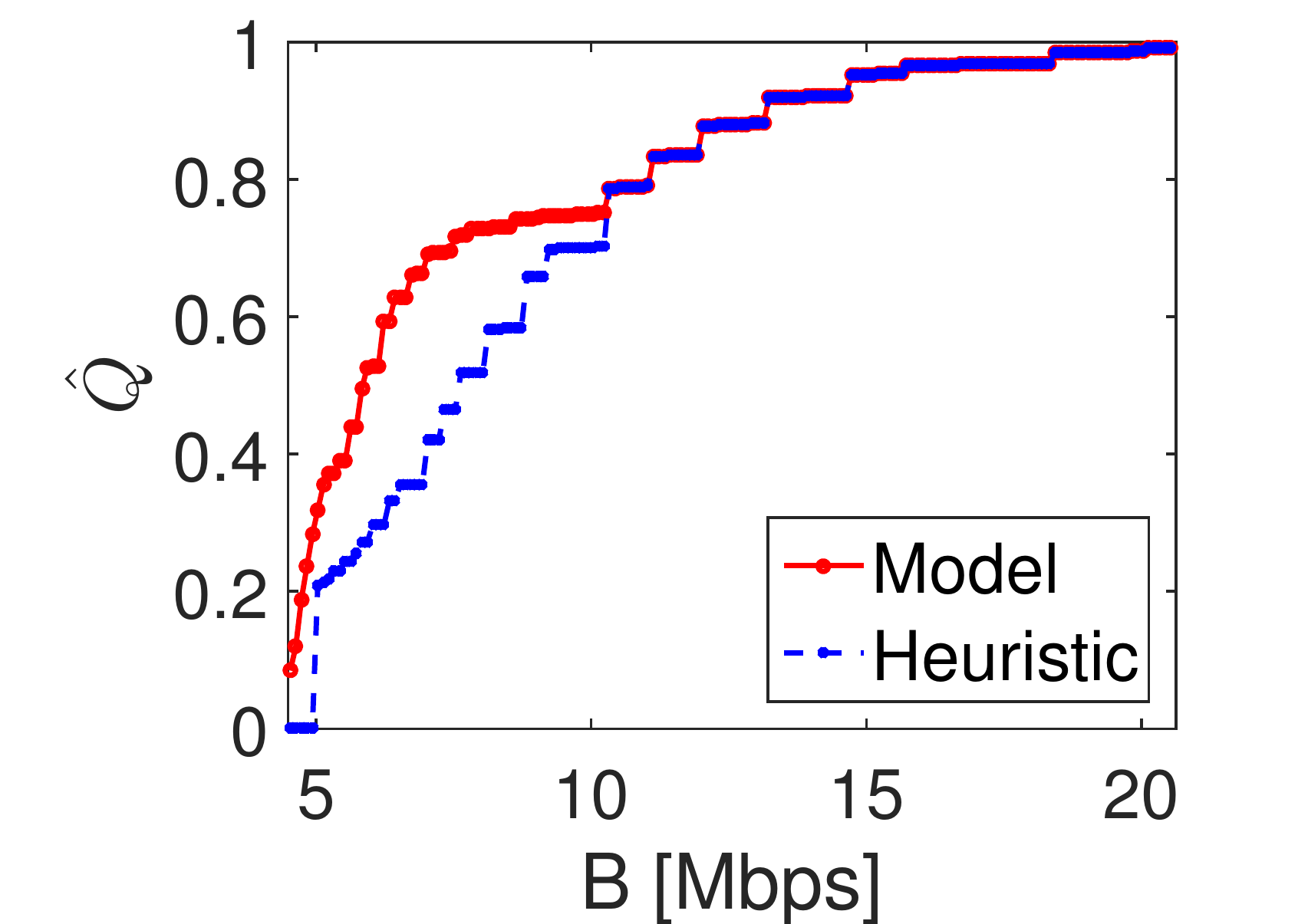}}
 \subfigure[Elephants2]{ \includegraphics[scale=0.23]{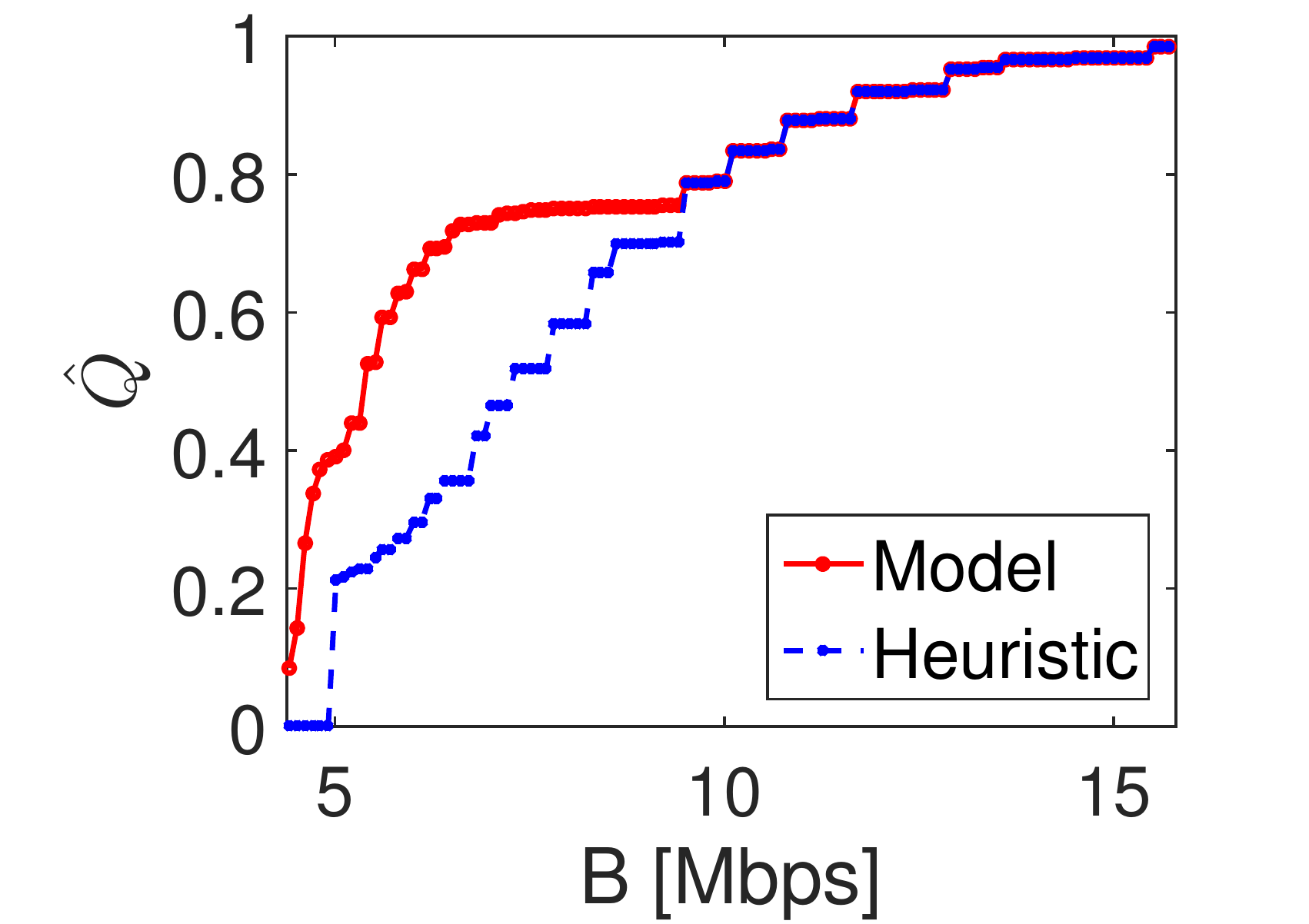}}
 \subfigure[NewYork]{ \includegraphics[scale=0.23]{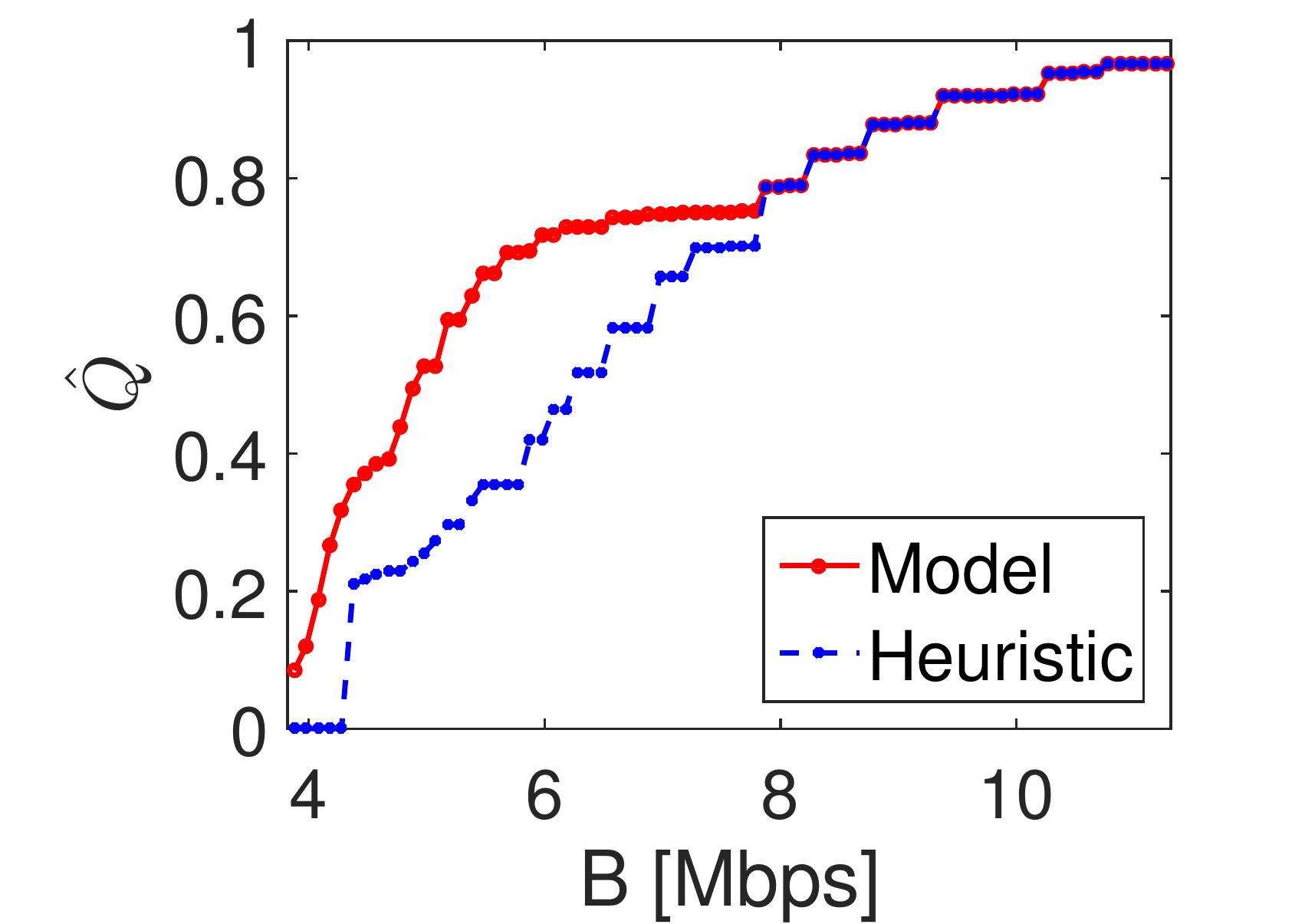}}
 \subfigure[Snowberg]{ \includegraphics[scale=0.23]{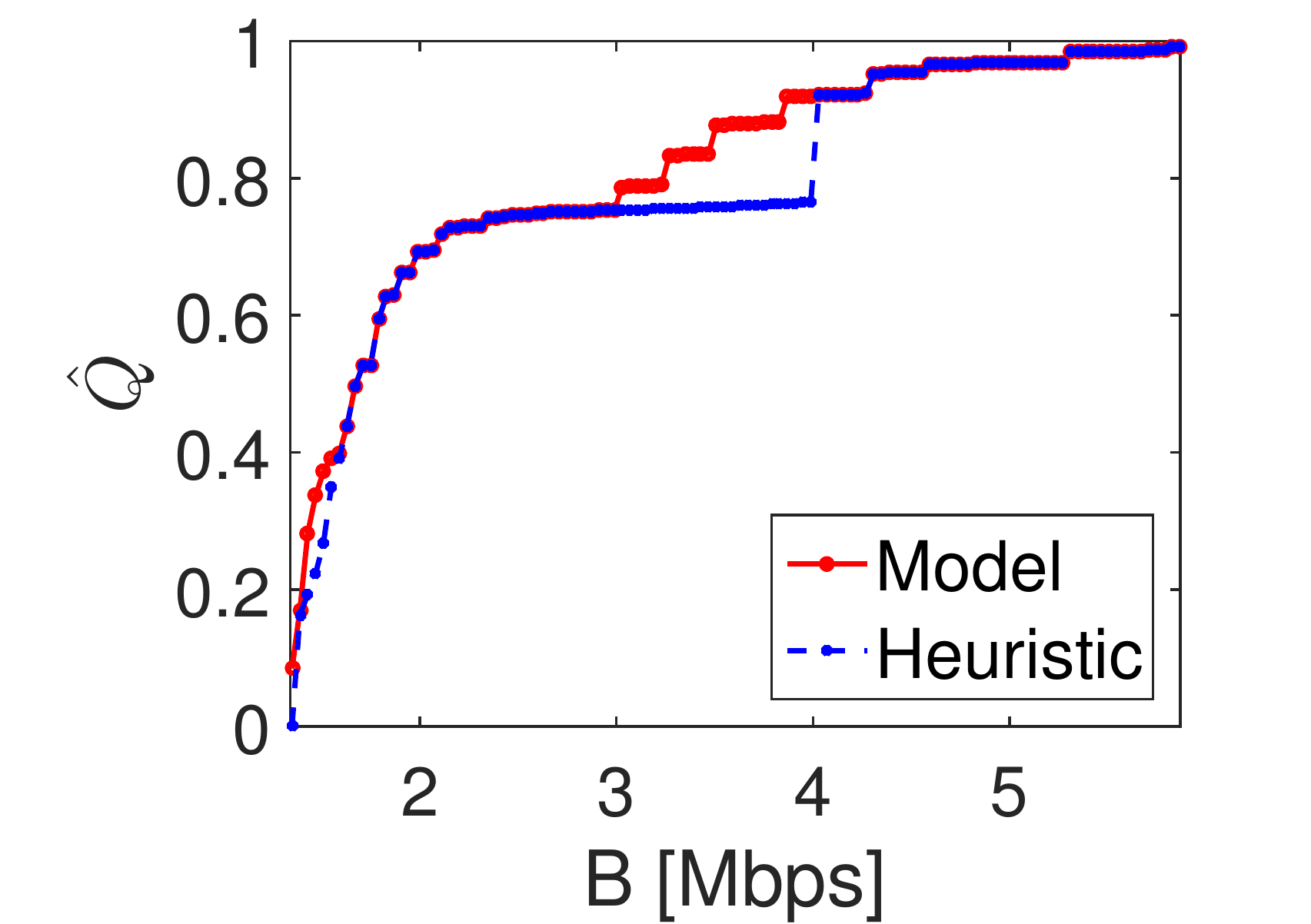}}
 \subfigure[Street2]{ \includegraphics[scale=0.23]{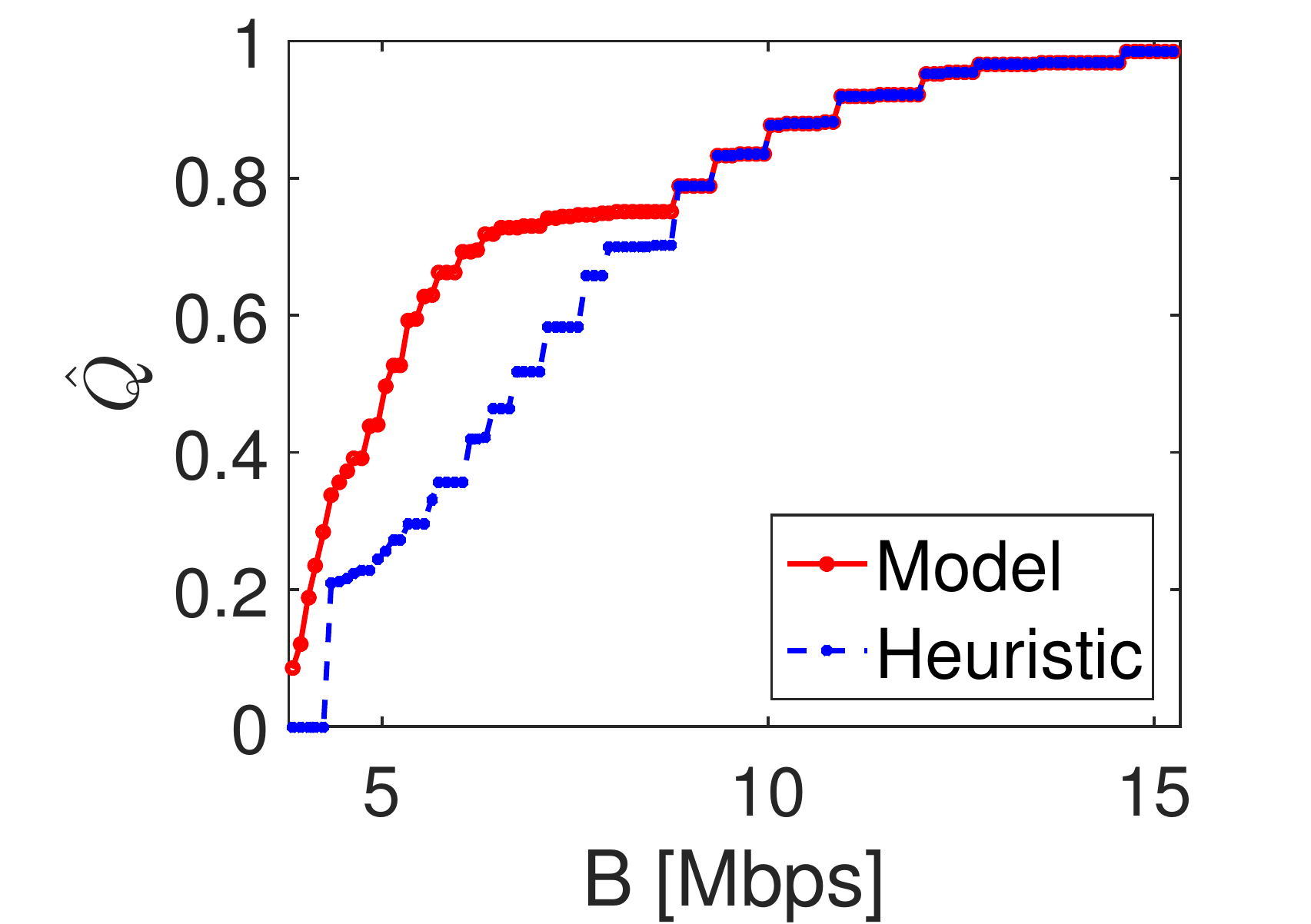}}
\caption{Perceptual quality comparison for our model based adaptation and a heuristic solution with corresponding ($q$, $\hat{s}$) annotated for each rate point. Overlapped points are annotated with one ($q$, $\hat{s}$). The initialization duration $T$ of a segment  is set to 5 seconds.}
\label{application_practical}
\end{figure}

\subsection{Optimal Solution Under Discrete $s$ and Continuous $q$}
\label{sec:DiscreteS}
In this section, we introduce the optimization when considering the variations from both spatial resolution and quantization.
Similar as the typical application, we enforce the spatial resolution with three possibilities, i.e., its native resolution with $\hat{s} = 1$,
and another two downscaled resolutions with $\hat{s} = 1/4$ and $\hat{s} = 1/16$ respectively. For the quantization stepsize,
we still assume that it can take continuous values.

To obtain the optimal solution under this scenario for each constrained bandwidth $B$, we determine the normalized quality values $\hat{Q}$ corresponding to all possible $\hat{s}$ and $q$ using \eqref{eq:overall_quality_model}, and choose the optimal combination of $\hat{s}$ and $q$ using \eqref{eq:OptimalProblem} that leads to the highest $\hat{Q}$.

In this case, the bitrate of a RL video is determined by
\begin{align}
R_{mi}^{\rm RL}(\hat{q},\hat{s}) = R_{\max}\cdot\hat{q}^\alpha\cdot\hat{s}^\beta. \label{eq:RateModel2}
\end{align}
The model parameter $\beta$ is also content dependent, as listed in Table~\ref{Rmax-BDrate} together with the content
dependent $\alpha$~\cite{Ma_RQModel,R-STAR}.

Fig.~\ref{application_qsQRopt} shows the optimal solutions versus an given $B$. Because the normalized spatial resolution $\hat{s}$ can only vary among discrete values (i.e., 1, 1/4 and 1/16), the corresponding optimal $q$ does not decrease monotonically with $B$. Instead, $q_{\rm opt}$ would jump to meet the rate constraint whenever $\hat{s}_{\rm opt}$ steps into the next higher value, and then decreases as $B$ increases while $\hat{s}_{\rm opt}$ keeps unchanged. Because of the limited options of the $\hat{s}$ and its interference with the $q$, when $q$ reaches $q_{\min}$ and $\hat{s}$ is still below $1$,  $\hat{s}$ will jump to the next level
as the $B$ continues to increase. More interestingly, when $\hat{s}$ stays with a smaller value,  $q$ decreases much faster and
so does the $\hat{Q}$ improve to the next step.

\subsection{Performance Evaluation for Practical Adaptation}
In practical streaming systems, each video is encoded with discrete quantization stepsize $q$ and discrete spatial resolution $s$ to produce various quality scales for adaptation in response to the time-varying network bandwidth. With our proposed model,
we could search for the combination of the $q$ and $\hat{s}$, so as to maximize the $\hat{Q}$, given any $B$ constraint.
Since we only need to evaluate at most 3$\times$51 = 153 possibilities, even with the brutal-force search, the overhead is
still neglectable on current GigaHertz computing chip.

We compare the quality achieved with our model (``Model") to an alternative solution (``Heuristic") where the streaming system chooses the $\hat{s}$ based on the constrained rate, following an heuristic rule: $\hat{s}=1/16$ when $B<1$ Mbps, $\hat{s}=1/4$ when $1\leq B<4$ Mbps, $\hat{s}=1$ when $B\geq4$ Mbps. The comparison results are shown in Fig.~\ref{application_practical}, where we can find that our model based solution can obtain obviously higher subjective quality at most rate points than the ``Heuristic". Additionally, due to the $\hat{s}$ chosen by the heuristic rule are sometimes same as the model derived
optimal $\hat{s}$ for a constrained rate, both solutions can achieve the same optimal quality $\hat{Q}$. Inspired by the
BD-Rate (Bjontegaard Delta Rate)~\cite{BDRate} widely adopted in evaluating the coding efficiency for video compression,
we measure the gain of the normalized perceptual quality $\hat{Q}$ with respect to the network bandwidth $B$. Ideally, we calculate the area difference between two curves.
Meanwhile, the efficiency of the ``Model'' based solution against the ``Heuristic '' solution is listed Table~\ref{Rmax-BDrate} for
each cross-validation video, with about 9.36\% BD-Rate improvement.

\section{Concluding Remarks} \label{sec:conclusion}
In this paper, we investigated the perceptual impact of the quality variations when performing the refinement within a period of time $\tau$. Usually, quality variation is determined by adapting the quantization stepsize $q$ and spatial resolution $s$. Therefore, the overall model was represented by a product of two exponential functions where each of them detailed the quantization and spatial resolution impact on the perceptual quality in terms of the refinement duration, respectively.
We finally reached at a closed-form model, producing very accurate quality estimation, even with all parameters fixed. We then randomly selected another set of data to perform the model cross-validation, where results had demonstrated the high accuracy of our model with both Pearson and Spearman's rank correlations close to 0.98 for all test videos.
Moreover, we provided the comprehensive examples to devise our proposed model to guide the FoV adaptive streaming, so as to maximize the perceptual quality under the rate constraint. With our model, we could resolve the optimization problem analytically. In a practical application, we could achieve about 9.36\% BD-Rate improvement in comparison with the heuristic solution where the spatial resolution is often predetermined by the network bandwidth.

We will focus on the FoV adaptation prediction and apply the proposed model in practical immersive streaming system to further evaluate the efficiency of our proposed model. We also would like to make our data public accessible at \url{http://vision.nju.edu.cn/immersive_video}.


\section*{Acknowledgement}
The authors would like to thank the human subjects participating in this experiment and the reviewers for their valuable comments.

\bibliographystyle{IEEEtran}
\bibliography{FoVAdaptModel_all}

\begin{thebibliography}{10}
\providecommand{\url}[1]{#1}
\csname url@samestyle\endcsname
\providecommand{\newblock}{\relax}
\providecommand{\bibinfo}[2]{#2}
\providecommand{\BIBentrySTDinterwordspacing}{\spaceskip=0pt\relax}
\providecommand{\BIBentryALTinterwordstretchfactor}{4}
\providecommand{\BIBentryALTinterwordspacing}{\spaceskip=\fontdimen2\font plus
\BIBentryALTinterwordstretchfactor\fontdimen3\font minus
  \fontdimen4\font\relax}
\providecommand{\BIBforeignlanguage}[2]{{%
\expandafter\ifx\csname l@#1\endcsname\relax
\typeout{** WARNING: IEEEtran.bst: No hyphenation pattern has been}%
\typeout{** loaded for the language `#1'. Using the pattern for}%
\typeout{** the default language instead.}%
\else
\language=\csname l@#1\endcsname
\fi
#2}}
\providecommand{\BIBdecl}{\relax}
\BIBdecl

\bibitem{david_giga}
D.~J. Brady, M.~E. Gehm, R.~A. Stack, D.~L. Marks, D.~S. Kittle, D.~R. Golish,
  E.~M. Vera, and S.~D. Feller, ``Multiscale gigapixel photography,''
  \emph{Nature}, vol. 486, no. 7403, pp. 386--389, Jun. 2012.

\bibitem{H264}
T.~Wiegand, G.-J. Sullivan, G.~Bjontegaard, and A.~Luthra, ``Overview of the
  {H.264/AVC} video coding standard,'' \emph{IEEE Trans. Circuits and Systems
  for Video Techn.}, vol.~13, no.~7, pp. 560--576, Jul. 2003.

\bibitem{jov_peri_vision_review}
H.~Strasburger, I.~Rentschler, and M.~Juettner, ``Peripheral vision and pattern
  recognition: {A} review,'' \emph{Journal of Vision}, vol.~11, no.~13, pp.
  1--82, May 2011.

\bibitem{DynamicVR}
Y.~Hu, S.~Xie, Y.~Xu, and J.~Sun, ``Dynamic {VR} live streaming over {MMT},''
  in \emph{Proc. IEEE Int. Symp. Broadband Multimedia Systems and Broadcasting
  (BMSB'17)}, Jun. 2017.

\bibitem{OptimizingCell}
F.~Qian, L.~Ji, B.~Han, and V.~Gopalakrishnan, ``Optimizing 360 video delivery
  over cellular networks,'' in \emph{Proc. ACM Workshop All Things Cellular:
  Operations, Applications and Challenges (ATC'16)}, Oct. 2016.

\bibitem{Two-tier360}
F.~Duanmu, E.~Kurdoglu, S.~A. Hosseini, Y.~Liu, and Y.~Wang, ``Prioritized
  buffer control in two-tier 360 video streaming,'' in \emph{Proc. ACM Workshop
  Virtual Reality and Augmented Reality Netw. (VR/AR Netw.'17)}, Aug. 2017, pp.
  13--18.

\bibitem{HEVCsr}
A.~Zare, A.~Aminlou, M.~M. Hannuksela, and M.~Gabbouj, ``{HEVC-compliant}
  tile-based streaming of panoramic video for virtual reality applications,''
  in \emph{Proc. ACM Multimedia Conf. (MM'16)}, Oct. 2016, pp. 601--605.

\bibitem{Viewportadaptive}
X.~Corbillon, G.~Simon, A.~Devlic, and J.~Chakareski, ``Viewport-adaptive
  navigable 360-degree video delivery,'' in \emph{Proc. IEEE Int. Conf. Commun.
  (ICC'17)}, May 2017.

\bibitem{AdaptiveStreaming}
M.~Graf, C.~Timmerer, and C.~Mueller, ``Towards bandwidth efficient adaptive
  streaming of omnidirectional video over {HTTP}: Design, implementation, and
  evaluation,'' in \emph{Proc. ACM Multimedia Systems Conf. (MMSys'17)}, Jun.
  2017, pp. 261--271.

\bibitem{UltraWideView}
R.~Ju, J.~He, F.~Sun, J.~Li, F.~Li, J.~Zhu, and L.~Han, ``Ultra wide view based
  panoramic {VR} streaming,'' in \emph{Proc. ACM Workshop Virtual Reality and
  Augmented Reality Netw. (VR/AR Netw.'17)}, Aug. 2017, pp. 19--23.

\bibitem{QuantizationScheme}
S.~Ma, W.~Gao, D.~Zhao, and Y.~Lu, ``A study on the quantization scheme in
  {H.264/AVC} and its application to rate control,'' in \emph{Proc. Pacific Rim
  Conf. Multimedia (PCM'04)}, Nov. 2004, pp. 192--199.

\bibitem{pv_mobileQSTAR}
Y.~Xue, Y.-F. Ou, Z.~Ma, and Y.~Wang, ``Perceptual video quality assessment on
  a mobile platform considering both spatial resolution and quantization
  artifacts,'' in \emph{Proc. IEEE Int. Packet Video Workshop (PVW'10)}, Dec.
  2010, pp. 201--208.

\bibitem{Ma_RQModel}
Z.~Ma, M.~Xu, Y.-F. Ou, and Y.~Wang, ``Modeling of rate and perceptual quality
  of compressed video as functions of frame rate and quantization stepsize and
  its applications,'' \emph{IEEE Trans. Circuits and Systems for Video Techn.},
  vol.~22, no.~5, pp. 671--682, May 2012.

\bibitem{VQMTQ}
Y.-F. Ou, Z.~Ma, T.~Liu, and Y.~Wang, ``Perceptual quality assessment of video
  considering both frame rate and quantization artifacts,'' \emph{IEEE Trans.
  Circuits and Systems for Video Techn.}, vol.~21, no.~3, pp. 286--298, Jun.
  2011.

\bibitem{RoIcrop}
N.~Quang Minh~Khiem, G.~Ravindra, A.~Carlier, and W.~T. Ooi, ``Supporting
  zoomable video streams with dynamic region-of-interest cropping,'' in
  \emph{Proc. ACM Multimedia Systems Conf. (MMSys'10)}, Feb. 2010, pp.
  259--270.

\bibitem{SpatialSeg}
R.~Van~Brandenburg, O.~Niamut, M.~Prins, and H.~Stokking, ``Spatial
  segmentation for immersive media delivery,'' in \emph{Proc. IEEE Int. Conf.
  Intelligence in Next Generation Networks (ICIN'11)}, Oct. 2011, pp. 151--156.

\bibitem{InteractiveOmni}
P.~Rondao~Alface, J.-F. Macq, and N.~Verzijp, ``Interactive omnidirectional
  video delivery: A bandwidth-effective approach,'' \emph{Bell Labs Technical
  Journal}, vol.~16, no.~4, pp. 135--147, Mar. 2012.

\bibitem{LiveOmni}
D.~Ochi, Y.~Kunita, A.~Kameda, A.~Kojima, and S.~Iwaki, ``Live streaming system
  for omnidirectional video,'' in \emph{Proc. IEEE Virtual Reality (VR'15)},
  Mar. 2015, pp. 349--350.

\bibitem{TwoTierSystem}
F.~Duanmu, E.~Kurdoglu, Y.~Liu, and Y.~Wang, ``View direction and bandwidth
  adaptive 360 degree video streaming using a two-tier system,'' in \emph{Proc.
  IEEE Int. Symp. Circuits and Systems (ISCAS'17)}, May 2017.

\bibitem{ShootingMove}
Y.~Bao, H.~Wu, T.~Zhang, A.~A. Ramli, and X.~Liu, ``Shooting a moving target:
  Motion-prediction-based transmission for 360-degree videos,'' in \emph{Proc.
  IEEE Int. Conf. Big Data (Big Data'16)}, Dec. 2016, pp. 1161--1170.

\bibitem{SaliencyinVR}
\BIBentryALTinterwordspacing
V.~Sitzmann, A.~Serrano, A.~Pavel, M.~Agrawala, D.~Gutierrez, and G.~Wetzstein,
  ``Saliency in {VR}: How do people explore virtual environments?''
  \emph{Computer Science}, vol. abs/1612.04335, Dec. 2016. [Online]. Available:
  \url{https://arxiv.org/abs/1612.04335}
\BIBentrySTDinterwordspacing

\bibitem{FixationPrediction}
C.-L. Fan, J.~Lee, W.-C. Lo, C.-Y. Huang, K.-T. Chen, and C.-H. Hsu, ``Fixation
  prediction for 360° video streaming in head-mounted virtual reality,'' in
  \emph{Proc. ACM Workshop Netw. and Operating Systems Support for Digital
  Audio and Video (NOSSDAV'17)}, Jun. 2017, pp. 67--72.

\bibitem{imageModel_rongbin}
R.~Zhou, M.~Huang, S.~Tan, L.~Zhang, D.~Chen, J.~Wu, T.~Yue, X.~Cao, and Z.~Ma,
  ``Modeling the impact of spatial resolutions on perceptual quality of
  immersive image/video,'' in \emph{Proc. IEEE Int. Conf. 3D Imaging
  (IC3D'16)}, Dec. 2016.

\bibitem{Xiaokai_TIP}
M.~Huang, Q.~Shen, R.~Zhou, Z.~Ma, X.~Cao, and A.~C. Bovik, ``Modeling the
  perceptual quality of immersive images rendered on head mounted displays,''
  \emph{submitted to IEEE Trans. Image Processing}, 2017.

\bibitem{peiyao_VCIP}
P.~Guo, Q.~Shen, M.~Huang, R.~Zhou, X.~Cao, and Z.~Ma, ``Modeling peripheral
  vision impact on perceptual quality of immersive images,'' in \emph{Proc.
  IEEE Int. Conf. Visual Communications and Image Processing (VCIP'17)}, Dec.
  2017.

\bibitem{peiyao_TIP}
P.~Guo, Q.~Shen, Y.~Meng, L.~Xu, X.~Cao, and Z.~Ma, ``Perceptual quality
  assessment of immersive images considering peripheral vision impact,''
  \emph{submitted to IEEE Trans. Image Processing}, 2017.

\bibitem{shaowei_ISCAS}
S.~Xie, Y.~Xu, Q.~Qian, Q.~Shen, Z.~Ma, and W.~Zhang, ``Modeling the perceptual
  impact of viewport adaptation for immersive video,'' in \emph{submitted to
  Proc. IEEE Int. Symp. Circuits and Systems (ISCAS'18)}, 2018.

\bibitem{JVET}
\BIBentryALTinterwordspacing
{JVET}. [Online]. Available: \url{ftp://ftp.ient.rwth-aachen.de}
\BIBentrySTDinterwordspacing

\bibitem{YouTube}
\BIBentryALTinterwordspacing
{YouTube}. [Online]. Available: \url{https://www.youtube.com/}
\BIBentrySTDinterwordspacing

\bibitem{HMVR}
E.~Upenik and T.~Ebrahimi, ``A simple method to obtain visual attention data in
  head mounted virtual reality,'' in \emph{Proc. IEEE Int. Conf. Multimedia and
  Expo Workshops (ICMEW'17)}, Jul. 2017, pp. 73--78.

\bibitem{ratecontrol}
H.~Hu, Z.~Ma, and Y.~Wang, ``Optimization of spatial, temporal and amplitude
  resolution for rate-constrained video coding and scalable video adaptation,''
  in \emph{Proc. IEEE Int. Conf. Image Processing (ICIP'12)}, Oct. 2012, pp.
  717--720.

\bibitem{BT500}
{Rec. ITU-R BT.500-11}, ``Methodology for the subjective assessment of the
  quality of television pictures,'' 2002.

\bibitem{num_scaling_SPIE}
A.~M. van Dijk, J.-B. Martens, and A.~B. Watson, ``Quality assessment of coded
  images using numberical category scaling,'' in \emph{Proc. SPIE Advanced
  Image and Video Commun. and Storage Techn.}, Feb. 1995, pp. 90--101.

\bibitem{spatial_tempral}
Y.-F. Ou, Y.~Xue, and Y.~Wang, ``{Q-STAR}: a perceptual video quality model
  considering impact of spatial, temporal, and amplitude resolutions,''
  \emph{IEEE Trans. Image Processing}, vol.~23, no.~26, pp. 2473--2486, Jan.
  2014.

\bibitem{VRU}
\BIBentryALTinterwordspacing
{VRU}. [Online]. Available: \url{ftp://vru@47.93.196.121/VRUProposals/seq/}
\BIBentrySTDinterwordspacing

\bibitem{Pearson_coeff}
\BIBentryALTinterwordspacing
{Pearson correlation coefficients}. [Online]. Available:
  \url{https://en.wikipedia.org/wiki/Pearson\_correlation\_coefficient}
\BIBentrySTDinterwordspacing

\bibitem{DASH}
S.~Lederer, C.~Mueller, and C.~Timmerer, ``Dynamic adaptive streaming over
  {HTTP} dataset,'' in \emph{Proc. ACM Multimedia Systems Conf. (MMSys'12)},
  Feb. 2012, pp. 89--94.

\bibitem{MMT}
Y.~Lim, K.~Park, Y.~L. Jin, S.~Aoki, and G.~Fernando, ``{MMT}: An emerging mpeg
  standard for multimedia delivery over the internet,'' \emph{IEEE Multimedia},
  vol.~20, no.~1, pp. 80--85, Feb. 2013.

\bibitem{HEVC}
K.~Misra, A.~Segall, M.~Horowitz, S.~Xu, A.~Fuldseth, and M.~Zhou, ``An
  overview of tiles in {HEVC},'' \emph{IEEE Journal of Selected Topics in
  Signal Processing}, vol.~7, no.~6, pp. 969--977, Jun. 2013.

\bibitem{R-STAR}
Z.~Ma, F.~C.~A. Fernandes, and Y.~Wang, ``Analytical rate model for compressed
  video considering impacts of spatial, temporal and amplitude resolutions,''
  in \emph{Proc. IEEE Int. Conf. Multimedia and Expo Workshops (ICMEW'13)},
  Jul. 2013.

\bibitem{Ma2012Rate}
\BIBentryALTinterwordspacing
Z.~Ma, H.~Hu, M.~Xu, and Y.~Wang, ``Rate model for compressed video considering
  impacts of spatial, temporal and amplitude resolutions and its applications
  for video coding and adaptation,'' \emph{Computer Science}, vol.
  abs/1206.2625, Jun. 2012. [Online]. Available:
  \url{https://arxiv.org/abs/1206.2625}
\BIBentrySTDinterwordspacing

\bibitem{BDRate}
G.~Bjontegaard, ``Calculation of average {PSNR} differences between {RD}
  curves,'' in \emph{Proc. ITU-T Q.6/SG16 VCEG}, Apr. 2001.

\end{thebibliography}

\end{document}